\documentclass[useAMS,usenatbib]{mn2e}

\newcommand{\etal}{{\em et al. }}
\usepackage{longtable,booktabs}
\usepackage[dvips]{graphicx}
\usepackage{epsfig}
\usepackage{multicol}
\usepackage[tight]{subfigure}

\title[A Comprehensive Study of the Radio Properties of Brightest Cluster Galaxies]{A Comprehensive Study of the Radio Properties of Brightest Cluster Galaxies}
\author[M. T. Hogan \etal ]{M. T. Hogan$^{1,2,3}$\thanks{E-mail:
m4hogan@uwaterloo.ca (MTH)},  A. C. Edge$^{1}$, J. Hlavacek-Larrondo$^{4,5,6}$, K. J. B. Grainge$^{7}$,  S. L. Hamer$^{8}$, \newauthor  E. K. Mahony$^{9}$, H. R. Russell$^{10}$, A. C. Fabian$^{10}$, B. R. McNamara$^{2,3}$ and R. J. Wilman$^{1}$  \\
\\
$^{1}$Centre for Extragalactic Astronomy, Department of Physics, Durham University, Durham, DH1 3LE UK \\
$^{2}$Department of Physics and Astronomy, University of Waterloo, Waterloo, ON, N2L 3G1, Canada \\
$^{3}$Perimeter Institute for Theoretical Physics, Waterloo, ON, N2L 2Y5, Canada \\
$^{4}$Kavli Institute for Particle Astrophysics and Cosmology, Stanford University, 382 Via Pueblo Mall, Stanford, CA 94305-4060, USA \\
$^{5}$Department of Physics, Stanford University, 452 Lomita Mall, Stanford, CA 94305-4085, USA \\
$^{6}$D\'{e}partement de physique, Universit\'{e} de Montr\'{e}al, C.P. 6128, succ. centre-ville, Montr\'{e}al (Qu\'{e}bec) H3C 3J7, Canada \\
$^{7}$Jodrell Bank Centre for Astrophysics, School of Physics and Astronomy, The University of Manchester, Manchester, M13 9PL, UK \\
$^{8}$LERMA Observatoire de Paris, CNRS, Universite Paris-Sud, 91405 Orsay, France \\
$^{9}$ASTRON, Netherlands Institute for Radio Astronomy, Postbus 2, NL-7990 AA Dwingeloo, the Netherlands \\
$^{10}$Institute of Astronomy, Madingley Road, Cambridge CB3 0HA, UK}

\begin{document}

\date{Accepted 2015 July 7.  Received 2015 June 21; in original form 2014 September 12}

\pagerange{\pageref{firstpage}--\pageref{lastpage}} \pubyear{2011}

\maketitle

\label{firstpage}

\begin{abstract}
We examine the radio properties of the Brightest Cluster Galaxies (BCGs) in a large sample of X-ray selected galaxy clusters comprising the Brightest Cluster Sample (BCS), the extended BCS (eBCS) and ROSAT-ESO Flux Limited X-ray (REFLEX) cluster catalogues.  We have multi-frequency radio observations of the BCG using a variety of data from the Australia Telescope Compact Array (ATCA), {\it Jansky} Very Large Array (VLA) and Very Long Baseline Array (VLBA) telescopes. The radio spectral energy distributions (SEDs) of these objects are decomposed into a component attributed to on-going accretion by the active galactic nuclei (AGN) that we refer to as `the core', and a more diffuse, ageing component we refer to as the `non-core'.  These BCGs are matched to previous studies to determine whether they exhibit emission lines (principally H-$\alpha$), indicative of the presence of a strong cooling cluster core. We consider how the radio properties of the BCGs vary with cluster environmental factors. Line emitting BCGs are shown to generally host more powerful radio sources, exhibiting the presence of a strong, distinguishable core component in about 60\% of cases. This core component more strongly correlates with the BCG's [OIII]5007\AA{} line emission.  For BCGs in line-emitting clusters, the X-ray cavity power correlates with both the extended and core radio emission, suggestive of steady fueling of the AGN over bubble-rise timescales in these clusters.
\end{abstract}


\section{Introduction}

Brightest Cluster Galaxies occupy an important parameter space within the study of galaxy formation and evolution.  Residing spatially and dynamically at the centres of galaxy clusters, they are the most massive galaxies observed and their growth and evolution is intrinsically linked to that of the host cluster \cite[e.g.][]{Tremaine90, Lin07}.  There is general consensus that a strong evolutionary connection exists between galaxies and the super-massive black holes (SMBHs) at their centres \cite[][and references therein]{Silk98, Magorrian98}.  The influence of BCG-hosted active galactic nuclei (AGN) on their surroundings (``AGN feedback'') persists long after the BCGs' formation and (major) growth.  Understanding the nuclear activity of these galaxies in relation to the wider cluster environment is therefore important.  This has important ramifications not only within cluster dynamics and BCG evolution but for comprehending AGN feedback processes globally. 

The radiative cooling time at cluster centres is often much less than the Hubble time \cite[e.g.][]{Peres98, Voigt04}.  This gives an expected cooling rate of 10$^{2}$-10$^{3}$ M$_{\odot}$yr$^{-1}$, giving an expected sink of cold material on the order of a few 10$^{11}$-10$^{12}$ M$_{\odot}$.  However, only $\sim$1-10\% of this amount is observed - the classical `cooling-flow problem' \citep{Fabian94}.  Star formation is seen in cooling clusters although typically only at the level required to account for a few percent of gas cooling \cite[][]{O'Dea08, Rafferty08}.  Similarly, much less cold molecular gas is observed than if cooling dominated \cite[e.g.][]{Edge01, Salome03}.  Additionally there is a deficit of gas seen at intermediate cooling temperatures \cite[e.g.][]{Peterson03, Sanders11} and the central gas temperature fails to cool below $\sim$30-40\% of the ambient temperature at greater radii \citep{Mittal09}.  The AGN action of the BCG is believed to counteract the expected cooling \cite[for reviews see:][]{McNamara07, McNamara12, Fabian12}.  This AGN feedback is also used to explain the high mass trunctation of the galaxy luminosity function \cite[e.g.][]{Bower06, Croton06, Benson03}. 

BCG-hosted AGN typically accrete inefficiently, hence their mode of feedback is via `maintenance-mode' (also called `low-excitation', `radio-mode') rather than `quasar-mode' (also `high-excitation') \cite[e.g.][and references therein]{Heckman14}.  Mechanical heating is the favoured method by which energy is coupled from the AGN to the intra-cluster medium (ICM).  Evidence for this AGN feedback is most readily observable through the inflation of cavities by relativistic jets, in the X-ray emitting plasma of many massive clusters \cite[e.g.][]{McNamara00, Fabian00, Hlavacek-Larrondo12a}.  These cavities, which are often co-spatial with extended radio emission, subsequently rise buoyantly increasing cluster entropy and re-distributing energy.  Detailed studies of local systems have shown that weak shocks \cite[e.g.][]{Randall11} and density sound-waves \cite[e.g.][]{Fabian05} contribute towards energising the ICM, though they cannot provide enough energy in all systems \cite[][]{Fujita07}. Observations of large-scale, low surface brightness structures such as radio mini-haloes around some BCGs suggest that cosmic rays may also play a role \cite[][]{Pfrommer03}. 

AGN energy from BCGs is sufficient to counteract runaway cooling in cluster cores globally \cite[e.g.][]{Dunn08}.  However, this is not always the case for individual sources and the details of the AGN activity cycle and fuelling are uncertain.  Often, direct observations of mechanical effects are observationally infeasible thus requiring proxy measures of AGN power.  

Radio power correlates with mechanical power for statistical samples, albeit with large scatter \cite[][]{Birzan08}.  However, both the ratio `k' \cite[e.g.][]{Willott99} of non-radiating particles to synchrotron-emitting electrons and the magnetic field are typically poorly constrained, which leads to large uncertainties in conversions between radio emission and mechanical power for individual sources.

An issue that often complicates the study of BCGs (and indeed all galaxies) using radio observations is that the observed emission can be separated significantly in both time and space from the event that created it \cite[e.g.][for a discussion of spectral ageing]{Harwood13}.  Single-band, often low spatial resolution observations are frequently the only radio information available for a BCG.  While these galaxies are then known to be radio loud, there is large uncertainty both in the physical scale and age of this emission, making multi-wavelength comparisons difficult.  Observations at a range of spectral frequencies and spatial resolutions are required to overcome this issue.  In this paper we use a wide range of data to decompose the radio emission of nearly 300 BCGs into an active component - attributable to current accretion, and an inactive component - attributable to historical accretion.  We use these decompositions to link the central AGN activity to the larger cluster environment.

Galaxy clusters can be broadly split into two types.  The first, known as `cool cores' (CCs), have short central cooling times (t$_{cool}$) and strongly peaked X-ray profiles \cite[e.g.][]{Million09}.  The second type are more dynamically disturbed, exhibit flatter X-ray profiles and are commonly known as `non-cool cores' (NCCs).  Directly measuring t$_{cool}$, or equivalently entropy, of a cluster is observationally difficult and strongly affected by resolution effects \cite[e.g.][]{Panagoulia14}.  The presence of line-emitting (predominantly H$\alpha$+[NII]) filamentary nebulae around BCGs is known to be strongly indicative of the dynamical state of the cluster.  Such lines indicate the presence of multi-phase gas, and are almost always present only when the central entropy drops below 30~keV~cm$^{2}$, equivalent to a t$_{cool}$ $<$5x$10^{8}$yr \cite[][]{Cowie83, Heckman89, Cavagnolo08, Rafferty08, Sanderson09}.  These filamentary structures are photoionised by neither the AGN itself nor star formation.  Instead they appear to be photoionised by either cosmic rays or hot X-ray plasma penetrating cold gas \cite[][]{Ferland09}.  Alternatively it has been suggested that they may be excited by the reconnection of magnetic field lines trailing in the wakes of rising cavities \cite[][]{Churazov13}. The presence of these lines can therefore be used as a proxy, with line-emitting (LE) clusters corresponding to strongly cooling cores and non-line emitters (NLEs) to either weak- or non-cooling core clusters \cite[see also][]{Crawford99}.  

Overall, this paper looks at the general radio properties of the BCG population as a whole.  It uses a large data volume to decompose the radio emission and better understand the radio-loudness of BCGs in relation to the wider environment.

The paper is arranged as follows.  In Section 2 we present the sample. Section 3 presents the observations and data.  We discuss the radio-SED decomposition in Section 4.  Section 5 considers the monochromatic (L-band, $\sim$1.4~GHz) radio properties of the sample.  Section 6 considers decomposed radio properties.  Sections 7 and 8 contain the discussion and conclusions respectively.  Comments on individual source decompositions can be found in Appendix A.  Data products for data presented for the first time in the current paper are tabulated beyond this and are available in the online version.  Unless otherwise stated, we refer to a BCG by use of its host cluster name.  We have used a standard $\Lambda$CDM cosmology with the following cosmological parameters unless otherwise stated: $\Omega_{m}$ = 0.3, $\Omega_{\Lambda}$ = 0.7, $H_{0}$ = 70 km s$^{-1}$ Mpc$^{-1}$.

\section{The Sample}

\subsection{Parent Cluster Sample}
Our parent sample is drawn from three ROSAT X-ray selected cluster catalogues - the BCS \citep{Ebeling98}, eBCS \citep{Ebeling00} and REFLEX \citep{Bohringer04} samples, which contain 206, 107 and 447 clusters respectively.  Since publication some catalogue entries have been re-classified, and there are also a small number of cross-catalogue duplicates.  We therefore remove a minority of sources, leaving us with a sample of 199, 104 and 417 sources in the BCS, eBCS and REFFLEX samples respectively.  Our total X-ray selected parent sample therefore consists of 720 clusters.

The BCG for each of these clusters has been optically identified and observed spectrally (\citealt{Crawford99}, Edge \etal {\it in prep.}).  The sample was split into those which exhibit optical emission lines (line-emitters, hereforth LEs) and those which do not (non line-emitters, hereforth NLEs) (see also Table \ref{eBCSREFLEXBreakdown}).  As mentioned in the introduction, the presence of these lines is strongly indicative of the cluster containing a strong cool-core.  It must be noted that their absence is not sufficient to rule out the presence of a weaker cooling core (5x$10^{8}$yr $<$ t$_{cool}$ $<$ t$_{Hubble}$) but does point towards these clusters being less dynamically settled.

An additional point to bear in mind is that not all clusters contain a single top-ranked BCG, and may instead have two or more similarly sized top-ranked galaxies.  In our Main Sample (see Section \ref{MS_SAMPLE}, roughly 5\% of our clusters are identified as having multiple BCGs (see Appendices).  This is less than the equivalent $\sim$15\% of the sample of 215 clusters in \cite{Crawford99} that are claimed to have multiple BCGs.  We could therefore be missing a co-dominant galaxy in some cases, though we note that all clusters were visually inspected in both optical and radio maps and no instances were found of clear unattributeds BCGs.  Quantifying the effect of unidentified secondary BCGs is difficult.  Of the 5\% of our sample with multiple BCGs, only one double-BCG system is hosted by a CC-cluster (A2627) - both BCGs in this cluster are found to be radio-loud.  Only around 30\% of the multiple-BCGs in NCC clusters host radio-loud AGN, which is less than the approximately 50\% of all NCC clusters in our sample that host radio-loud BCGs (see Section \ref{LE_NLE_Match_Rates}).  Whilst clearly circumstantial, and affected by small number statistics, this actually points towards the differences in the radio-loud fraction between CCs and NCCs being higher in systems with multiple BCGs, which would strengthen the results within this paper.  The fraction of clusters with multiple BCGs is expected to increase with redshift \cite[e.g][]{Brough02}, so should be less of an effect for our sample, which is at low redshift.  Overall, we do not believe we are overlty affected by any potential small fraction of unidentified co-dominant BCGs, although we do caution that this caveat should be considered when interpreting the results.

The BCS/eBCS sky coverage overlaps fully the NRAO VLA Sky Survey (NVSS) radio survey \cite[][]{Condon98} (DEC $>$ -40$^{\circ}$, complete to $\sim$2.5 mJy at 1.4~GHz).  Of the 417 REFLEX clusters in our sample, 297 fall within the NVSS region and 156 within the southern sky region covered by the Sydney University Molonglo Sky Survey \cite[SUMSS:][]{Mauch03} (DEC $<$ -30$^{\circ}$, complete to $\sim$6 mJy for DEC$<$-50$^{\circ}$ and $\sim$10 mJy for DEC$<$-30$^{\circ}$ at 0.843~GHz).  These sub-samples are hereon referred to as the REFLEX-NVSS and REFLEX-SUMSS samples.  

Thirty-six REFLEX sources fall in the overlap between SUMSS and NVSS.  These sources are left in both the REFLEX-NVSS and REFLEX-SUMSS samples when these samples are treated independently.  Combining the BCS and eBCS catalogues provides a 303 source subsample that we henceforth refer to as the (e)BCS.  The comparable number of clusters in (e)BCS and REFLEX-NVSS allows for self-consistency checks of any found results since the samples were independently compiled.

\begin{table}
\caption{BCGs within the (e)BCS and REFLEX galaxy cluster samples (after removal of `contaminant sources').  The samples are broken down both by radio detection in NVSS and/or SUMSS and also the presence of extended optical emission lines around the, which are indicative of the presence of a strong cool core.} \label{eBCSREFLEXBreakdown}
\begin{minipage}[b]{\linewidth}\centering
\begin{tabular}{|c|c|c|}
\hline\hline
 &  Lines & No Lines \\
 \hline\hline
 BCS &  &  \\
 Detected     & 55 & 67 \\
 Non-detected & 8  & 69 \\
 \hline
 eBCS &  &  \\
 Detected     & 19 & 44 \\
 Non-detected & 6  & 35 \\
 \hline
 REFLEX - Total &  &  \\
 Detected     & 91 & 161 \\
 Non-detected & 17 & 148 \\
 \hline
 REFLEX - NVSS &  &  \\
 Detected     & 67 & 119 \\
 Non-detected & 10 & 101 \\
 \hline
 REFLEX - SUMSS &  &  \\
 Detected     & 36 & 58 \\
 Non-detected &  7 & 55 \\
 \hline\hline
\end{tabular}
\end{minipage}
\end{table}

\subsection{Main Sample} \label{MS_SAMPLE}

We have targeted follow-up observations of 246 of these sources at either C- or X- ($\sim$5 or 8~GHz) band (or both) using the ATCA or VLA (see Appendices \ref{APPENDIX_NOTES} and \ref{APPENDIX_TABLES}). These sources constitute the `Main Sample' of this paper.  They comprise 106 + 64 BCGs (detections + non-detections respectively) in the BCS, 13 + 2 in the eBCS, 39 + 7 in REFLEX and 9 + 6 additional sources that since publication of the catalogues have been found to fulfil the detection criteria but were not initially classified as clusters. These extra sources are - Non-detections: A7, A2552, Zw15, Zw5029 and Z7833. Detections: A11, A291, A1664, A2228, and 4C-05.84). This constitutes follow-up completenesses of 85.4\%, 14.4\% and 11.0\% for the BCS, eBCS and REFLEX catalogues respectively. 

Whilst only a minority of eBCS clusters are included in the Main Sample, these were randomly selected from the eBCS prior to its publication and should constitute a fair addition to the Main Sample. The minority of clusters followed up from within the REFLEX sample were selected to be the radio-brightest (L-band) line-emitting BCGs.  The high completeness for BCS provides us with a radio-observed, X-ray selected cluster sub-sample that is un-biased by any radio priors on the BCGs, which can be tested.

\subsubsection{Main Sample+}
The Main Sample is supplemented using literature values to build SEDs for all BCGs in the Parent Sample that are radio-matched to at least one NVSS or SUMSS source and have L-band fluxes $>$15 mJy.  We note that the 843~MHz observing frequency of SUMSS is technically below the traditional L-band range of 1-2~GHz, however we refer to both SUMSS and NVSS as `L-band' for succinctness.  Whilst without targeted follow-up, the spectral decompositions of these sources (see Section \ref{SEDDecomp} and Appendix \ref{APPENDIX_MSP_DECOMPOSITIONS}) contain a higher fraction of upper limits, the addition of these sources does ensure that we are drawing conclusions on the radio behaviour of BCGs from a complete flux-limited sample and immunises us against missing extreme (fainter {\it or} brighter) objects that may oppose any trends found from the targeted campaign.  We refer to this flux-limited sample as the Main Sample+ (hereforth MS+).

\section{Observations}

This work has utilised data from a variety of observing campaigns, supplemented by literature measurements.  

Both pre and post-CABB \cite[Compact Array Broadband Backend;][]{Wilson11} data from the ATCA were analysed, as was VLA data from five observing campaigns (see Table \ref{obs_table}).  All of these observations followed a standard `snapshot' schedule, with multiple visits to each target source separated by several hours to ensure good hour-angle (HA) coverage.  Primary flux calibrators were observed at the beginning and end of each run.  Each visit to a science target was sandwiched between short observations of a nearby standard source for phase calibration. Exact calibrators for each source can be found in the online archives.

ATCA and VLA data were reduced and fluxes measured for the detected sources using the Multichannel Image Reconstruction Image Analysis and Display \cite[MIRIAD:][]{Sault95} and the Astronomical Image Processing System \cite[AIPS:][]{Greisen03} packages respectively, following the standard reduction procedures.  

For the ATCA data, the standard flux calibrator 1934-638 was used in all instances for both absolute gain and bandpass calibration.  For the VLA observations, three of the VLA standard flux calibrators (namely 3C286, 3C48 and 3C147) were used, with the actual source used for each observing block dependent on observability. For VLA datasets with more than one frequency channel per spectral window, these sources could also be used for bandpass calibration in most cases.  However, in a couple of instances the primary calibrator appeared slightly resolved (specifically 3C147 for AE117 and AE125 at X-band) and so an alternate, unresolved bandpass calibrator was chosen from amongst the phase calibrators for that particular run.  

\begin{table}
\caption{Breakdown of the ATCA and VLA observations used in this paper that are not previously published elsewhere (All: P.I. Edge).  Note that several BCGs are repeated, so these numbers do not correspond to the total number of clusters.  See Appendices for full list of sources.  This data was used when compiling our SEDs and is presented here for the first time.} \label{obs_table}
\begin{minipage}[b]{\linewidth}\centering
\begin{tabular}{|c|c|c|c|c|}
\hline\hline
   Proposal ID & Year & Array & Frequency & BCGs Observed \\
               &      &       &   (GHz)   &                  \\
\hline\hline
 C1958 & 2011 & ATCA-6A & 5.500 & 14 \\
 C1958 & 2011 & ATCA-6A & 9.000 & 14 \\
 C1958 & 2008 & ATCA-6A & 4.800 & 31 \\
 C1958 & 2008 & ATCA-6A & 8.640 & 31 \\
 AE125 & 1999 & VLA-C   & 8.435 &  4 \\
 AE125 & 1999 & VLA-C   & 4.835 & 80 \\
 AE125 & 1999 & VLA-C   & 1.465 &  3 \\
 AE117 & 1998 & VLA-A   & 1.385 & 12 \\
 AE117 & 1998 & VLA-A   & 8.435 & 32 \\
 AE110 & 1997 & VLA-C   & 4.835 & 86 \\
 AE107 & 1996 & VLA-C   & 4.835 & 28 \\
 AE099 & 1994 & VLA-C   & 4.835 & 24 \\
\hline\hline
\end{tabular}
\end{minipage}
  \label{obs_table}
\end{table}

Literature searches were performed, and SEDs populated, for all sources in the Main Sample and its extension.  Data were taken from the major radio catalogues (e.g. Australia Telescope 20~GHz Survey (AT20G), \citealt{Murphy10}; NVSS and Faint Images of the Radio Sky at Twenty-cm (FIRST) at 1.4~GHz, \citealt{Condon98, White97}; SUMSS at 843~MHz, \citealt{Mauch03}; Molonglo Reference Catalogue (MRC) at 408~MHz, \citealt{Large81}; Texas Survey of Radio Sources (TEXAS) at 365~MHz, \citealt{Douglas96}; Westerbork Northern Sky Survey (WENSS) and Westerbork In the Southern Hemisphere (WISH) at 325~MHz, \citealt{Rengelink97, DeBreuck02}; TIFR GMRT Sky Survey (TGSS) at 150~MHz; VLA Low-Frequency Sky Survey (VLSS) at 74~MHz, \citealt{Cohen07}).  Additional fluxes found by searches around the radio-peak coordinates in the NASA/IPAC Extragalactic Database (NED), High Energy Astrophysics Science Archive Research Center (HEASARC) database and/or the National Radio Astronomy Archive (NRAO) were individually scrutinised to ensure matches. 

\subsubsection{A Note on Flux Scales}
Our ATCA data were calibrated to the \cite{Reynolds94} flux scale for the standard flux calibrator 1934-638, as calculated in the MIRIAD version 20110127 task `mfboot'.  VLA data were scaled to the standard AIPS version 31DEC11 Perley-Butler scale using the task `setjy' \cite[see][]{Perley13}.  Both of these scales are very similar to, but have small improvements over the older Baars scale \cite[][]{Baars77}.  The wide range of literature values use a range of similar flux standards.  This represents an added uncertainty to our SEDs.  For data pre-dating the Baars scale (very small number of instances), we have used later determinations re-scaled to the Baars scale.  The uncertainties introduced by this range of flux scales should be considered, however they are much less than those associated with intrinsic source variability and the absolute flux uncertainties, hence should be negligible.

\section{SED Decomposition} \label{SEDDecomp}
\subsection{Core versus `Non-core'}

The total radio SED of each BCG can be decomposed into two major components - an active, typically flat-spectrum (spectral index\footnote{ Where Flux $\propto$ $\nu$$^{-\alpha}$} $\alpha$$<$0.5) component attributed to current activity within the AGN and a typically steeper spectrum component ($\alpha$$>$0.5) that comprises all other emission.

The steeper component is most likely to be due to lobe emission and so traces past activity of the central engine.  Alternatively it could be due to other acceleration processes towards the centre of the galaxy cluster in which the BCG resides, such as sloshing giving rise to a mini-halo or phoenix relics \cite[e.g.][and references therein]{vanWeeren11, Zuhone13, Walker14}.  The specific morphological and energetic nature of this component is often difficult to determine and a large variety of nomenclature for the various source-types has arisen \cite[see][for a detailed review]{Kempner04}.  The steep spectrum means that this component becomes less prominent towards higher frequencies and is often faint at L-Band ($\sim$1.4~GHz) and above, rendering it difficult to detect if an active core component is present. Conversely, current AGN activity manifesting itself as a flat-spectrum core component can be swamped at lower frequencies.  High resolution observations are required at low frequency to properly characterise these components.  However, in this work we wish to address the difference in radio emission between ongoing core activity and older, steep-spectrum emission.  We use `non-core' as an umbrella term to refer to the steeper spectrum component not associated with ongoing core activity, regardless of physical origin and spatial resolution.  

The core component may show a synchrotron self-absorption turnover at $\sim$few GHz but tends to remain flat (or, less commonly, rising) out to several tens of GHz.  Similarly the `non-core' component may show a turnover, albeit at much lower ($\textless$ a few hundred MHz) frequencies.  This is more likely to be attributable to free-free absorption.  Additionally, this component may exhibit a steepening index at higher frequencies due to synchrotron ageing of the underlying electron population.  

\subsubsection{A Note on Non-AGN Related Cluster Radio Emission} \label{other_structures_note}
When present, both cluster scale radio halos and peripheral relics \cite[a.k.a. `radio gischt'; e.g.][]{Giacintucci11} are excluded from our non-core component, as these are both unrelated to the current AGN activity of the BCG \cite[e.g.][]{Feretti12, Brunetti14.}  There are however a number of other radio structures seen in cluster cores such as mini-haloes and phoenix relics whose relationship to the BCG and the cooling core is more uncertain \cite[e.g.][]{Gitti02, Gitti06, Gitti07a}.  There is debate in the literature as to the true nature of so-called `mini-halos' and these present a somewhat contentious middle-ground between cluster related and AGN-related radio-emission.  True mini-halos tend to be reasonably spherically symmetric, steep spectrum, low surface brightness sources that are a few 100~kpcs in diameter and centred on the BCG \cite[e.g.][]{Burns92, Owen00, Giacintucci14b}.  The electron diffusion lifetime is so long as to exclude acceleration by the central AGN. The emitting particles must therefore be accelerated in-situ and appear to be related to the cooling flow.  It is however possible (and indeed likely) that the seed population originates in the AGN.  

Confusingly, it appears that BCGs often contain a smaller ($<$ few 10s kpc), amorphous halo or {\it `confined cluster core'} \cite[e.g.][]{Venturi04} that appears to be related directly to confined, ageing AGN ejecta.  This is also occasionally referred to as a mini-halo.  
As the exact nature of the non-core emission is not paramount for our purposes, we may include in our non-core measures emission that could reasonably be called `mini-halo' in the literature, if it contributes a portion of the flux in low-resolution observations that provide such a measurement as the `BCG flux'.

\subsection{SED Fits} \label{SED_Fits}

The majority of sources were unresolved at the resolution limit of our observations (typically around 4'' at C-band, equivalent to $\sim$8~kpc at our median redshift).  Where core and non-core components were morphologically distinct, individual SEDs were produced and fitted directly for each component.

Sources that remain unresolved on few arcsecond scales could consist of an active core only or small scale yet ageing non-core emission.  When Very Long Baseline Interferometry (VLBI) observations were available, the ratio of milli-arcsecond to arcsecond scale flux could be used to determine if any flux was resolved out on intermediate scales, and what fraction of the flux was truly associated with the core. For sources that are unresolved on few-arcsecond scales but for which VLBI observations are not available, variability and spectral shape were used to perform a breakdown of the SEDs on a case-by-case basis (see below).  Where a strong case could not be made for unresolved sources being either core or non-core dominated, conservative assumptions were employed.  

\subsubsection{Main Sample}
Four simple models were considered for each source SED - a single power law of the form:
\begin{equation}
S = A_{0} \nu^{-\alpha} 
\end{equation}
a split power law of the form:
\begin{equation}
S = A_{0} \nu^{-\alpha_{1}}  + A_{1} \nu^{-\alpha_{2}}
\end{equation}
a `dropline' model, which allows for a high frequency rollover to simulate synchrotron ageing:
\begin{equation}
S = A_{0} \nu^{-\alpha}(1 - A_{1}e^{-\frac{\nu_{0}}{\nu}}) 
\end{equation}
or a Gigahertz Peaked Source (GPS)-like model that allows for a spectral peak to simulate self-absorption \citep{Orienti14}
 \begin{equation}
Log(S) = A_{0} + Log(\nu)(A_{1} + A_{2}Log(\nu))
\end{equation}
where S denotes measured flux and `A's are constants to be determined.

These fits are not intended as physical models.  Indeed it is clear that for the dropline model, extrapolation outside of the region being considered would lead to negative predicted flux.  Instead they are merely phenomenological fits to the known data that provide a decomposition of the two radio components - core and non-core.

The single power law provides the simplest possible fit, and often had to be employed for poorly sampled SEDs.  The index of this power law was considered, with a flat or inverted slope leading to the detection being attributed to a core component, or a steep slope meaning the detection was classed as a non-core.  The nominal cut-off between these distinctions was taken as $\alpha$ = 0.5.  However, anything flatter than $\alpha$ = 0.8 or steeper than $\alpha$ = 0.2 was further scrutinised.  Maps were visually searched for physical extension at each observed frequency.  Variability was considered to indicate the presence of a powerful core, with the proviso that this component is more likely to vary on shorter timescales.  In situations where there was strong evidence for one component dominating the observed flux and hence spectral shape, then  the value of this component was taken as a measurement of that component.  A limit was placed on the other component by extrapolating with a representative steep ($\alpha$$_{non-core}$ = 1.0, see section \ref{index_section}) index from the lowest observed frequency or with representative flat ($\alpha$$_{core}$ = 0.2, see section \ref{index_section}) index from the highest observed frequency for non-core and core limits respectively.

\begin{figure*}    
  \centering
    \subfigure[{\it RXJ0439.0+0520}]{\includegraphics[width=8cm]{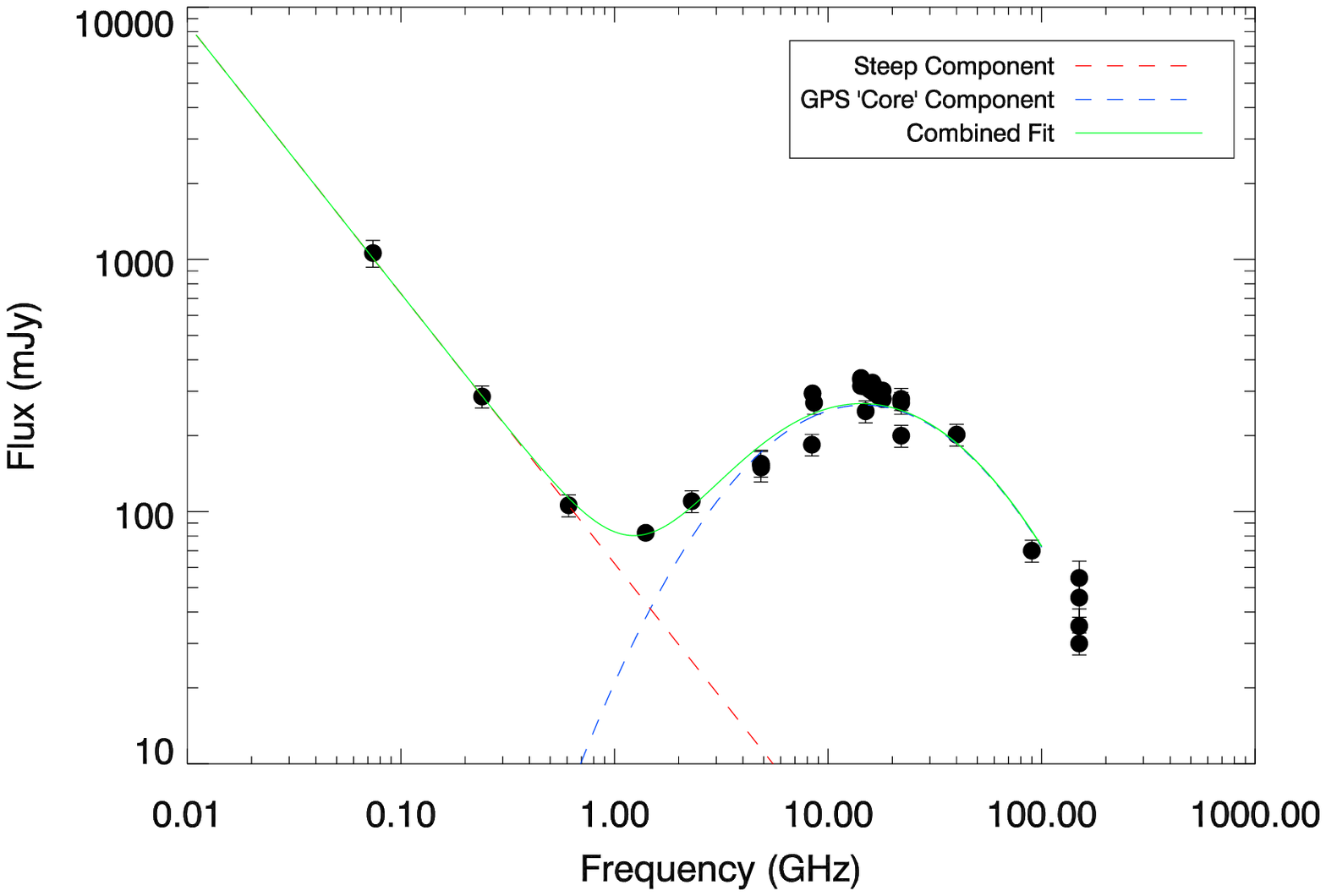}}
    \subfigure[{\it Z8276}]{\includegraphics[width=8cm]{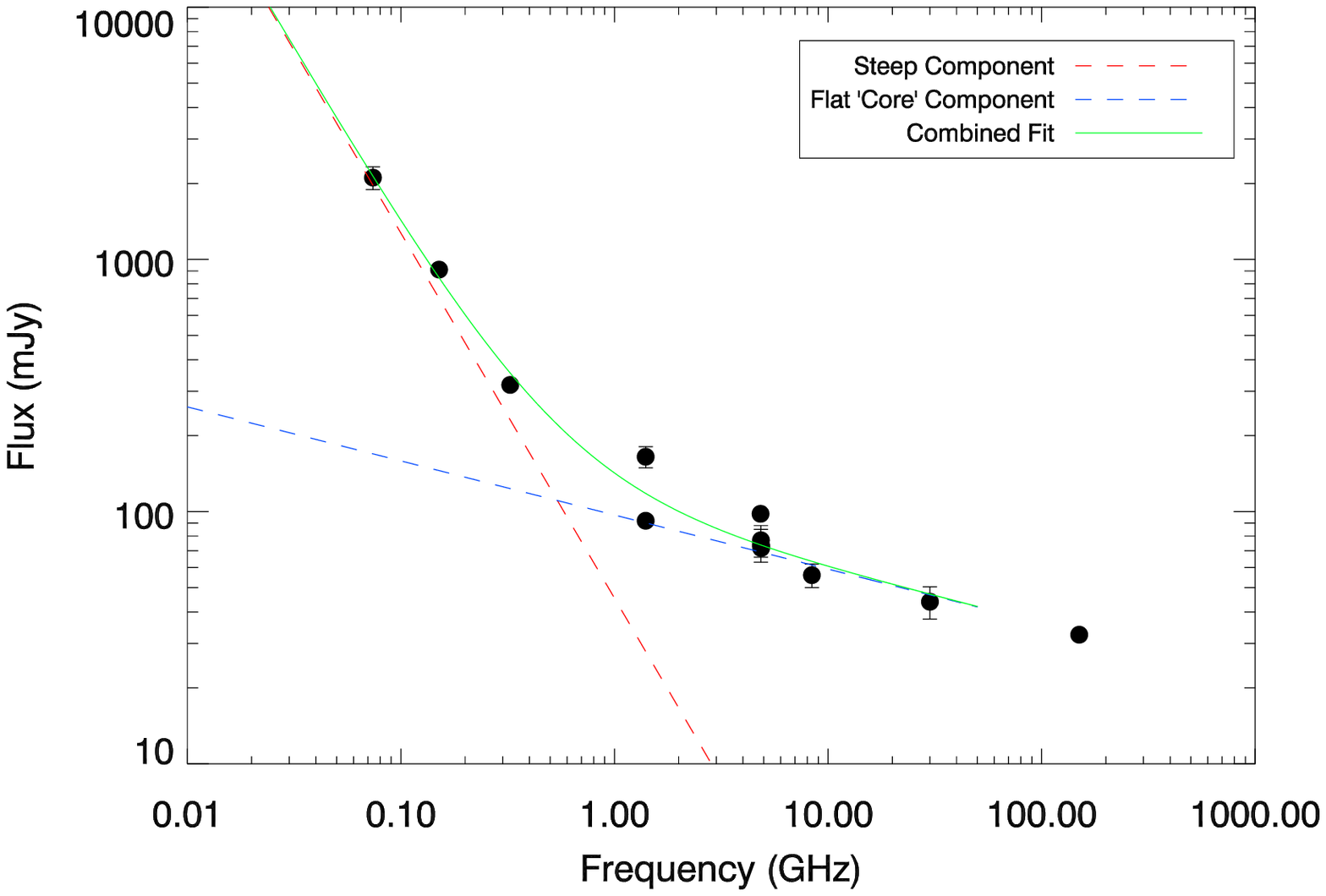}}
    \subfigure[{\it RXJ0747.5-1917}]{\includegraphics[width=8cm]{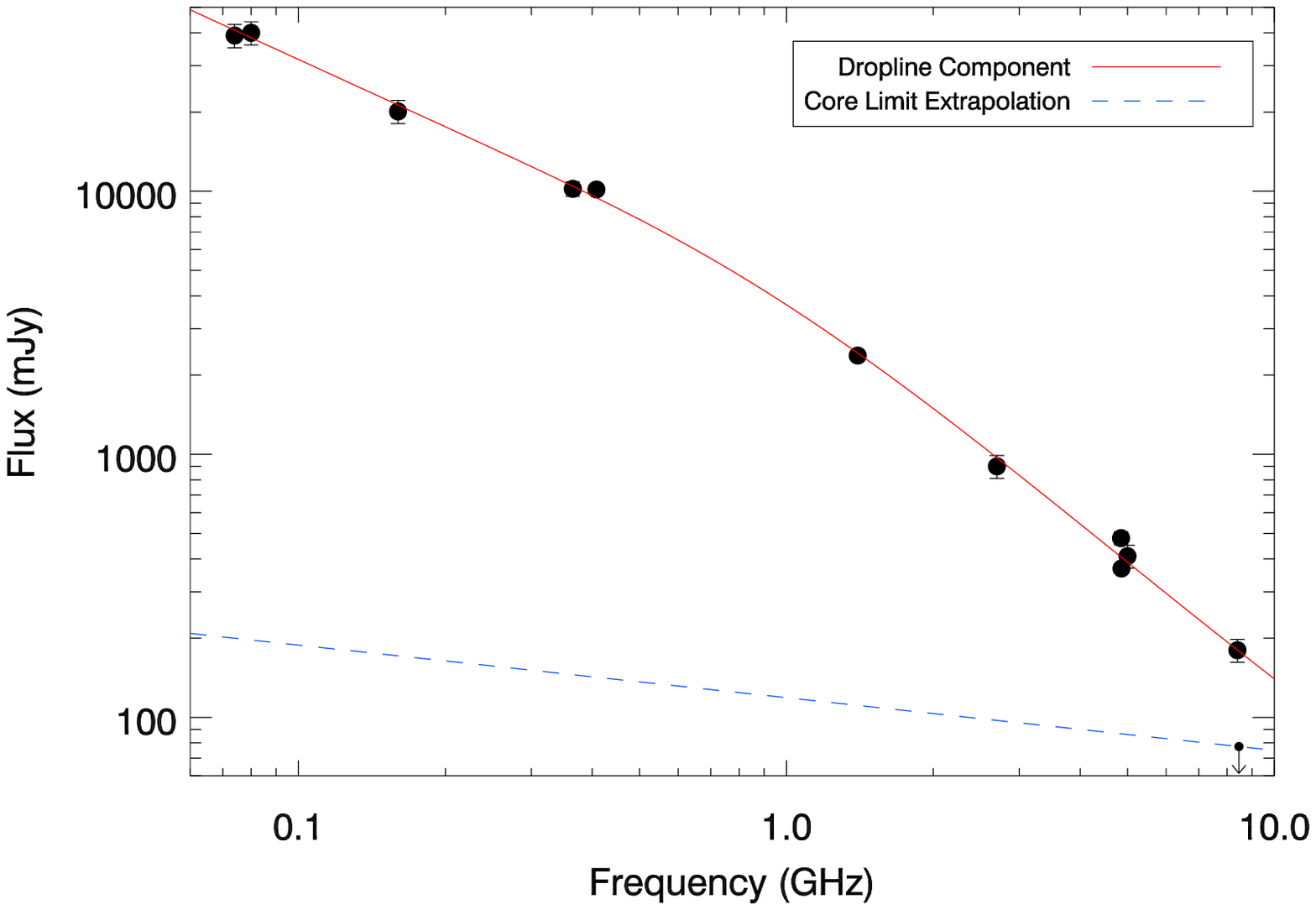}}
    \subfigure[{\it RXJ1315.4-1623 (NGC5044)}]{\includegraphics[width=8cm]{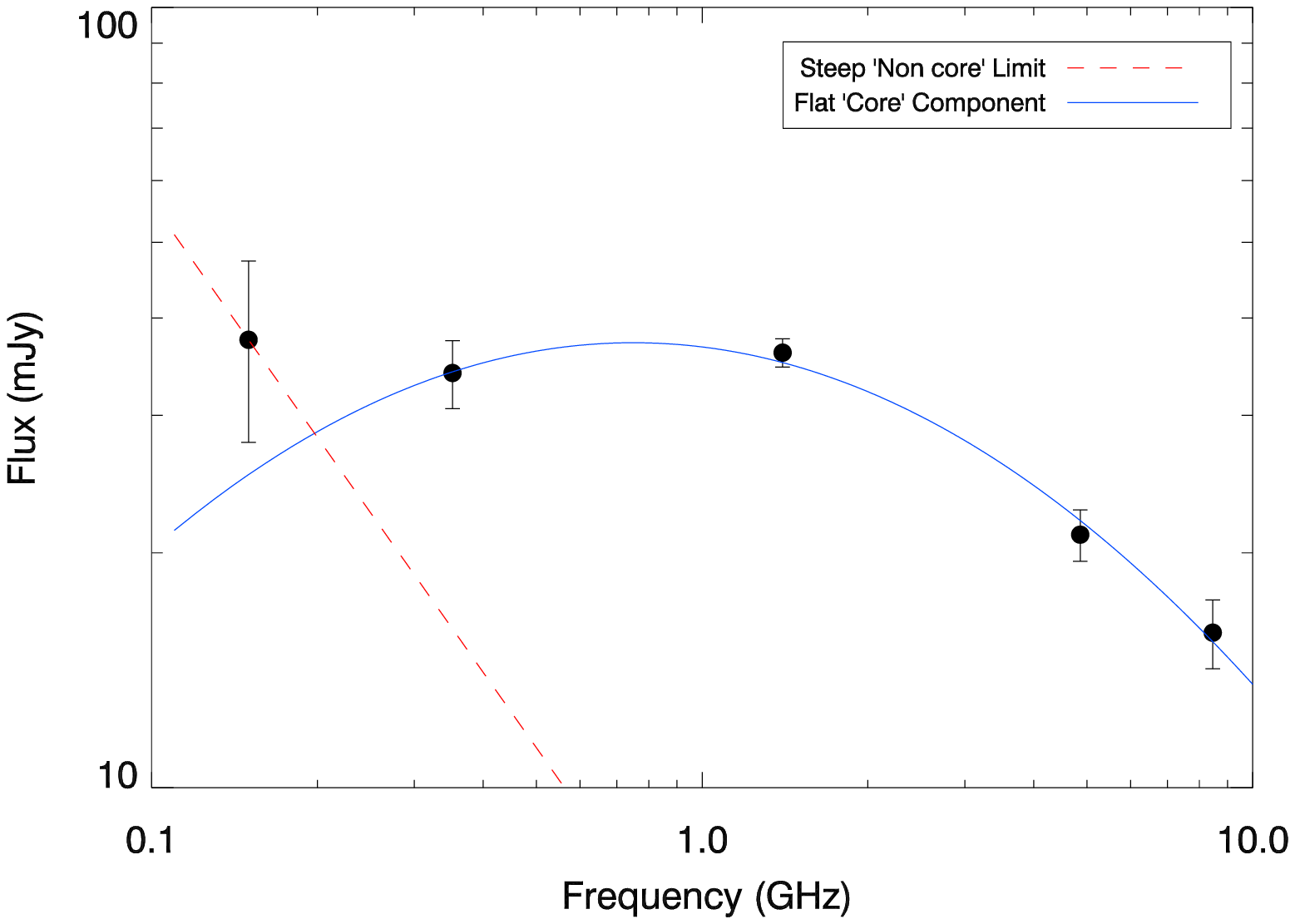}}
  \caption{Example SEDs for four different source-types.  In panel a, RXJ0439.0+0520 is dominated by a GPS-like core component at frequencies above $\sim$1~GHz although a clear steep spectrum component is seen at lower frequencies.  Panel b shows Z8276 where such a dominant inverted core is not present yet there is still a distinct flattening of the spectral index to higher frequency suggestive of a significant active component in the system.  This appears well represented by a split power-law.  Not all sources can be fit with two distinct components.  Panel c for example shows RXJ0747.5-1917, which appears to be well fit by a steep power law with a rollover to higher frequencies (i.e. a {\it `dropline'}, see text).  This source is slightly resolved at X-band with the VLA-A, hence extrapolating from the peak of this observation allows an upper limit to be placed on a core contribution.  Alternatively, panel d shows RXJ1315.4-1623 whose BCG is unresolved at all observed frequencies and exhibits a peaked SED, indicative of it being dominated by an active, self-absorbed core.  There is a map detection at 150~MHz from TGSS that may indicate the presence of a weak non-core component although the uncertainty is too large to derive a high-quality measurement and hence a limit on the non-core is derived by extrapolating from this low frequency point.} 
 \label{exampleSEDs}
\end{figure*}

Where a clear spectral break could be seen (e.g. see Panel b, Figure \ref{exampleSEDs}), a split power law was employed to fit the two components. Highly variable sources could display an apparently split spectrum if observations taken at different frequencies were substantially separated in time.  Timescales and any variations between observations at different frequencies were used to help distinguish between consistent and apparent spectral breaks.   

SEDs that exhibited the high frequency rollover typical of an ageing electron population were fit with a dropline model.  When only this component was apparent in an SED, the value derived was attributed to the non-core component and a limit placed on the core (e.g. Panel c, Figure \ref{exampleSEDs}).  

Somewhat unexpectedly, a minority of sources exhibited a strongly inverted, peaked spectrum.  These spectral components were fitted using the GPS-like model.  This spectral shape is typical of a GPS source, commonly interpreted to be either a young or recently re-triggered AGN and as such is indicative of a strong, active core \cite[][]{O'Dea98}.  However in a couple of cases a steep-spectrum, low frequency `tail' of emission can also be seen in the SED, allowing the non-core component to also be measured (e.g. Panel a, Figure \ref{exampleSEDs}). Where only a GPS-like component was apparent in an SED, a value was taken for the core and a limit derived for the non-core component (e.g. Panel d, Figure \ref{exampleSEDs}).

During the fitting, consideration was also given to the possibility of flux on spatial frequencies not sampled by observations and hence effectively `resolved out'. We checked the ratio of peak flux to flux extracted using a single Gaussian fit (AIPS task JMFIT) for sources that appeared unresolved.  Furthermore, where possible the flux ratio between two unresolved observations of a source at the same frequency but different spatial sampling was taken (e.g. FIRST/NVSS at 1.4~GHz). In truly unresolved sources, both of these ratios should give unity.  Of course it should be noted that variability could also affect these ratios.  Again, caution was employed to ensure that such effects did not overly bias our breakdowns.

Full details of the breakdowns for each source in the Main Sample can be found in Appendices \ref{APPENDIX_NOTES} - \ref{APPENDIX_DECOMPOSITIONS}.  All SEDs are available online \cite{Hogan14}.

\subsubsection{Main Sample+}

For some brighter members of the MS+ extension (e.g. Hydra-A and Hercules-A) clear morphological decompositions were possible. Where this was not the case, spectral breakdowns were performed using the same criteria as for the Main Sample with the exception that, due to the typically much lower spectral coverage and less available data, GPS-like and dropline models were not used as they were poorly constrained.  Often a single index fit around the normalisation frequency had to be taken.

For many of the sources in the MS+ extension, radio coverage was limited to only the L-band detection.  In these instances the detection thresholds of higher frequency large-scale surveys, predominantly the Green Bank 6-cm at 4.85~GHz \cite[GB6:][]{Condon94}, the Parkes-MIT-NRAO surveys at 5~GHz \cite[PMN:][]{Griffith93} and AT20g were considered to determine whether the source has a flatter spectrum component to high frequencies or whether the L-band flux is attributable to a dominant non-core.  Where a limit on a higher frequency flux determined that the L-band detection must be due to a source with spectral index $\alpha$$>$0.8 then it was attributed to dominant non-core emission and an estimate of this made using a short extrapolation with $\alpha$$_{Steep}$=1.0.  Core limits were then drawn using $\alpha$$_{Flat}$=0.2.  These indices were chosen on the basis of being typical of the $\alpha$s found for clearly identifiable components. 

When the higher frequency limits did not constrain spectra to be flatter than $\alpha$=0.8 caution was employed and limits taken on both components, extrapolating from the L-band detection with indices $\alpha$$_{Steep}$=1.0 and $\alpha$$_{Flat}$=0.2 for the steep and flat components respectively  (see also Section \ref{Core_notcore_section}).  Similarly, lower frequency survey limits were considered to ascertain whether a lower limit of $\alpha$$_{Flat}$$<$0.5 could be determined for any sources, so ruling out a source being steep spectrum dominated.  

Notes for the decomposition of each SED in the MS+ can be found in Appendices \ref{APPENDIX_MSP} and \ref{APPENDIX_MSP_DECOMPOSITIONS}. All SEDs are available online \cite{Hogan14}.

\subsection{Measured Indices} \label{index_section}

\begin{figure}
  \begin{minipage}[b]{0.5\linewidth}
  \centering  
  \includegraphics[width=9cm]{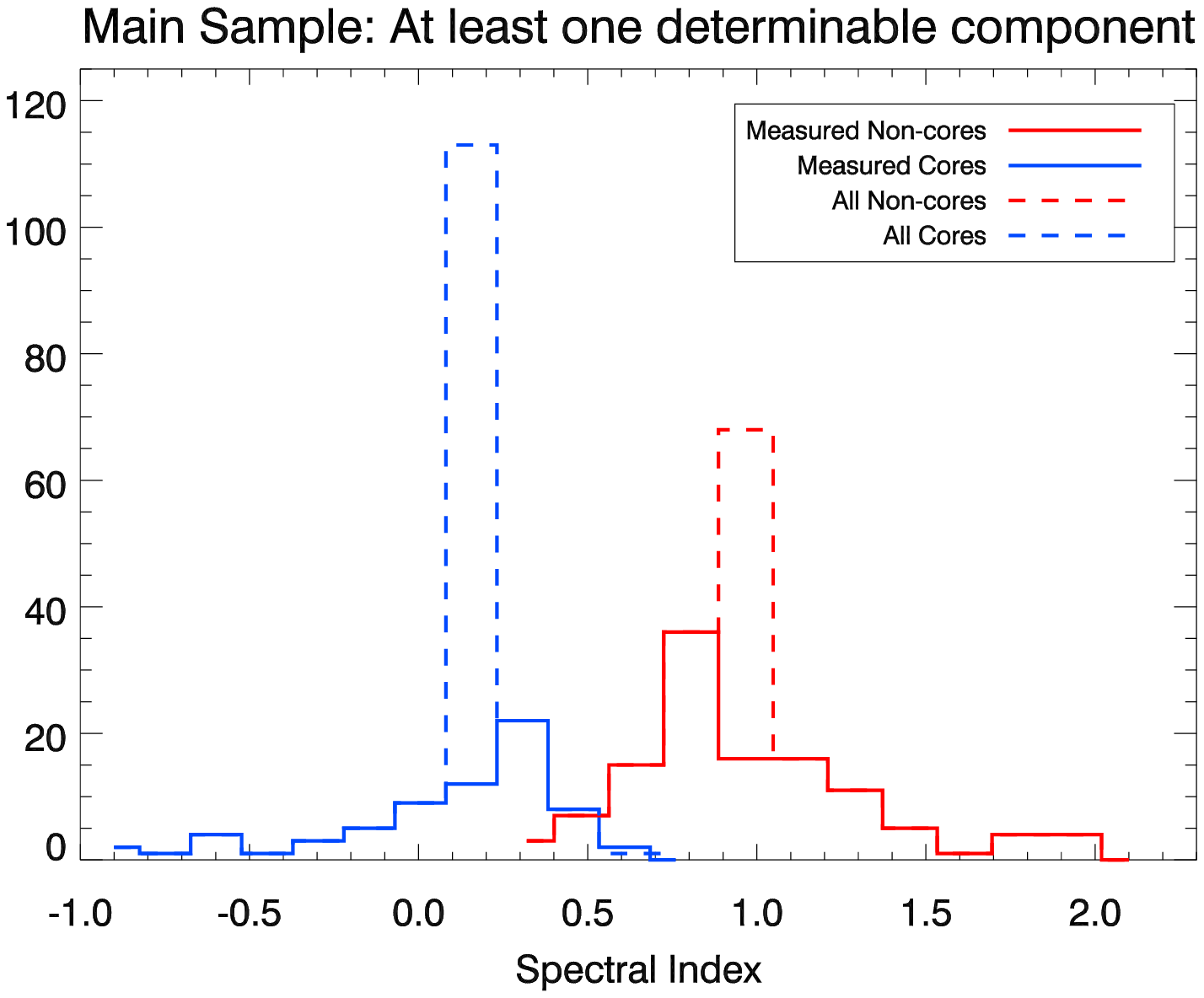}  
  \end{minipage}
  \begin{minipage}[b]{0.5\linewidth}
  \includegraphics[width=9cm]{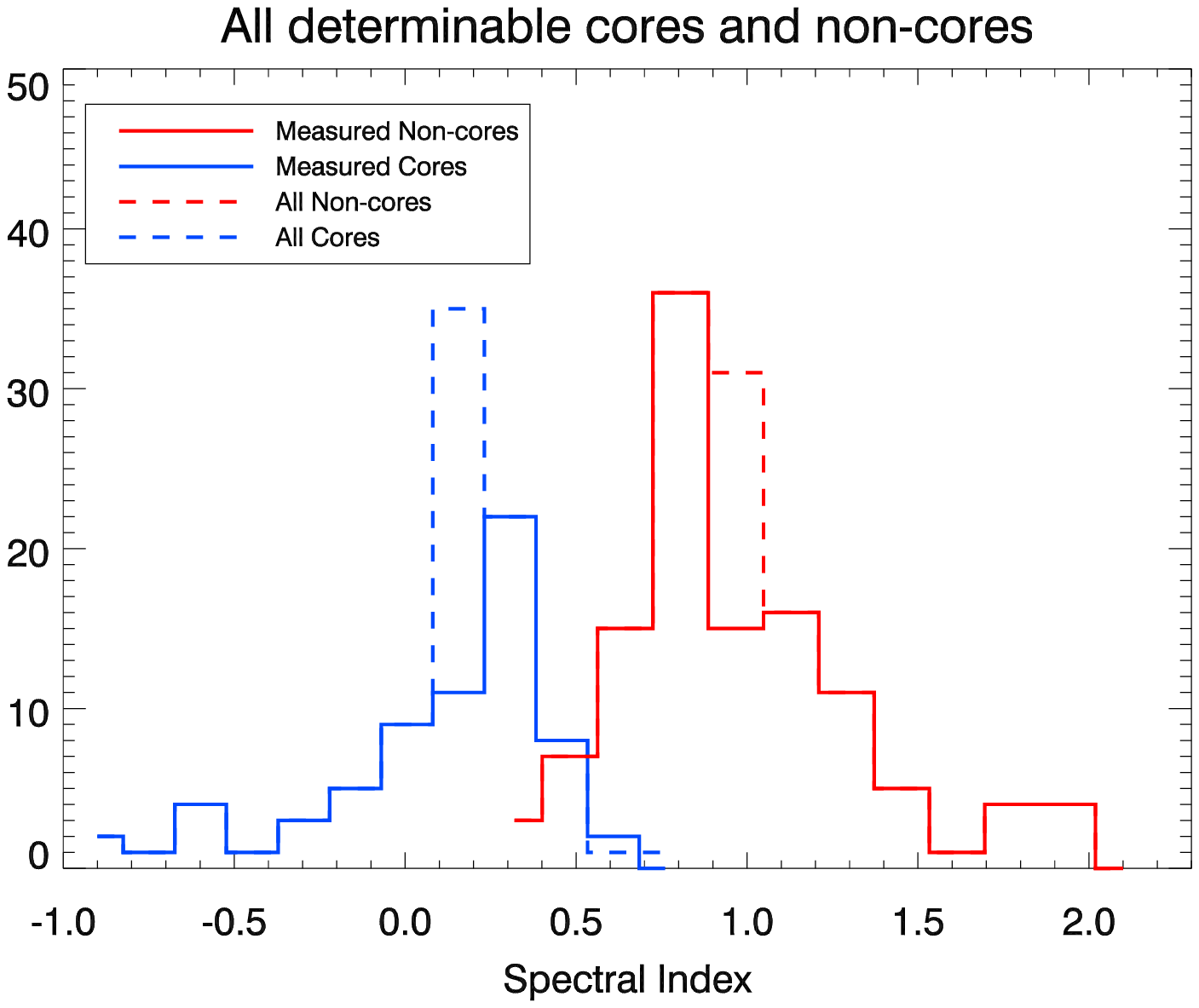}  
  \end{minipage}
  \caption{Incidence of fitted indices for the core and non-core fitted components to our Main Sample. In both panels, blue lines represent the fitted index of the core component and red that of the non-core.  Solid lines are instances where the relevant component could directly be determined whereas the dotted lines include the components where a representative index had to be assumed ($\alpha$$_{Steep}$=1.0 and $\alpha$$_{Flat}$=0.2, see text for justification).  For the top panel, all sources in the Main Sample are included for which at least one component could be measured, hence the high incidence of assumed indices as required for extrapolated limits. In the the lower panel, the red lines trace all instances where a core value could be determined and the blue lines trace all instances where a non-core value could be determined.  Note that limits remain in the lower panel for when a component could be resolved and had a measurement near to either normalisation (e.g. a VLBI core measurement) and only a small but secure extrapolation was required, nonetheless needing an index to be assumed.  See also Appendix \ref{APPENDIX_NOTES}.}  
  \label{Indices_MS}
\end{figure}

\begin{figure}
  \begin{minipage}[b]{0.5\linewidth}
  \centering  
  \includegraphics[width=9cm]{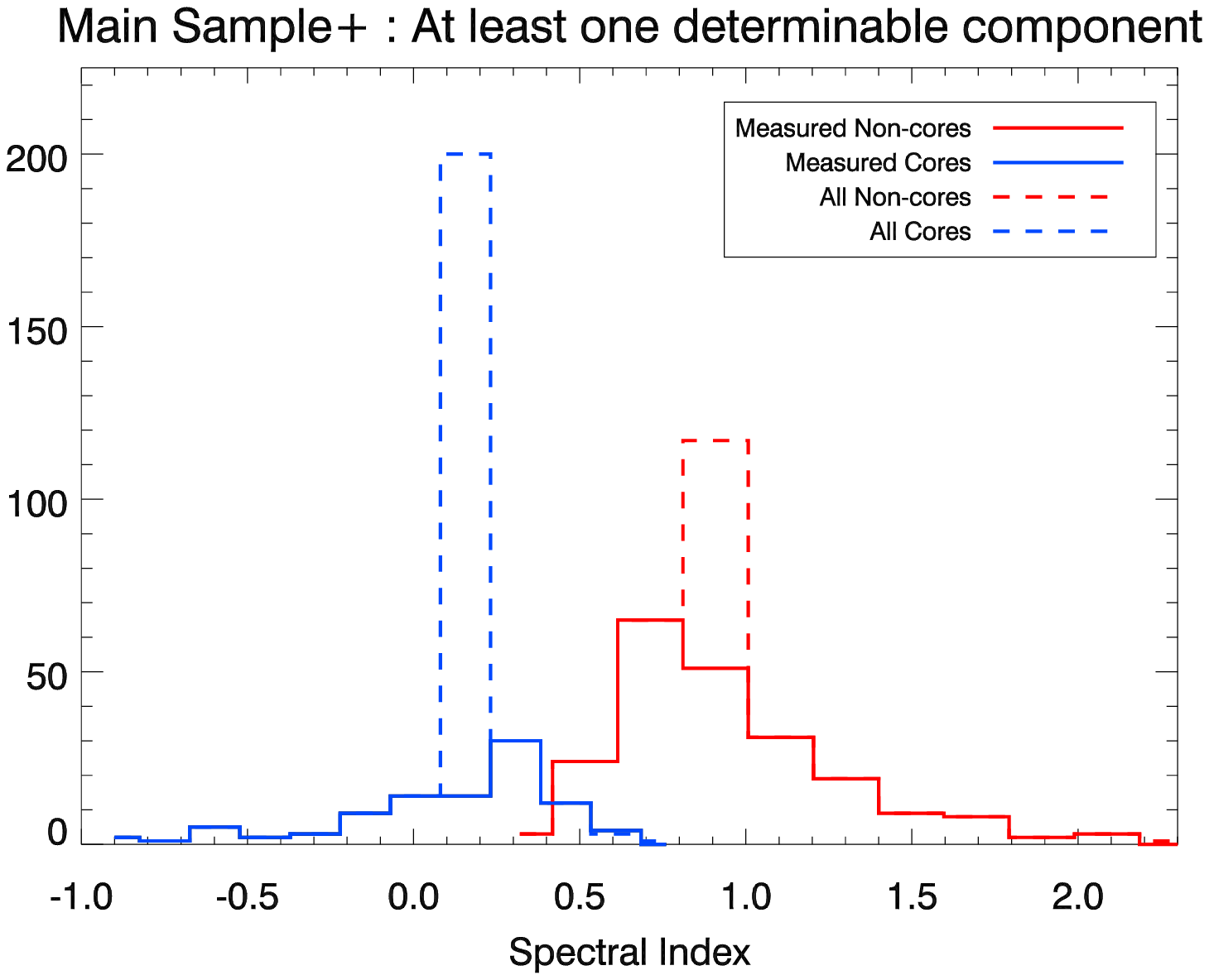}  
  \end{minipage}
  \begin{minipage}[b]{0.5\linewidth}
  \includegraphics[width=9cm]{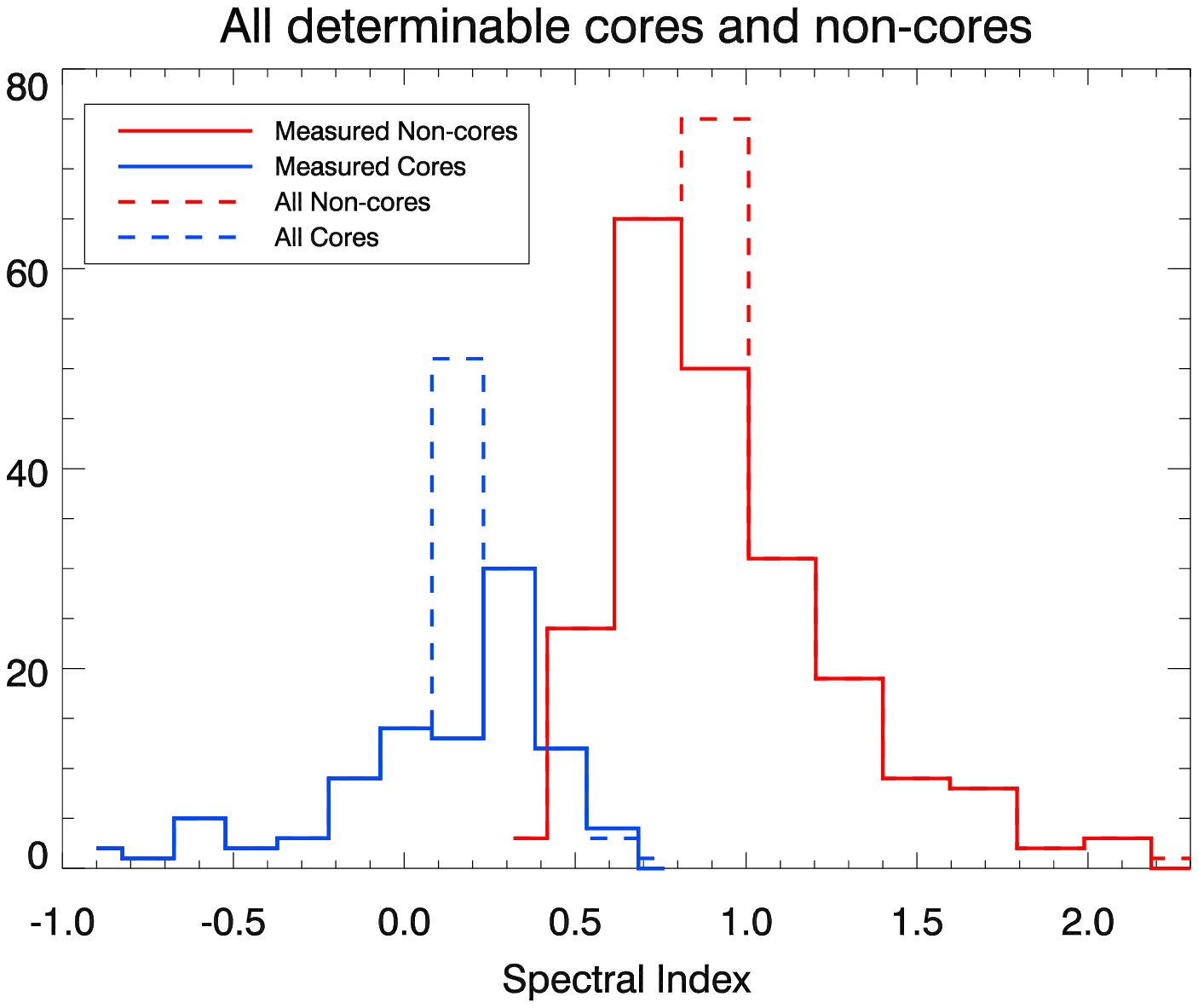}  
  \end{minipage}
  \caption{Same as for Figure \ref{Indices_MS} except now extended to include the full flux limited MS+.  We again include sources with an assumed index as these highlight the ratio of instances where an index could be directly determined to when, even with our relatively good spectral coverage, an index had to be assumed for an individual component.}  
  \label{Indices_MSp}
\end{figure}

Figures \ref{Indices_MS} and \ref{Indices_MSp} show the distributions of measured indices for our Main Sample and Main Sample+ respectively.  Whilst (partly by nature of the differentiation criteria) there is a clear split between the flat spectrum core component and the steep spectrum non-cores, some crossover of indices is seen.  For the core components, not all will necessarily be flat and exhibit self-absorption.  Indeed, for a `naked core' where the base of a recently triggered jet can be seen then an injection index of $\sim$0.6 may be expected \cite[][]{Laing14}.  Alternatively, extended emission can have a relatively flat spectrum ($\alpha$$\leq$0.6).  A tail to inverted core spectra is seen, mirrored by a tail to ultra-steep components for non-cores.    

Our choice of a representative indices of $\alpha$$_{core}$=0.2 and  $\alpha$$_{non-core}$=1.0 for extrapolations initially appear to be slightly too flat and too steep respectively when considering Figures \ref{Indices_MS} and \ref{Indices_MSp}.  These choices are however governed by conservatism.  If we consider the median index for the core and non-core when all measured indices for each of these components are included, we arrive at $\alpha$$_{core}$=0.33 and $\alpha$$_{non-core}$=0.96 (considering the Main Sample only, as this has most reliable index measurements).  However, if we restrict our sample to the subset of sources where both a measure of the core and non-core were obtainable our median indices are $\alpha$$_{core}$=0.28 and $\alpha$$_{non-core}$=1.08.  Often when a source is detectable out to frequencies above 10~GHz it either contains a clearly identifiable active core component or is a bright source, which in most cases have VLBI measurements.  For these reasons in most cases where core extrapolation is required it is from below the 10~GHz normalisation.  Therefore a flatter index provides a more conservative upper limit, hence our choice of $\alpha$$_{core}$=0.2

On the other hand, the majority of extrapolations for the non-core component were from below the 1~GHz normalisation frequency and hence a flatter index is more conservative.  We therefore choose $\alpha$$_{non-core}$=1.0 as our representative index.  For cases where an extrapolation of a given component is required we employ uncertainties of ${\alpha}$$_{error}$=$\pm$0.2.

The presence of a significant tail of ultra-steep non-cores suggests that such emission may be fairly common around BCGs.  Amorphous and mini-halo emission is often found to be very steep ($\alpha$$>$1.5), suggesting there may be a link between these structures and emission from a persistent AGN.

For sources not detected in our pointed observations, literature searches were performed to determine whether there was any weak, steep spectrum emission present at lower frequencies.  Limits on both the core and non-core components were determined by extrapolating with representative indices from the relevant survey limits.

\section{Monochromatic L-Band Radio Properties}  \label{L_band_section}

Initially we consider the flux limited monochromatic radio properties of our Parent Sample around 1.4~GHz, to gain an overview of the general radio properties.

\subsection{Radio Matching}
The optical BCG positions of the entire parent sample were cross-referenced with the NVSS and SUMSS radio catalogues, initially searching within their respective positional uncertainties ( 1-7'' for NVSS, 1-10'' for SUMSS). Additional uncertainty for the radio source position arises due to the relatively large beam sizes ($\sim$ 45''). Where a radio source was matched to within one beam size but greater than 2'' from the optical position, the sources were inspected visually to ascertain whether the match was robust.  Additionally, where multiple matches fell within the beam size they were visually inspected to ensure multiple components of individual sources were appropriately included.   

A sizeable proportion of the (e)BCS clusters ($\sim$73\%) fall within the FIRST \citep{Becker95} survey region (matched to SDSS, DEC limit $>$ -10), which has a higher resolution ($\sim$ 5'') and lower flux limit ($\sim$ 2 mJy) than NVSS.  A much smaller portion ($\sim$26\%) of sources within the REFLEX-NVSS sample also fell within the FIRST region. Sources within this overlap region were additionally cross-matched with the FIRST survey and radio-optical overlays were inspected.  With the higher resolution data of FIRST, we find that $\sim$15\% of sources tagged as matches in NVSS were likely not associated with the BCG. It should therefore be noted that the absence of a similarly high resolution southern survey is likely to introduce a bias in that mis-attributions of a radio-source to the BCG are less likely to be noticed in the southern hemisphere.  However, we also find that $\sim$10\% of sources in the FIRST footprint that were not detected by NVSS were either detected with the fainter threshold of FIRST or shown to be radio-loud independently of another nearby source and hence this bias is likely to be tempered by the lower detection threshold and higher resolution of the FIRST survey.

\begin{figure}
  \begin{minipage}[b]{0.5\linewidth}
  \centering  
  \includegraphics[width=9cm]{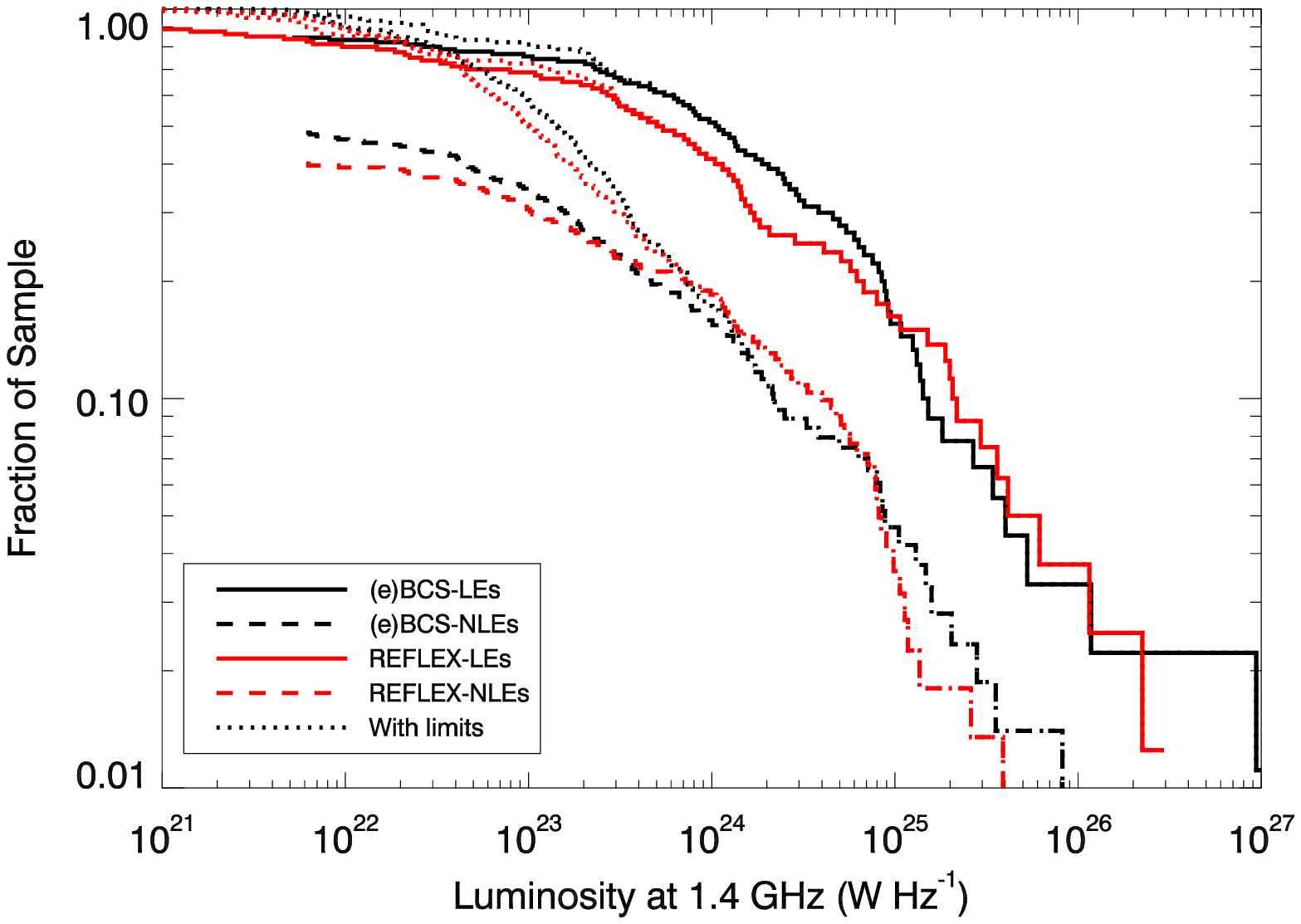}  
  \end{minipage}
  \begin{minipage}[b]{0.5\linewidth}
    \includegraphics[width=9cm]{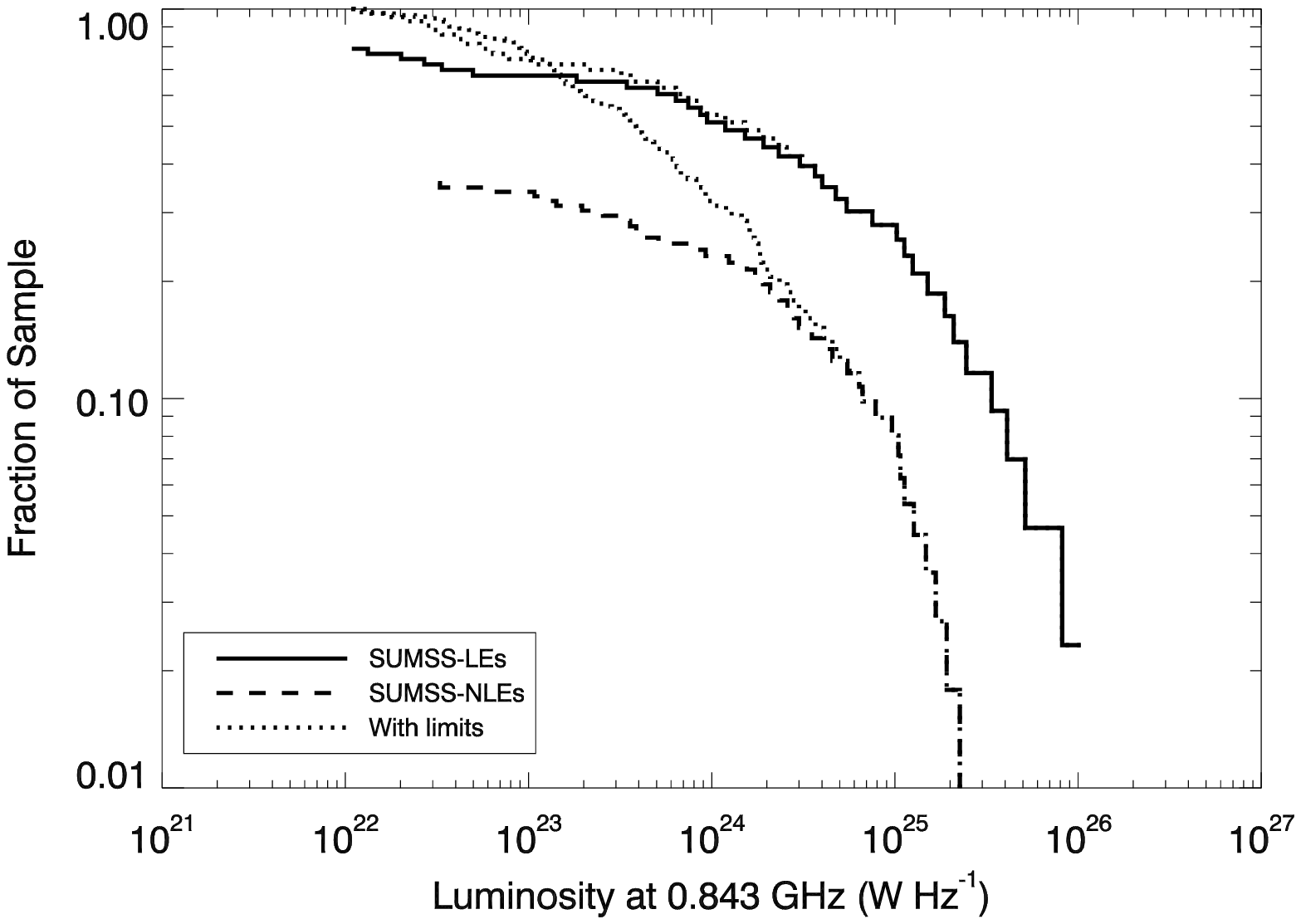}  
  \end{minipage}
  \caption{Radio luminosity functions (RLF) for all BCGs within the (e)BCS and REFLEX-NVSS (top panel) and REFLEX-SUMSS (bottom panel) X-ray selected catalogues, separated by the presence of extended optical emission lines. In both figures, solid lines trace LEs, dashed lines trace NLEs and dotted lines trace upper limits for galaxies that are radio-faint to the detection limit of the relevant survey.  A clear environmental dependence is seen for LEs to host more luminous sources. The top panel shows the 1.4~GHz RLF for all clusters that lie within the NVSS survey region for both the (e)BCS catalogue and REFLEX catalogue, with the relations for the two surveys showing good agreement and repeatability of the result.  The bottom panel shows the 0.843~GHz RLF for clusters in the REFLEX catalogue which fall within the SUMSS survey region.}
\label{fig:L_LFs}
\end{figure}

\subsection{LE and NLE Match Rates} \label{LE_NLE_Match_Rates}

The number of BCGs matched to radio-sources can be seen in Table \ref{eBCSREFLEXBreakdown}, along with a breakdown of whether these BCGs are optically line-emitting or not (LE or NLE).  We find detection percentages of 61.1$\pm$5.5\%, 62.6$\pm$5.5\% and 60.3$\pm$7.7\% for (e)BCS, REFLEX-NVSS and REFLEX-SUMSS respectively, assuming simple binomial errors with a 95\% confidence level.  

These BCG radio-detection rates are slightly elevated compared to other X-ray selected samples.  For example, \cite{Sun09} found a 50.3\% detection rate for a luminosity cut of L$_{1.4GHz}$ $>$ 10$^{23}$WHz$^{-1}$ for their sample of 43 nearby galaxy groups with Chandra archival data. A $\sim$50\% detection rate was found for both the Brightest 55 \cite[B55,][]{Edge90, Peres98} and The HIghest X-ray FLUx Galaxy Cluster Sample (HIFLUGCS) \cite[][]{Reiprich02, Mittal09} samples.  \cite{Ma13} studied a combined sample of 685 X-ray selected galaxy clusters.  Matching to the NVSS, they found a matched detection rate between the cluster coordinates and radio-sources $>$3 mJy of $\sim$52.1\% (or $\sim$48.5\% accounting for expected background contamination).  Where possible we have used the lower flux detection threshold of the FIRST survey and also matched the optical BCG positions rather than cluster X-ray positions.  Considering these factors, we have reasonable agreement with previous radio-BCG detection rates in X-ray selected clusters. 

Our detection rate is significantly higher than for otherwise derived samples \cite[e.g. $\sim$30\% for optical selection,][with BCG mass-dependency]{Stott12, Best07}.  This reflects the fact that X-ray selected samples are biased towards selecting the most massive, settled clusters which are more likely to host a radio-loud BCG \cite[][]{Burns90, Mittal09}.  However, a fraction of our non-detections are likely to be radio-loud but below our detection threshold and so our detection rate can only provide a lower limit on the overall duty cycle of radio-loudness of BCGs. The mis-identification of some clusters that contain central radio sources, due to X-ray emission being incorrectly attributed to the AGN rather than the ICM, may also contribute.  We discuss possible cluster survey mis-identifications in Section \ref{SurveySection}.

More interesting is the difference in the detection rate between LE and NLE BCGs.  We detect 84.1$\pm$7.6\%, 87.0$\pm$7.5\% and 83.7$\pm$11.0\% of the LEs and 51.6$\pm$6.7\%, 54.7$\pm$6.6\% and 51.3$\pm$9.2\% of the NLEs in (e)BCS, REFLEX-NVSS and REFLEX-SUMSS respectively, again assuming a simple binomial errors with a 95\% confidence level.  Note that these detection fractions are affected by inconsistent flux limits and that if we instead consider only our C-band pointed observations of the Main Sample BCGs then we recover detection fractions of 90.8$\pm$5.1$\%$ for LEs and 48.9$\pm$8.8$\%$ for NLEs down to our typical detection limit of $\sim$0.3 mJy.  These detection fractions compare favourably with those of \cite{Mittal09}, who report detection fractions of 45$\%$, 67$\%$ and 100$\%$ for non-cool cores, weak cool cores (1Gyr $<$ t$_{cool}$ $<$ 7Gyr) and strong cool cores (t$_{cool}$ $<$ 1Gyr) respectively, for the HIFLUGCs sample of the 64 X-ray brightest galaxy clusters. 

If the simplest case is taken whereby the detection rate is taken as a proxy for the duty cycle of AGN activity \cite[e.g.][]{Lin07} then this provides clear evidence that the duty cycle of BCGs in strong cooling flow environments is higher than for BCGs in less extreme cluster centres.

\subsection{Luminosity Functions}

A further clear distinction in the radio properties between LEs and NLEs becomes apparent by considering their luminosity distributions (see Figure \ref{fig:L_LFs}).  Radio sources in LE BCGs are on average more powerful than for NLEs.  The distributions in Figure \ref{fig:L_LFs} are effectively complete to $\geq$few 10$^{23}$WHz$^{-1}$.  Although the distributions in the directly comparable (e)BCS and REFLEX-NVSS samples (Figure \ref{fig:L_LFs} top panel) are not perfectly aligned (see below), the same general result is seen.

\subsubsection{Redshift Considerations} \label{RED_CONS}

\begin{figure}
  \begin{minipage}[b]{0.5\linewidth}
  \centering  
  \includegraphics[width=9cm]{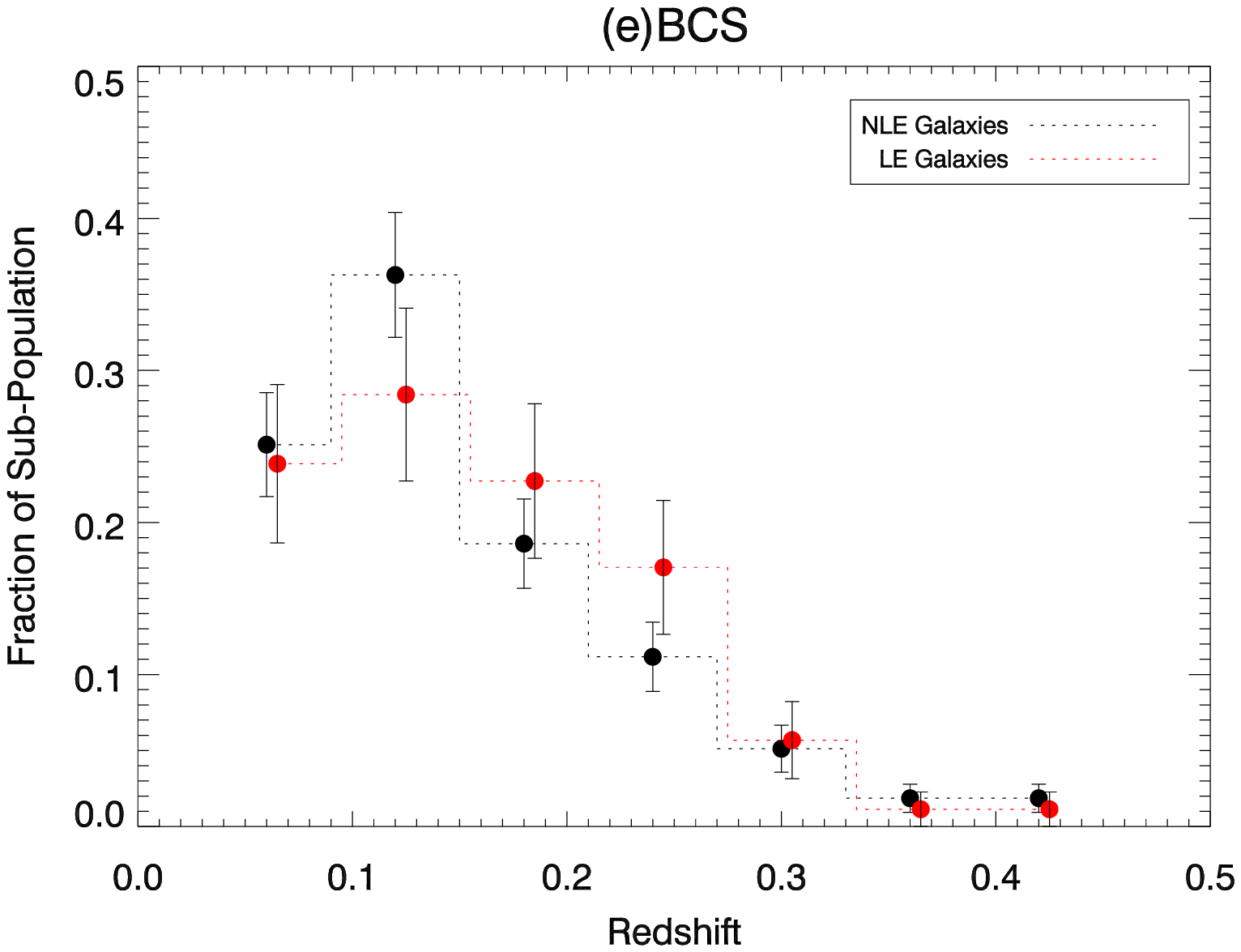}  
  \end{minipage}
  \begin{minipage}[b]{0.5\linewidth}
    \includegraphics[width=9cm]{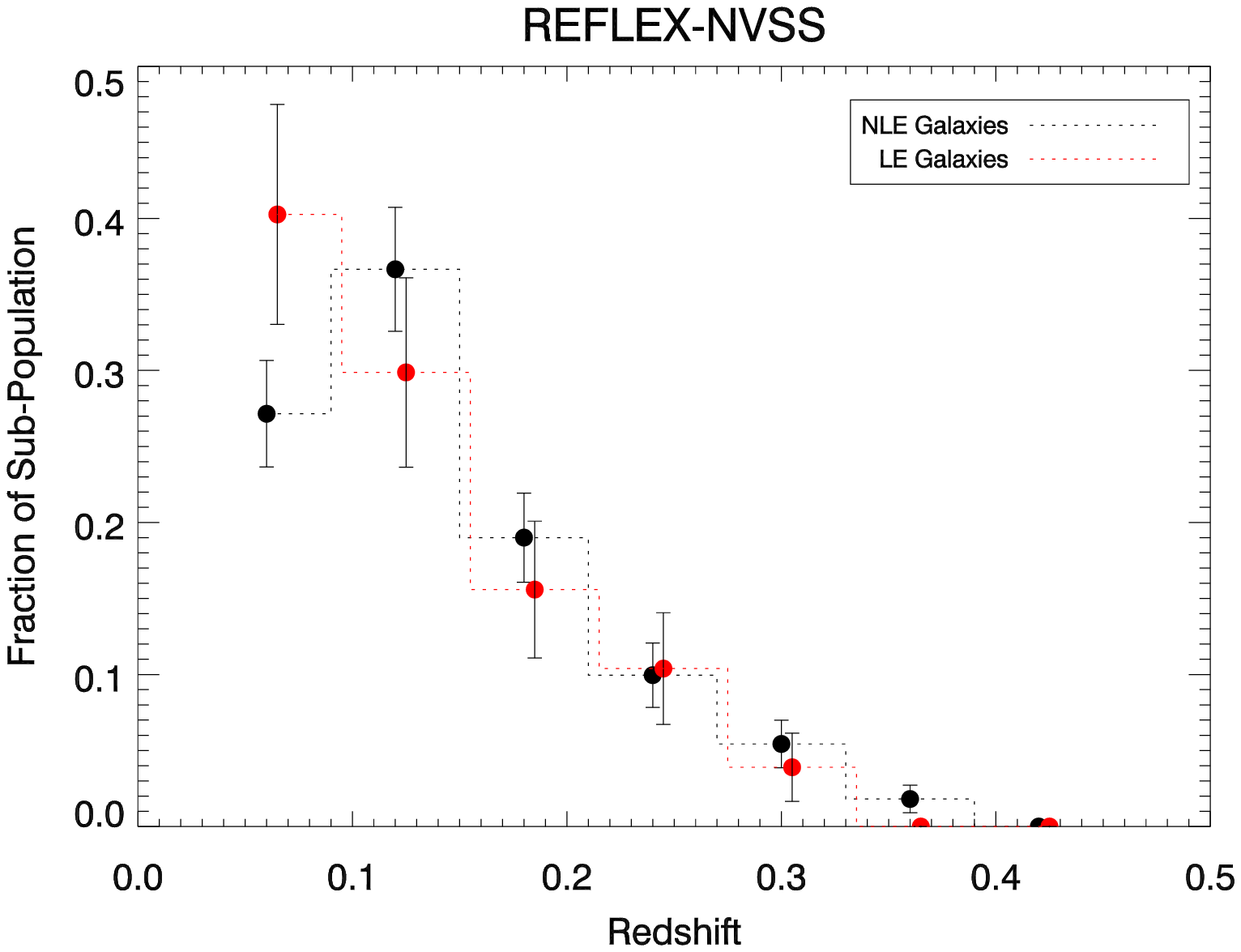}  
  \end{minipage}
  \begin{minipage}[b]{0.5\linewidth}
    \includegraphics[width=9cm]{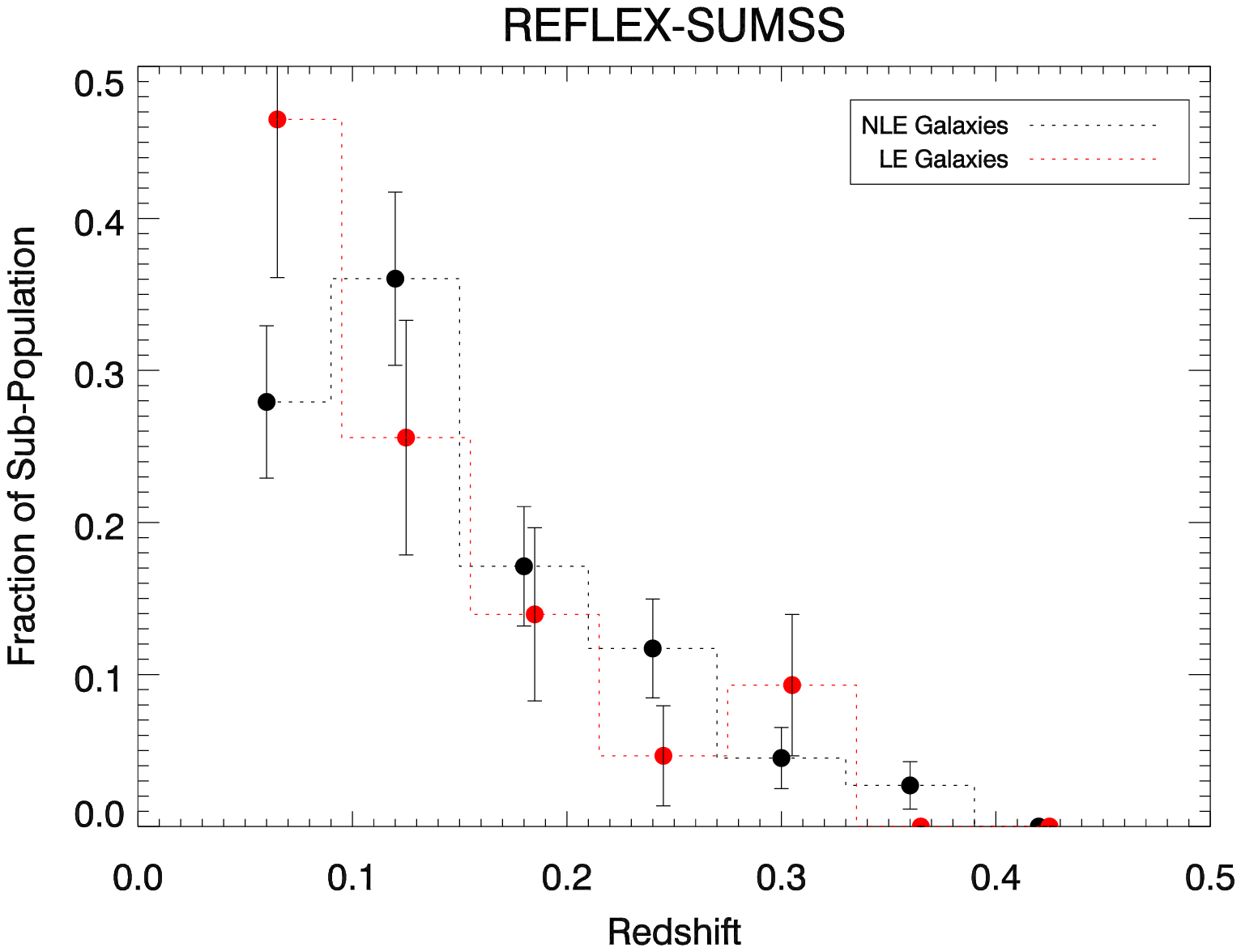}  
  \end{minipage}
  \caption {Binned redshift distributions for the radio-matched BCGs within the (e)BCS (top panel), REFLEX-NVSS (middle panel) and REFLEX-SUMSS (bottom panel) samples.} 
  \label{fig:redshiftdistributions}
\end{figure}

Differences in the redshift distributions of clusters in the two samples may account for some of this scatter.  Malmquist bias would give preference for detecting higher luminosity sources in a sample with a higher redshift distribution.  The redshift distributions of the clusters with radio-loud BCGs for the three sub-samples are shown in Figure \ref{fig:redshiftdistributions}, with 1$\sigma$ uncertainties assuming Poissonian errors.  There is no significant difference in the redshift distribution of our line emitting and non-line emitting BCGs.  For the REFLEX sample, to low significance there are proportionally more LE BCGs in the lowest redshift bin (z$<$0.1).  However, this would suggest a higher redshift distribution for our NLEs than LEs and should therefore be expected to bias our NLE sample to be more luminous than LEs. Hence such a bias, if present, actually serves to strengthen our result.

\begin{figure}
  \begin{minipage}[b]{0.5\linewidth}
  \centering  
  \includegraphics[width=9cm]{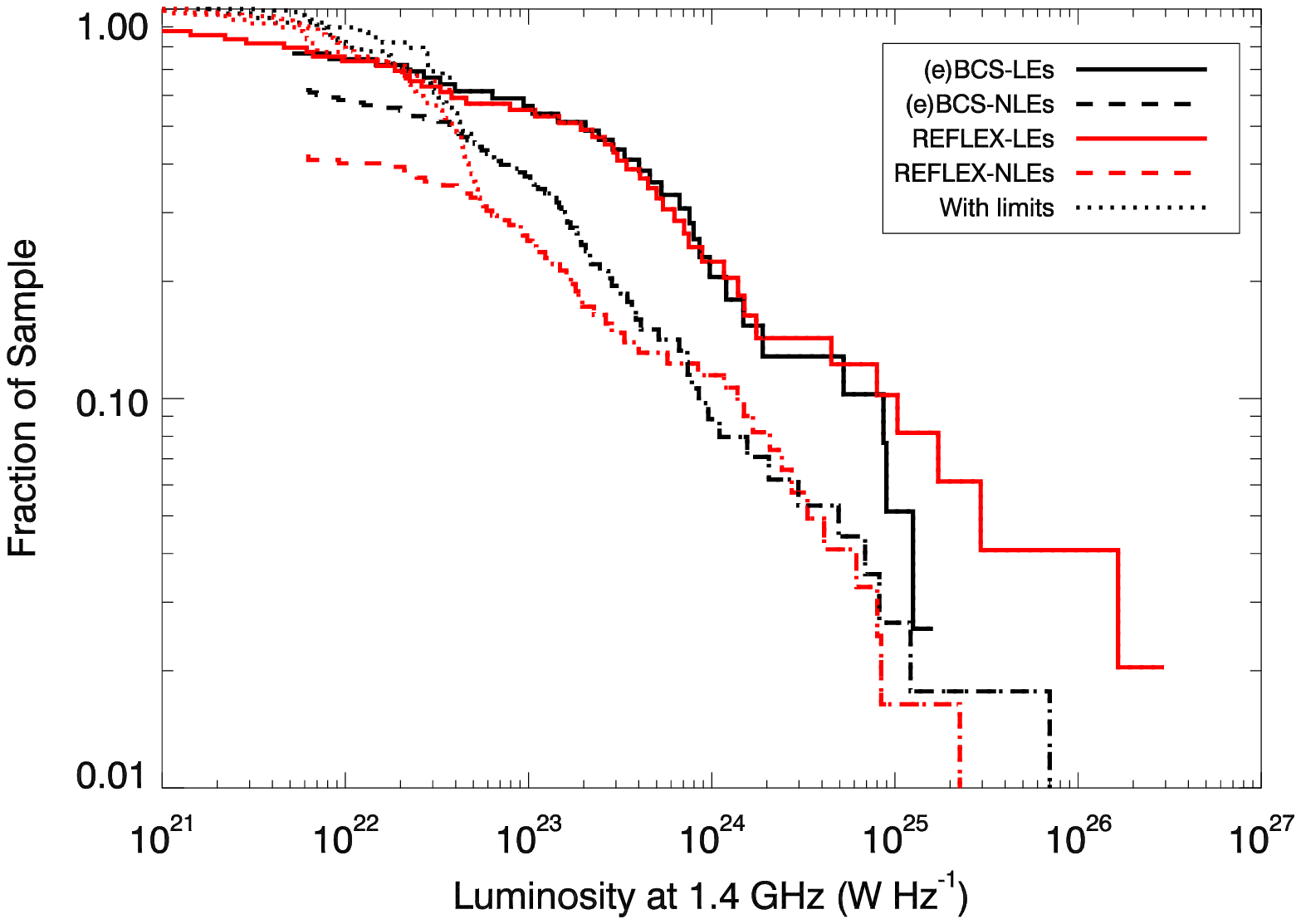}  
  \end{minipage}
  \begin{minipage}[b]{0.5\linewidth}
    \includegraphics[width=9cm]{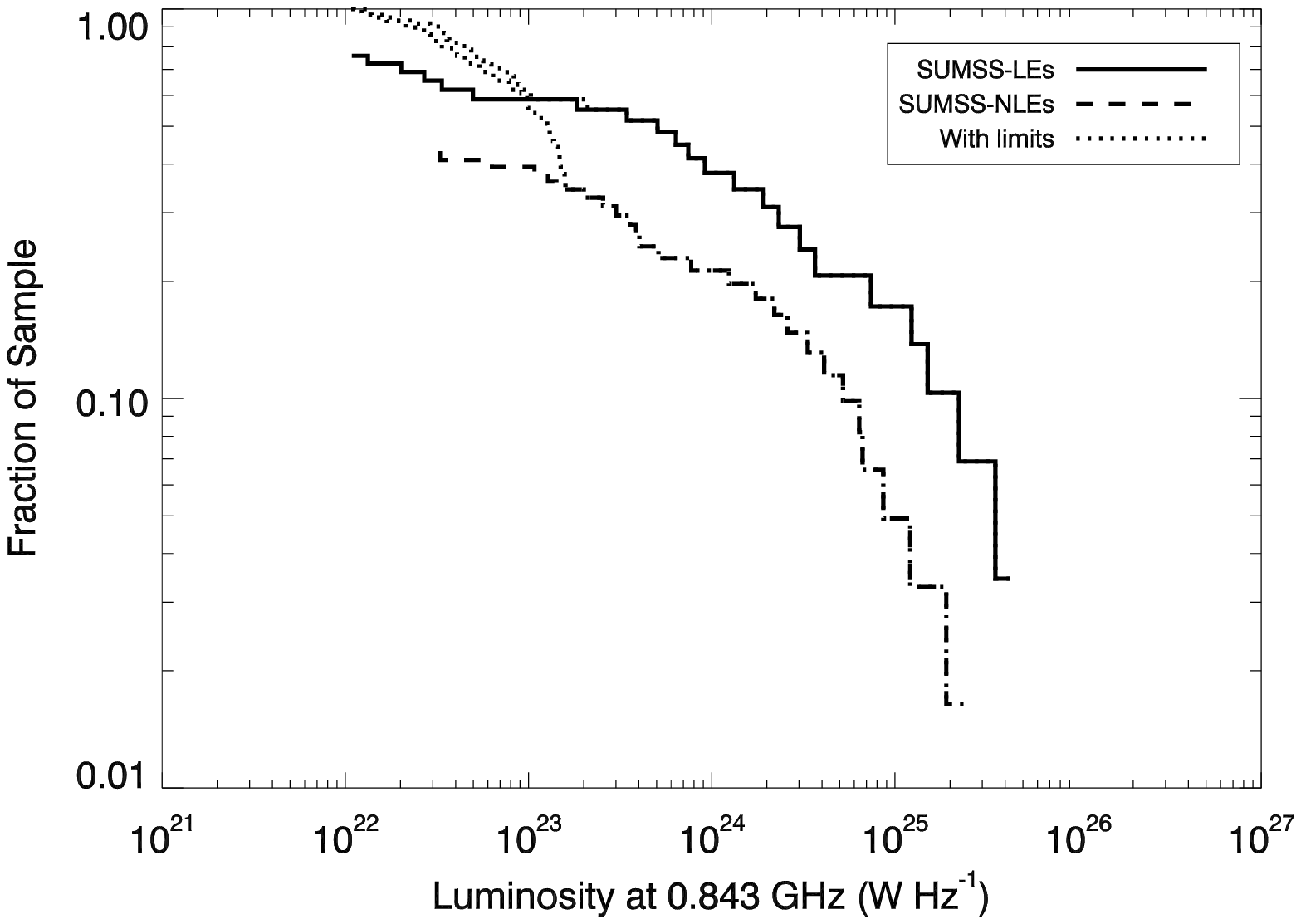}  
  \end{minipage}
  \caption{As for Figure \ref{fig:L_LFs} although restricted to z $<$ 0.1 clusters.  We see that the result remains, showing that it is a true physical result and not merely due to redshift bias.  In particular note that the relations for (e)BCS and REFLEX-NVSS now appear to align better.  Better identification of sources in REFLEX covered by NVSS but not FIRST at lower redshifts could account for this.}
\label{fig:LOWZ_L_LFs}
\end{figure}

To minimise any redshift related uncertainty, the distributions were replotted for only those clusters at z $<$ 0.1 (see Figure \ref{fig:LOWZ_L_LFs}).  The distributions between the (e)BCS and REFLEX-NVSS samples now appear to agree with less scatter and the trend for LEs to host more powerful radio sources remains.  

It should be considered that there is a decreasing fraction of strong cool cores observed to higher redshift, although this evolution is not believed significant over the redshift range of our sample \cite[][]{Santos10, Samuele11, McDonald13b}.

There were 36 matched BCGs within our sample which lie in the overlap of the NVSS and SUMSS regions.  This relatively small sample was found to have a reasonably uniform range of $\nu_{0.843GHz}$ - $\nu_{1.4GHz}$ indices of $\alpha$ = -0.27 - 2.12.  This range of indices highlights the wide diversity of radio activity found within the BCG population.  It also highlights the large uncertainties that are invoked by assuming a single spectral index for a population of sources when extrapolating too far from a single frequency observation.  As such we do not extrapolate the REFLEX-SUMSS sample to 1.4~GHz for direct comparison (Figures \ref{fig:L_LFs} and \ref{fig:LOWZ_L_LFs}) and instead note only that the luminosity distinction between LEs and NLEs remains.


\subsection{Cluster X-Rays to L-band Radio Connection} \label{Cluster_Xray_section_Lband}

\begin{figure}
\begin{center}
\includegraphics[width=9cm]{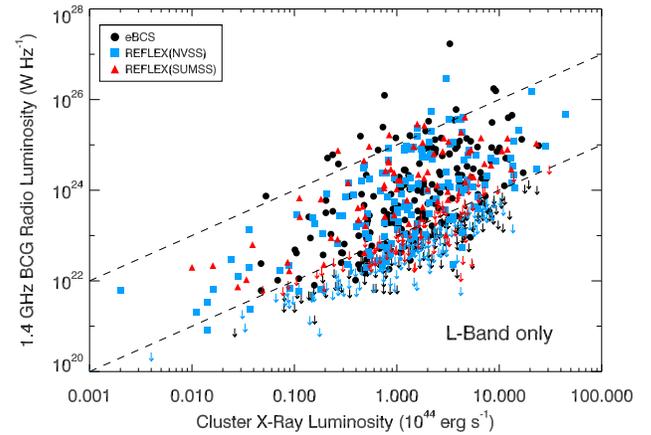}
\caption{Monochromatic (1.4~GHz) radio luminosity of the BCG as a function of its cluster's luminosity.  The empty corner to the bottom right reflects the detection limit of the radio surveys rather than a true correlation whereas the bound to the upper-left shows that more radio-luminous BCGs tend to inhabit more X-ray luminous clusters.  Note that REFLEX-SUMSS radio fluxes were extrapolated using an index of -0.8 for inclusion/completeness.  See Section \ref{RED_CONS} for why this extrapolation is uncertain.  X-ray luminosities were corrected to the same cosmology.  The dashed lines are illustrative only and represent one-to-one normalisation with a two order-of-magnitude separation.}
\label{fig:L_BAND_XRAY}
\end{center}
\end{figure}

To further investigate how the radio-behaviour of the BCG connects with the global properties of the host cluster we retrieve the integrated cluster X-ray emissions from the BCS, eBCS and REFLEX catalogues.  Table 6a of the REFLEX catalogue contains the X-ray luminosities calculated using our cosmology.  However, the BCS and its extension both give X-ray luminosities for an Einstein-de-Sitter universe with H$_{0}$=50.  We correct this to our cosmology, with typical conversion factors of 0.5-0.7 for our redshift range.
We show in Figure \ref{fig:L_BAND_XRAY} the monochromatic L-band radio powers for all sources in our Parent Sample as a function of the X-ray luminosity of their host clusters.  A general trend for the highest radio luminosity sources to inhabit the most X-ray luminous clusters is seen, albeit with significant scatter. The extended nature of clusters means that the catalogue observations sample regions of much greater physical extent than just the cluster core. The integrated X-ray luminosity is therefore driven predominantly by the cluster mass, with more massive clusters corresponding to more X-ray luminous systems.  The correlation observed in Figure \ref{fig:L_BAND_XRAY} is therefore indicative of the most massive clusters tending to host the most powerful radio-AGN, which may itself be due to more massive clusters hosting more massive BCGs and in turn more massive central black holes.  A secondary effect, for cool-core clusters to have elevated X-ray luminosities due to the higher core densities could also be contributing.  However this is likely to be a much more subtle effect than the mass dependency and is not the dominant driver of the correlation.

\section{Core and Non-core Emission} \label{Core_notcore_section}

As stated in the introduction, the true nature of any observed radio activity is difficult to understand when observing a sample at only one frequency.  We now consider the properties of the decomposed core and non-core radio components, their relationship to each other and their separate and linked relationships to other cluster and AGN properties.

\subsection{Linked Emission}

\begin{figure*}
  \begin{minipage}{\linewidth}
  \centering
  \includegraphics[width=16cm]{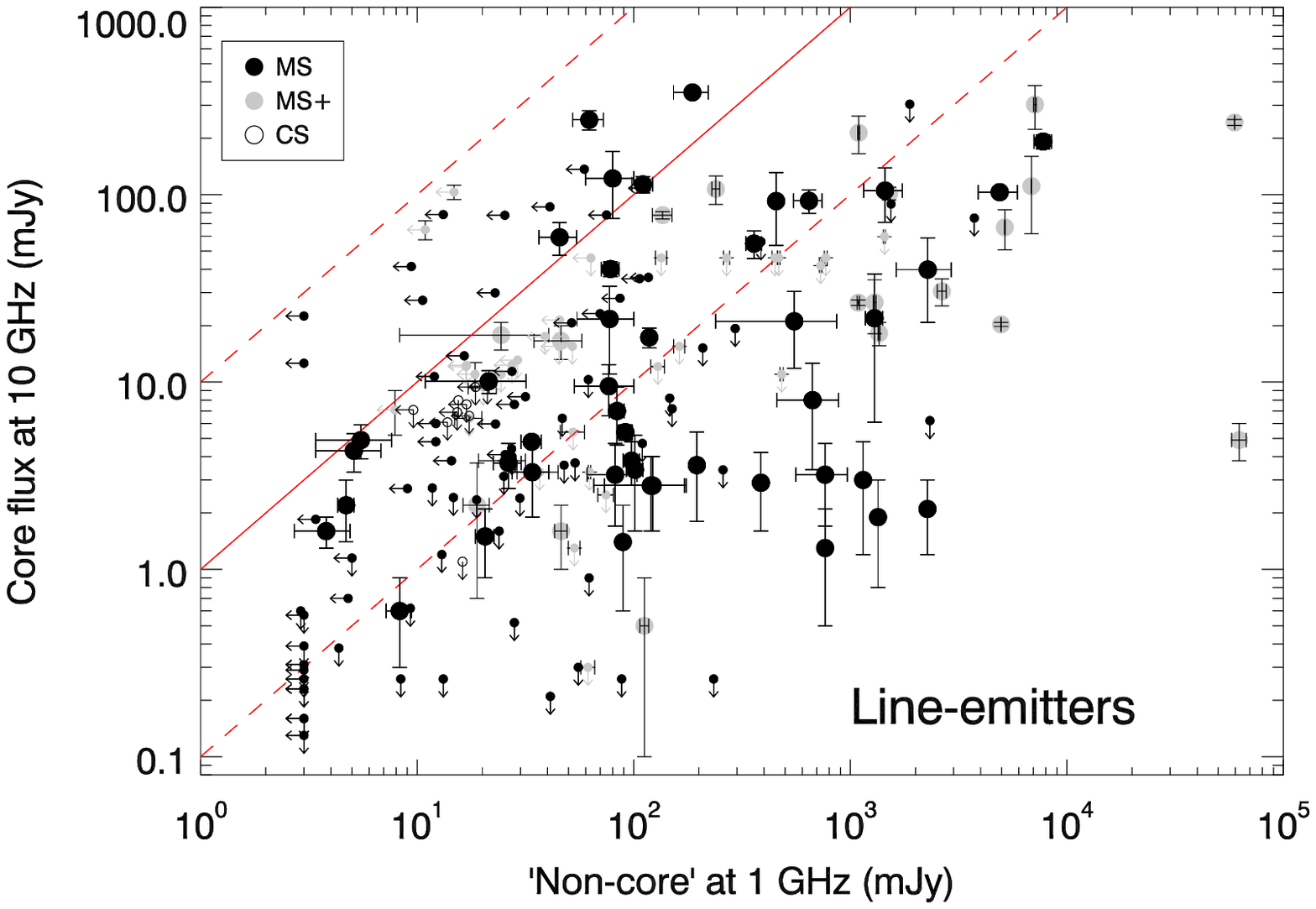}
  \end{minipage}
  \begin{minipage}{\linewidth}
  \centering
  \includegraphics[width=15cm]{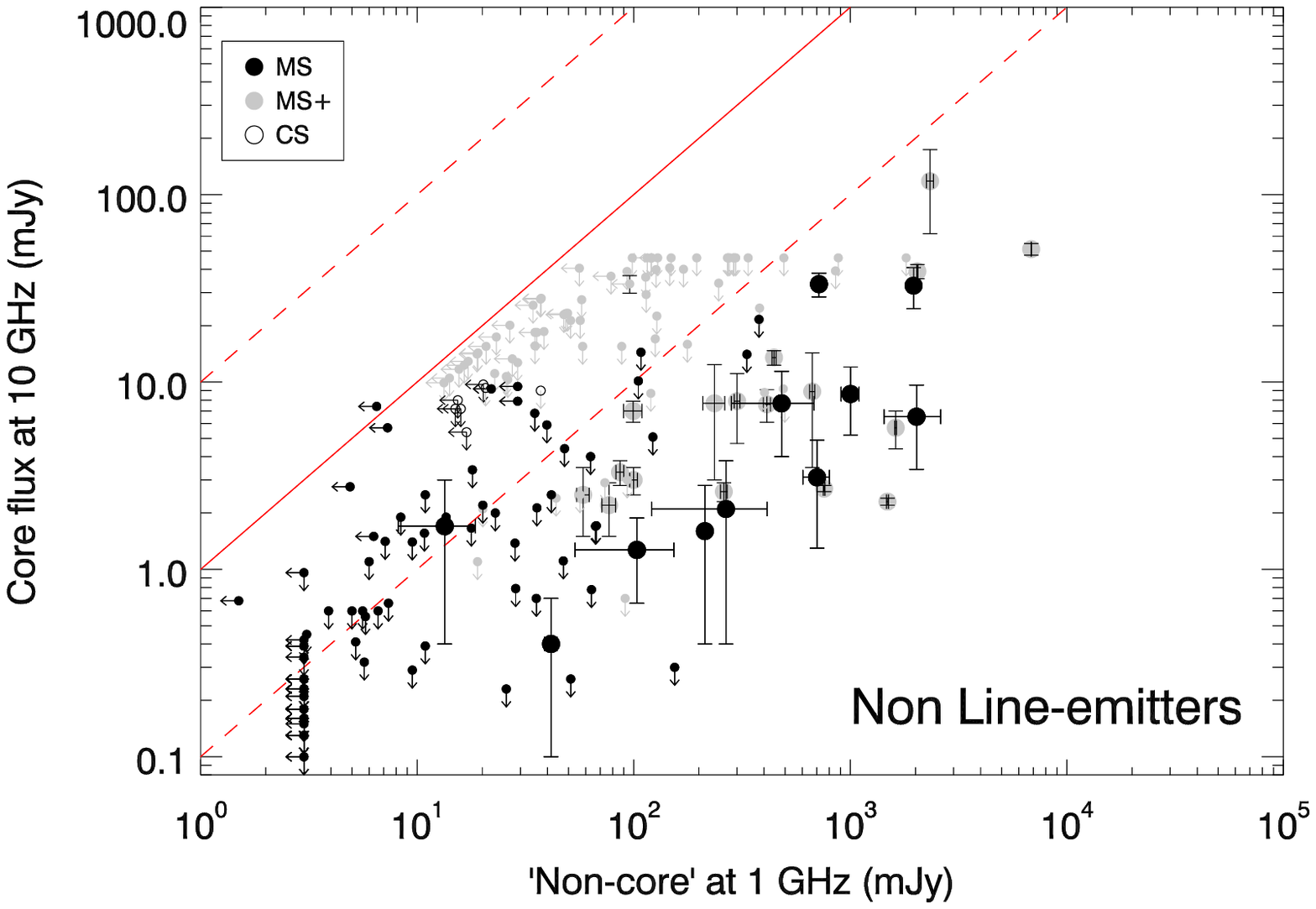}  
  \end{minipage}
  \caption{Comparison of the flux density of the radio core component (normalised at 10~GHz) to that of the more diffuse radio component
(normalised at 1~GHz) for those objects that exhibit optical line emission (top) to those that do not (bottom). The solid line
marks equal normalisation of both components, the dashed lines a factor of ten either side. It appears that LE clusters are significantly more likely
to be core dominated radio sources whilst also undergoing the most powerful events. Solid grey points in both plots are sources in the MS+ (L-band $>$ 15 mJy) which are not covered by the Main Sample (see text). Open symbols complete the radio coverage down to the flux limit of the SUMSS survey (15 mJy $>$ S$_{L-band}$ $>$10 mJy) and hence occupy only a narrow range in flux density.}
  \label{THE_DIAGRAM}
\end{figure*}

For each of the sources within the Main Sample, we take monochromatic proxy measures of the core and non-core components at 10 and 1~GHz respectively.  Ideally, the cores would be normalised at a higher frequency and non-cores at a lower frequency.  However, often our radio SEDs are more sparsely sampled far away from the traditional cm-range wavelengths.  Hence, our choice of 10~GHz and 1~GHz for the normalisations are a compromise between ensuring the relevant components can be separated whilst still being within the best-sampled section of the radio waveband.  We plot a radio `colour-colour diagram' of the 10~GHz core normalisation against the 1~GHz non-core normalisation in Figure \ref{THE_DIAGRAM}.

\begin{figure}
\begin{center}
\includegraphics[width=9cm]{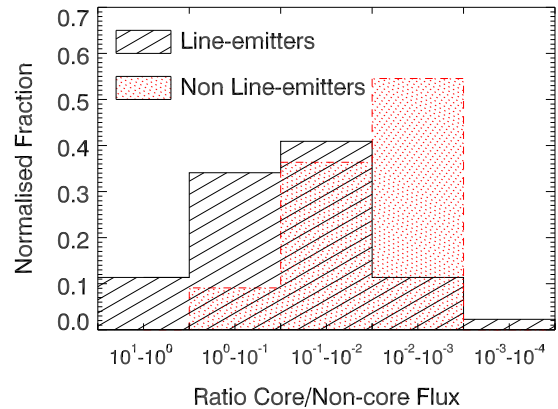}
\caption{Distribution of the core to non-core ratio of normalisations for the sources in the Main Sample only, that have determined values for both components.  This sub-sample of data is incomplete yet represents the most well-sampled of the data displayed in Figure \ref{THE_DIAGRAM} and highlights that LE-BCGs are on average much more likely to host core dominated radio sources than NLEs.}
\label{MS_HISTOGRAM}
\end{center}
\end{figure}

Both cluster types can host powerful radio sources, yet BCGs in strongly cooling clusters are much more likely to be core dominated. To further elucidate this difference in distribution, in Figure \ref{MS_HISTOGRAM} we plot a histogram of the ratio of the 10~GHz core measure to the 1~GHz non-core measure for the sources in the Main Sample that have well determined values for {\it both} components.  As in Figure \ref{THE_DIAGRAM}, we see a tendency for LEs to host much more core dominated radio-sources.

Strong core components only appear to be found in LEs.  The presence of very powerful non-cores in NLEs, normally associated with classical Fanaroff-Riley type I (FR-I) and type II (FR-II) sources, belies that there must at some times be powerful core activity in these objects.  It stands to reason that in a much larger sample we would find such events.  However, that they are not present in the current sample suggests that such core outbursts must be short-lived.  Indeed we find that for the 48.8\% of sources that are identified as LEs in our Main Sample, 60.2\% have a determinable core component of which 83.1\% are more powerful than 10$^{23}$WHz$^{-1}$ at 10~GHz (see also Section \ref{LEtoNLE_differences}).  However, of the remaining 51.2\% of sources tagged as NLEs only 11.6\% have an identifiable core component.  Only 5\% of these NLE core components are more powerful than 10$^{23}$W~Hz$^{-1}$ at 10~GHz and none has a core power above 5$\times$10$^{23}$WHz$^{-1}$ (see also Section \ref{DISCUSSION}).  This suggests an upper limit to the {\it core} duty cycles of NLEs of $<$1\% with core powers greater than 5$\times$10$^{23}$WHz$^{-1}$ whereas the prevalence of powerful core activity is much higher and more persistent, in LEs.  It is the presence (or lack) of an active {\it core} component that appears to be the most important distinction between the radio properties of LE and NLE BCGs, even more so than their overall incidence of radio loudness (see Section \ref{LE_NLE_Match_Rates}). 

The wide range of ratios of these two components has important implications for the variety of sources that will be contained within any flux selected radio sample, with samples selected at different frequencies containing vastly different source types.  This is discussed in Section \ref{SurveySection}

\subsubsection{Complete Sample}

To ascertain whether our Main Sample is representative of the full parent sample, in Figure \ref{THE_DIAGRAM} we also show (in light grey) the decompositions for all sources in the Main Sample+ that do not fall within the Main Sample.

Considering the top panel of Figure \ref{THE_DIAGRAM} we see that for the LEs, the Main Sample (black points) well samples the parent sample.  For the NLEs (bottom panel) it initially appears that many grey points lie above the black points. Almost all of these are shallow upper limits on the core component for sources where there is no indication of a strong active core.  Sources in the Main Sample are similarly spread in non-core flux but have deeper limits on the core components from the targeted observations at higher frequencies.  It therefore appears reasonable to assume that the advent of deeper higher frequency surveys \cite[e.g. AT20G deep,][]{Franzen14} will push the MS+ core limits down.  We therefore believe that the Main Sample for NLEs is indicative of the true distribution of the parent sample.  

Note that although the limits on NLEs in the Main Sample+ core components do tend to lie above the corresponding points in the Main Sample, they are still all below the brightest cores seen in LEs. 

For completeness, we also include in Figure \ref{THE_DIAGRAM} {\it approximate} positions for the remaining sources in the Parent Sample that have L-band radio fluxes between 10 mJy and 15 mJy.  This is the level to which {\it all} the L-band surveys used are complete.  This is refered to as the Completeness Sample (CS) in the image, and these points do not appear to lie anywhere special on the diagram.

\begin{figure}
  \begin{minipage}[b]{0.5\linewidth}
  \centering  
  \includegraphics[width=9cm]{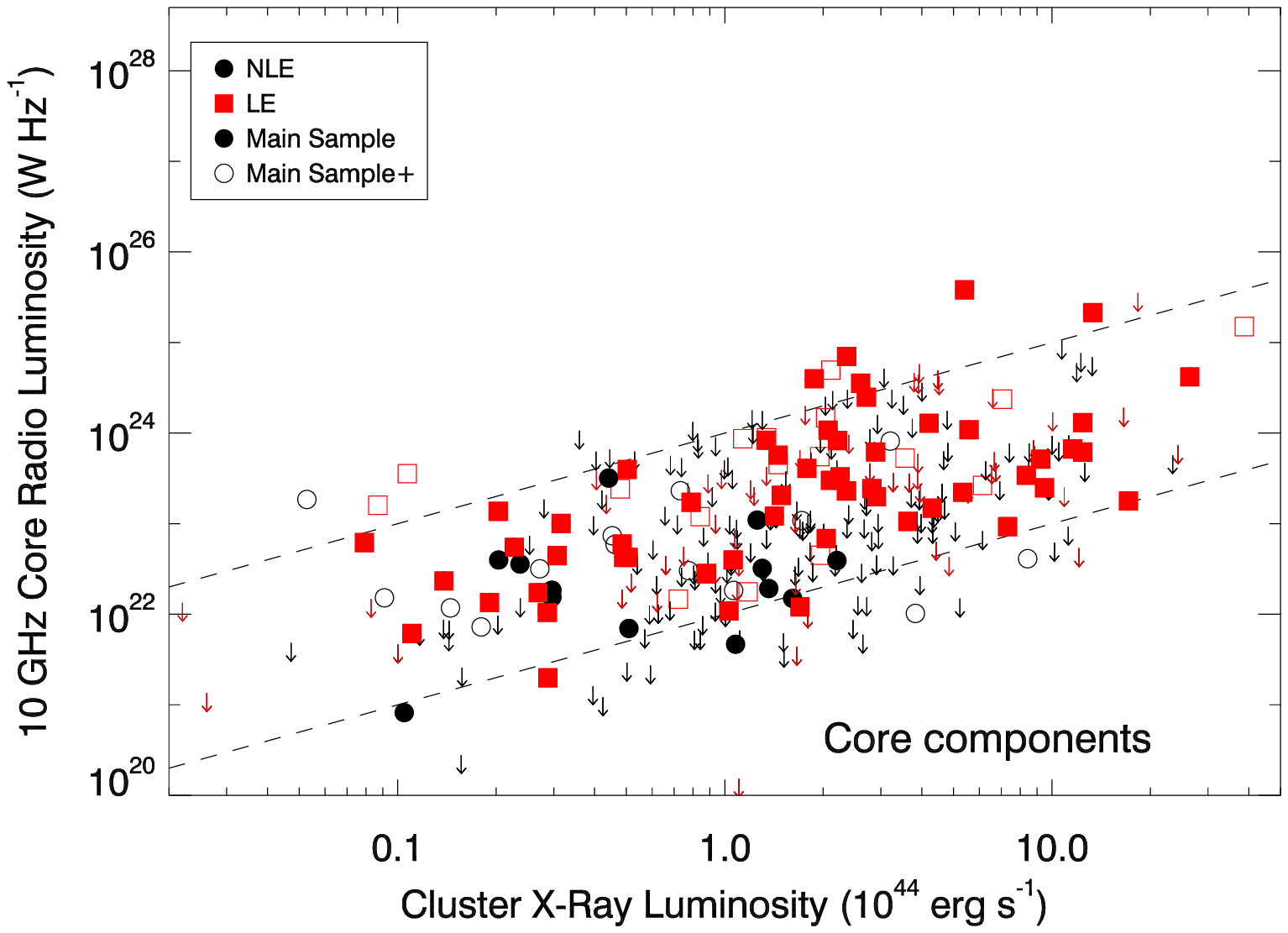}  
  \end{minipage}
  \begin{minipage}[b]{0.5\linewidth}
    \includegraphics[width=9cm]{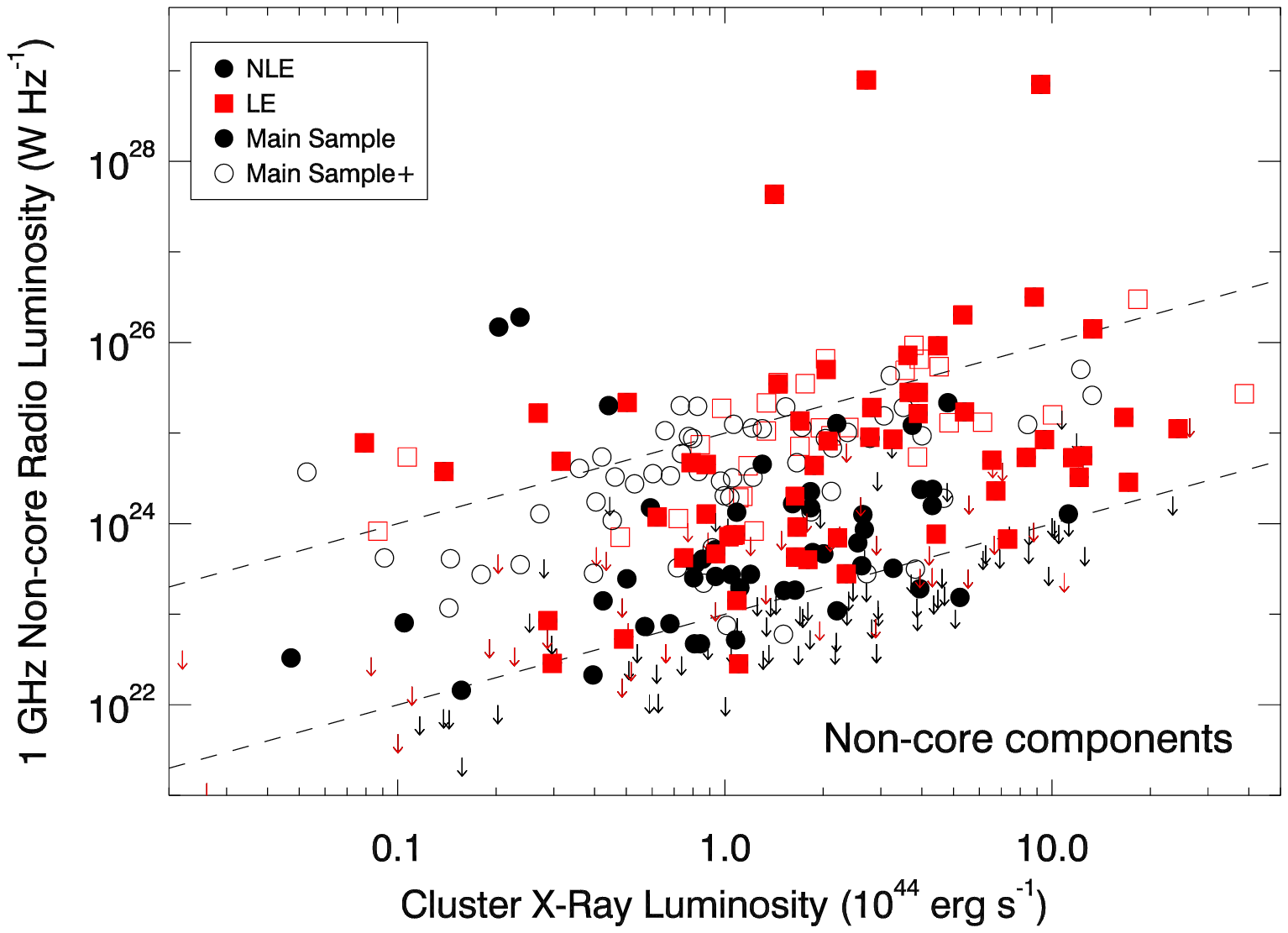}  
  \end{minipage}
  \caption{Relationship between the core (top) and non-core (bottom) radio components of the BCG to the overall X-ray luminosity of the host cluster.  Note that the strongest cores appear only to occur in the most X-ray luminous and hence most massive clusters whereas strong non-cores, attributable to single outburst lobe-dominated systems may also be found in low X-ray luminosity clusters.  Black points are non-line emitters (NLEs) whereas red points are line-emitters (LEs).  The dashed lines are illustrative only and represent one-to-one normalisation with a two order-of-magnitude separation.} 
  \label{CORE_PED_XRAY}
\end{figure}

\subsection{Cluster X-Rays to Decomposed Radio Connection} \label{Cluster_Xray_section_splits} 
For the full Main Sample+ sources, we plot the luminosity of their core and non-core radio components against their cluster X-ray luminosities in Figure \ref{CORE_PED_XRAY}.  Whilst there appears to be a general tendency that the most powerful cores only occur in the most X-ray luminous clusters, there is a population of powerful non-cores ($>$10$^{24}$ W Hz$^{-1}$) found in low X-ray luminosity ($<$5$\times$10$^{43}$ erg s$^{-1}$) clusters.  In the simplest scenario it would be expected that bright cores should evolve into bright non-cores.  The lack of bright cores alongside the presence of bright non-cores in low X-ray luminosity clusters is inconsistent with this simple picture, hence we must consider alternate explanations for the bright non-cores.

\subsubsection{Activity Cycle Considerations}
Whilst strong non-cores can be the result of either the superposition of multiple previous periods of activity or the fading of a single powerful outburst, a powerful core can only be observed as the result of strong, current activity.  Powerful non-cores in low X-ray luminosity clusters could therefore be explained by a couple of different scenarios.  Individual periods of core activity may be genuinely less powerful in less X-ray luminous clusters but happen relatively frequently, thus leading to a build-up of material.  Alternatively, powerful core outbursts may occur in low X-ray luminosity clusters yet be relatively rare and simply missed by our sample.

\begin{figure}
\begin{center}
\includegraphics[width=9cm]{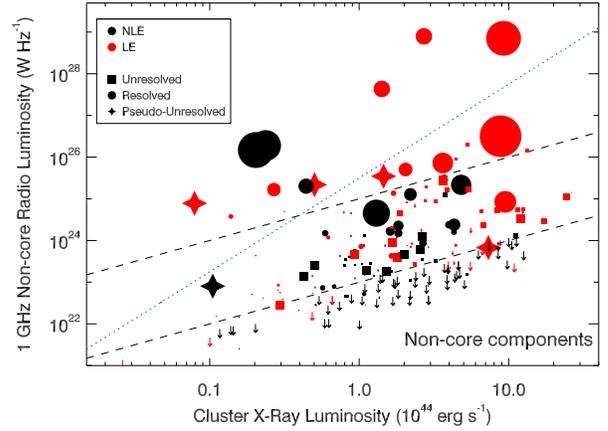}
\caption{As for bottom panel of Figure \ref{CORE_PED_XRAY} except only for the Main Sample and now symbols are sized by physical size of the non-core component, as seen at C-band with either the VLA C-Array or ATCA.  Note that stars represent sources where extended emission is resolved out at this resolution but where lower resolution and/or lower frequency observations show distinctive lobed features - i.e. they are resolved-out with our sampling but there is known extended emission in the systems, so they are effectively `pseudo-unresolved'.  Colours again represent the state of the cluster (black NLEs, red LEs) although circles now show sources that are resolved (at this resolution) and squares those that are unresolved and hence represent upper limits.  Sizes were measured as Largest Angular Size (LAS).  For unresolved sources, the major axis of the beam was taken as the size limit.  The dashed lines are illustrative only and represent one-to-one normalisation with a two order-of-magnitude separation.  The dotted line represents a normalisation power-law of order three and is used only as a differentiator (see text).}
\label{Physical_Sizes}
\end{center}
\end{figure}

The former of these two possible explanations could potentially explain the observed `amorphous haloes' observed around many BCGs (e.g. A2052, \citealt{Blanton01, Venturi04}), and of which our current sample contains several candidates (e.g. A2627, A407, AS753, AS861, RXJ1720.1+2638; also see Appendix \ref{APPENDIX_NOTES}).  This halo would contain multiple distinct populations of electrons from a number of previous outbursts, all at different stages of ageing but crucially none of which would be undergoing injection and hence manifest as an active, flatter spectrum core.  In this scenario, the superposition of weak populations would build to give the overall bright non-core.  High resolution spectral index maps of amorphous haloes could support this, if the distribution of the spectral index across the haloes was found to be clumpy rather than smoothly varying. Resolution effects at the varying frequencies would however make such an endeavour difficult.  

To attempt to break the degeneracy between single outburst and accumulated non-cores, we considered the physical sizes of the sources in the Main Sample.

\subsubsection{Size Considerations} \label{SIZE_SECTION}
We measured the largest angular size (LAS) at C-band ($\sim$4.8~GHz) of each source in the Main Sample using VLA C-array band imaging for the northern sources and ATCA-6km for the southern sources.  There is a small bias here in that the VLA-C is slightly more sensitive to resolving larger scale structures.  Sizes were extracted using the AIPS task \textsc{TVDIST} for resolved sources or alternatively, the major-axis of the Gaussian beam as fit by the AIPS task \textsc{JMFIT} was taken as an upper-limit on the LAS for unresolved sources.  Angular sizes were converted to physical scales using the Interactive Data Language (IDL) routine, {\it `zang'}.  In Figure \ref{Physical_Sizes} we replot the bottom panel of Figure \ref{CORE_PED_XRAY} but now weight the symbol size by the physical size of the non-core that it represents (LAS in kpc scaled by constant but arbitrary factor).  We do not include here the MS+ extension sources, in order to keep as much consistency as possible in the array configurations from which our LASs where measured.  However, we note that archival images of the sources in the MS+ extension show similar size distributions to those observed in the MS.

The most powerful non-cores at any given X-ray luminosity - those sources found above the dotted line in Figure \ref{Physical_Sizes} - are typically large resolved sources.  Specifically, in order of increasing cluster X-ray luminosity these sources are: A2634, RXJ2214.7+1350, A160, RXJ0058.9+2657, A407, A3581, Z1261, RXJ1320.1+3508, A2627a, A2055 and A1763.  All of these except A3581 are large jetted systems, of which A2627(a) shows one sided jetted structure and all others show classical FRI/II type morphology.  A3581 has only a hint of a resolved jet with ATCA-6km C-band observations. \cite{Canning13} however use the hybrid VLA-AnB array at L-band to show this source has radio-lobes that fill the X-ray cavities in this system.  The small extent and currently highly active core suggests the non-core emission here is recently injected and may perhaps be expected to expand outwards in future, which could classify this source as a young FRI/II precursor.  That the majority of the most powerful non-cores at any given X-ray luminosity are large FRI/II type sources supports the idea that powerful individual periods of activity may be responsible for the brightest non-cores and that in a larger sample we may expect to find bright cores in low X-ray luminosity clusters.

However, the general distribution shown by the bulk of the population below the dotted line in Figure \ref{Physical_Sizes}, is comprised of a mixture of large extended resolved and smaller unresolved sources over the full range of X-ray luminosities.  The extended sources here represent a mixture of FR-type and more amorphous sources.  That there is a general trend for brighter non-cores to reside in more X-ray luminous clusters regardless of their size but that the brightest tend to all be large extended sources suggests there may be two different activity types occuring in BCGs. 

Continuous (or repeated) core activity may lead to confused, fairly amorphous small scale structures comprised of accumulated material from extended periods of core activity.  The tendency for more powerful cores to inhabit more X-ray luminous clusters would then naturally account for brighter non-cores inhabiting these same clusters.   However, any BCG (indeed, any AGN) can experience a single strong outburst, powerful enough to eject material well-beyond the immediate BCG-vicinity and manifest as a lobe-dominated FR-type source, potentially explaining the powerful non-cores in less X-ray luminous clusters.

The overall brightness of the non-core will thus be a superposition of the accumulated material and the single powerful outburst, which could explain the extremely bright emission for the most-powerful non-cores in X-ray luminous clusters - if they are powerful FR-type non-cores superposed on top of a pedestal of accumulated emission.

In between large outbursts, the AGN remains persistently active albeit at a lower radio power, which is reflected in the high portion of (especially) LE BCGs that have an active core component.  This suggests that when studying AGN it is important not only to look at the time a source is radio-loud for, but also the fraction of this time for which the AGN is actively launching jets.


\subsection{[OIII]5007\AA{} Correlations} \label{OIII_section}

\begin{figure}
  \begin{minipage}[b]{0.5\linewidth}
  \centering  
  \includegraphics[width=9cm]{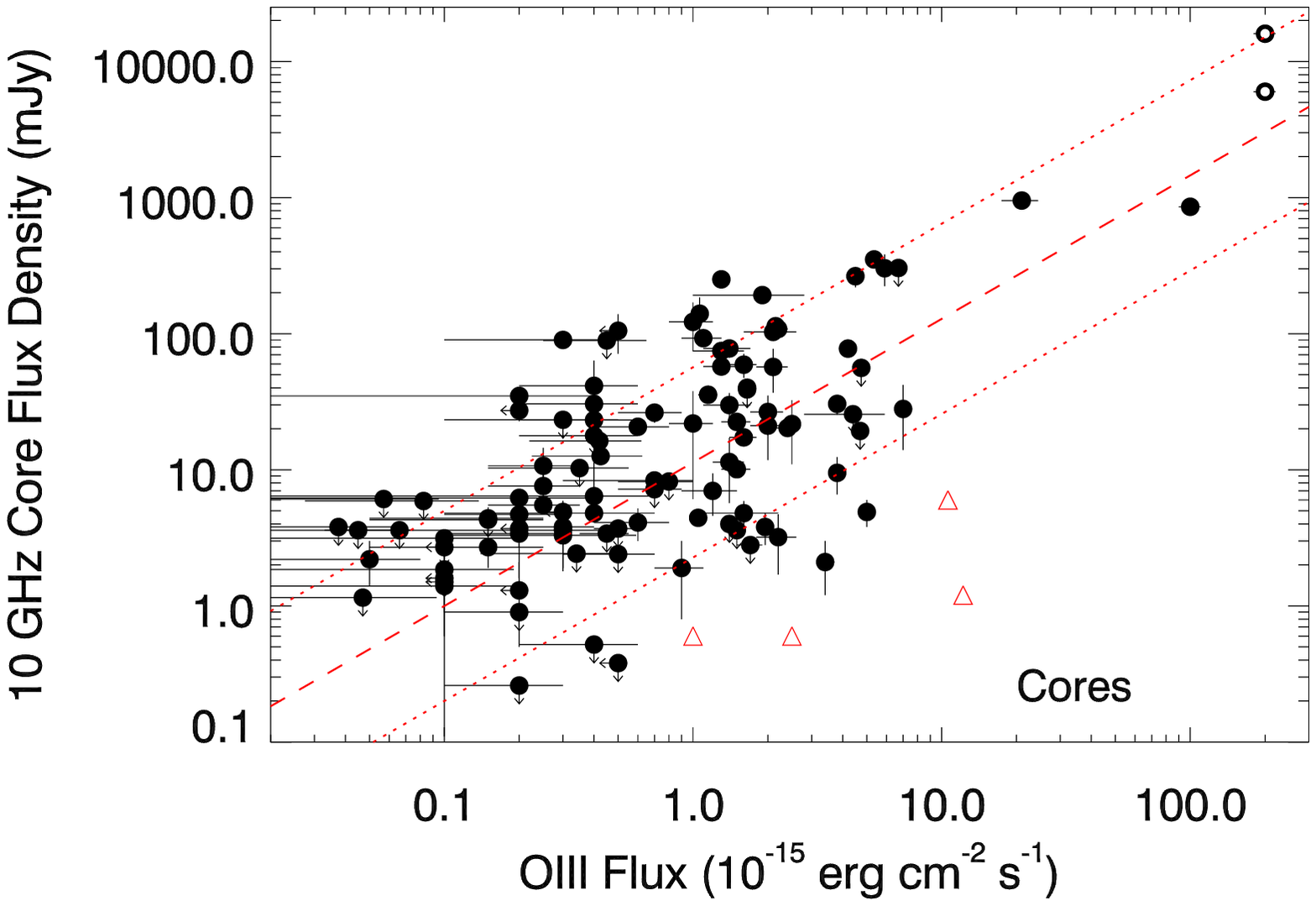}  
  \end{minipage}
  \begin{minipage}[b]{0.5\linewidth}
    \includegraphics[width=9cm]{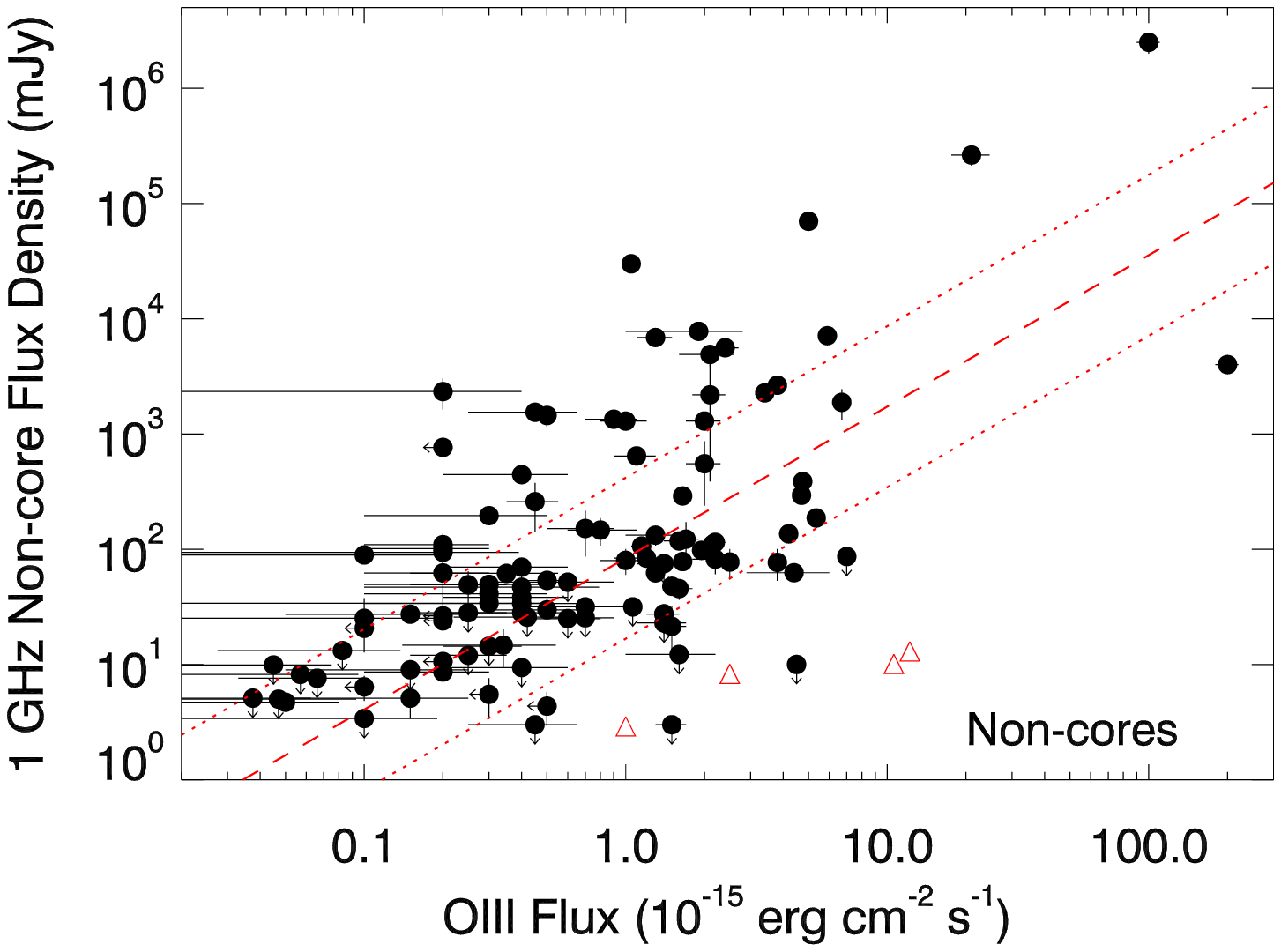}  
  \end{minipage}
  \caption{The 10~GHz and 1~GHz normalised respectively core and non-core radio flux components plotted against the [OIII]$\lambda5007$\AA{} flux.  The dashed line in the top panel marks a linear correlation between the [OIII]5007\AA{} and core radio flux which spans four orders of magnitude in each parameter (see Section \ref{OIII_section}).  The dotted lines mark a factor of twenty in scatter either side of the relation.  A correlation is also found for the non-core (bottom panel), although with significantly increased scatter, as highlighted by the inclusion of the dotted lines again indicating a factor of 20 scatter in either direction. Red triangles are sources significantly contaminated by a mid infra-red (MIR)-dominated AGN and are not included in the fit (see text). NGC1275 is shown twice in the upper panel (open circles) to highlight the variability of its core over $\sim$decade timescales.  Such variation in the sample as a whole is expected to account for a significant portion of the scatter.}  
  \label{OIII_Correlations}
\end{figure}

To investigate the relationship of radio-emission to other tracers of AGN activity, we consider the high ionisation [OIII]5007\AA{} forbidden-line, which can be used as a canonical tracer of current AGN activity \cite[][]{Maiolini95}. Combining longslit observations taken with the New Technology Telescope (NTT) and Very Large Telescope (VLT) (Edge \etal in prep.) with those in \cite{Crawford99}, we have [OIII] fluxes for 97 of the sources within our Main Sample+ (of which 13 are in the MS+ extension). 

These [OIII] fluxes are all from long slit observations of 1.2--1.5$''$ width so there will be some flux lost outside the slit. However, on the whole the nuclear [OIII] emission will dominate over the more extended, non-AGN emission so these fluxes should be representative of any underlying correlation.  Four of these sources (A2146, A1068, RXJ0821.0+0752, Z3146) are believed to host strong type-II AGN as a result of high MIR fluxes, high [OIII]/H$\beta$ ratios and very red 4.5/3.6~$\mu$m colours \cite[][]{Egami06, O'Dea08, Quillen08}.  The presence of this strong MIR-AGN component will boost the [OIII] fluxes within these objects above our typical sample and hence they do not represent a fair comparison.  We therefore highlight these objects in Figure \ref{OIII_Correlations} and remove them from the subsequent fits.

To increase our coverage, we identify an additional 16 systems for which equivalent [OIII] measurements are available that do not fall within our parent sample yet have comparable properties and hence are included within this analysis.  For each of these additional 16 sources, we populate their radio SEDs from the literature and perform decompositions (see Section \ref{SEDDecomp} and Table \ref{OIII_EXTRA_FITS}).  This creates a total [OIII]-sample of 113 BCGs.

Using the {\it bhkmethod} tool within the Image Reduction and Analysis Facility (IRAF) shows that [OIII] flux is correlated with {\it both} the core and non-core radio flux. Both return a probability of no correlation $<$0.001 (the limit) with calculated Kendall's-$\tau$ coefficents of 0.7533 and 0.7272 respectively.  We use the {\it buckleyjames} linear regression tool within IRAF to compute best-fits to the data.  This algorithm can handle censors only in the dependent variable of a dataset.  Neither of our variables is truly independent. We assign the [OIII] line-flux as our independent variable, since we have selected this sub-sample on the availability of [OIII] data and also it requires the smallest removal of data.  Nine sources with [OIII] limits are removed.  The re-calculated Kendall's-$\tau$ coefficients are 0.7661 and 0.7508 for the cores and non-cores respectively, so the removal has a minimal effect.

Between the [OIII] (in erg cm$^{-2}$ s$^{-1}$) and core radio (mJy at 10~GHz) fluxes we find a best fit power-law: 
\begin{equation}
log_{10}(S_{Core}) = (1.05\pm0.10) \hspace{0.5mm} log_{10}(F_{OIII}) + (1.05\pm0.01) 
\end{equation}
The equivalent correlation between [OIII] and the non-core radio component (normalised at 1~GHz) is: 
\begin{equation}
log_{10}(S_{NC}) = (1.31\pm0.16) \hspace{0.5mm} log_{10}(F_{OIII}) + (1.92\pm0.06) 
\end{equation}

Although it appears in Figure \ref{OIII_Correlations} that the correlations are driven largely by the bright local sources NGC1275, M87 and Cygnus-A, if these sources are removed then IRAF returns consistent fits, still with a probability of no correlation $<$0.001.  We therefore conclude that the presence of these correlations is robust.

\subsubsection{[OIII] Discussion}

The correlations between the radio component's flux and [OIII] flux (See Figure \ref{OIII_Correlations}) support previously found correlations between radio and [OIII] luminosities (e.g. \cite{Rawlings89} at 178~MHz, \cite{deVries07} at 1.4~GHz, \cite{Tadhunter98} at 2.7-5~GHz).  Interestingly, Rawlings \etal found a correlation between the [OIII] and total radio power but only low-significance indication for a correlation between the resolved `core' radio and the [OIII] in their sample.  Here we see such a correlation, with reduced scatter over that of the more extended emission.  

Our results show that whilst both radio components correlate with the [OIII], there is less scatter when considering only the radio core component (see Figure \ref{OIII_Correlations}).  Along with a loose intrinsic link between core and non-core power (see also Section \ref{Core_notcore_section}), this suggests that the primary correlation is with the core component.  There is tentative evidence ($<$2$\sigma$) that the index for the correlation with the non-core component is steeper than for the core component.  We interpret this as evidence for the core component driving the correlation, as the [OIII] forbidden line is collisionally de-excited and so traces recent AGN {\em core} activity.  Conversely, if the non-core emission is due to the cumulative effect of long-term activity then it should show more radio emission for a given [OIII] flux.  

\cite{Labiano09} found a correlation between the [OIII] and 5~GHz radio power that had increasing gradient when considering only GPS, GPS+CSS (Compact Steep Spectrum), only CSS and then combined GPS+CSS+FR-type radio sources.  Assuming the commonly quoted interpretation of these source types as a loose evolutionary sequence, GPS sources are expected to be core-dominated whereas the large, extended FR-type sources are most likely to be dominated by their extended emission at 5~GHz.  Indeed, the sources with the highest F$_{NC}$/[OIII] flux ratios in the bottom panel of Figure \ref{OIII_Correlations} correspond to the most physically extended sources (as seen in Figure \ref{Physical_Sizes}).  Under such assumptions this then supports our interpretation that the correlation is core driven. It follows that a significant portion of the scatter found in relations between radio-activity and AGN activity at other wavelengths is due to the varying timescales on which the emission was produced.  

\subsection{Cavity Correlations}
\subsubsection{Insight into the Activity Cycle}

\begin{figure}
  \begin{minipage}[b]{0.5\linewidth}
  \centering  
  \includegraphics[width=9cm]{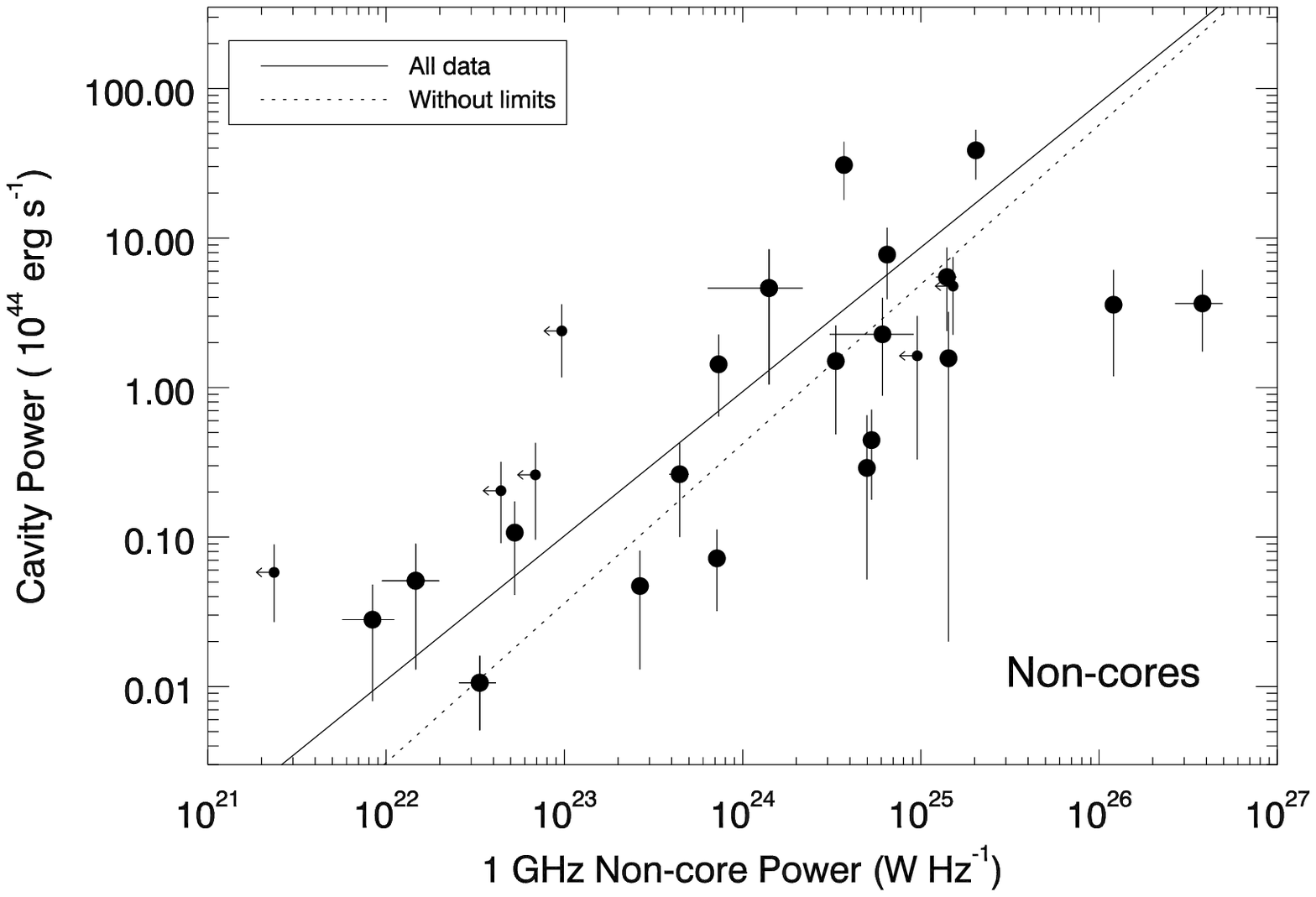}  
  \end{minipage}
  \begin{minipage}[b]{0.5\linewidth}
    \includegraphics[width=9cm]{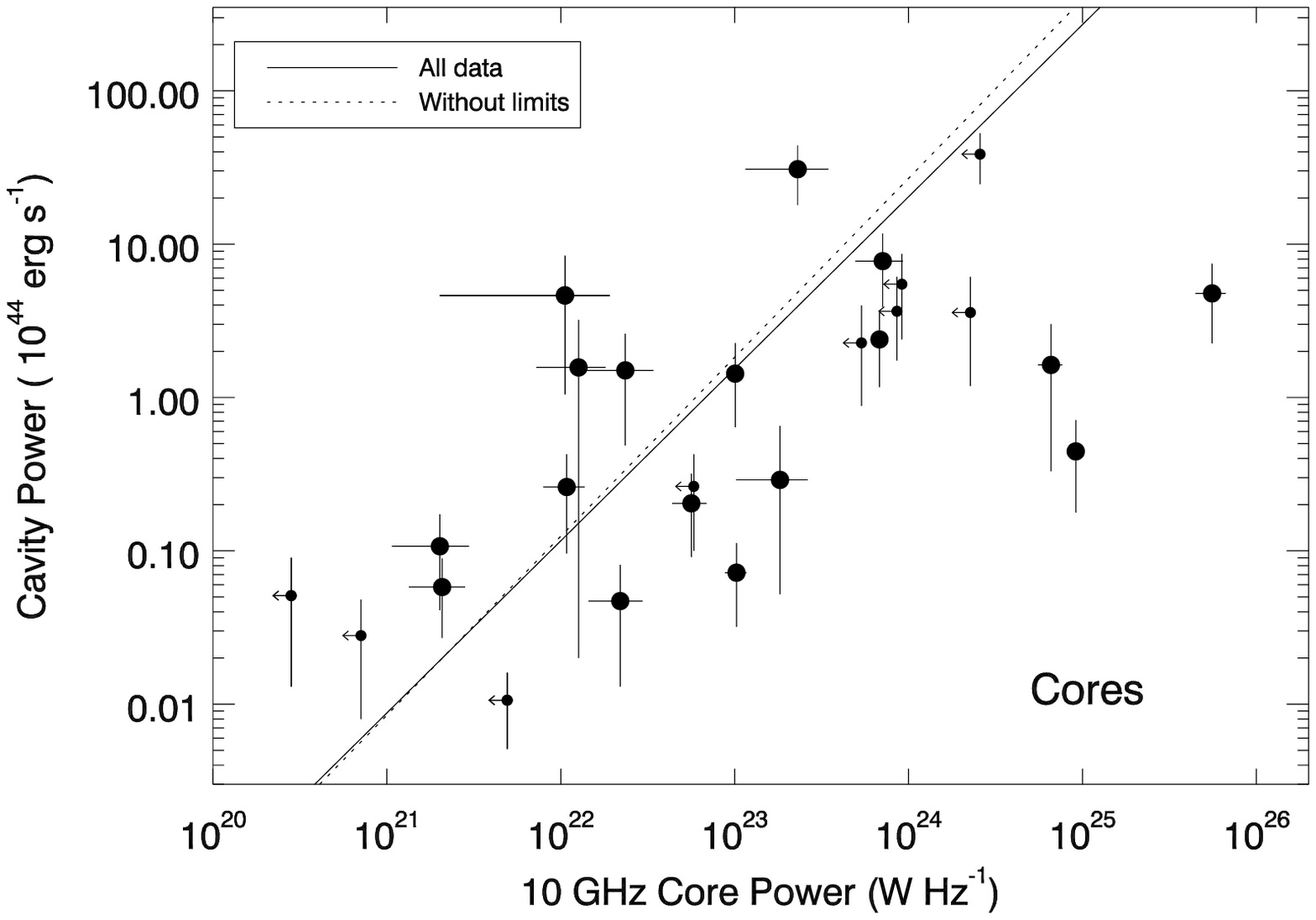}  
  \end{minipage}
  \begin{minipage}[b]{0.5\linewidth}
    \includegraphics[width=9cm]{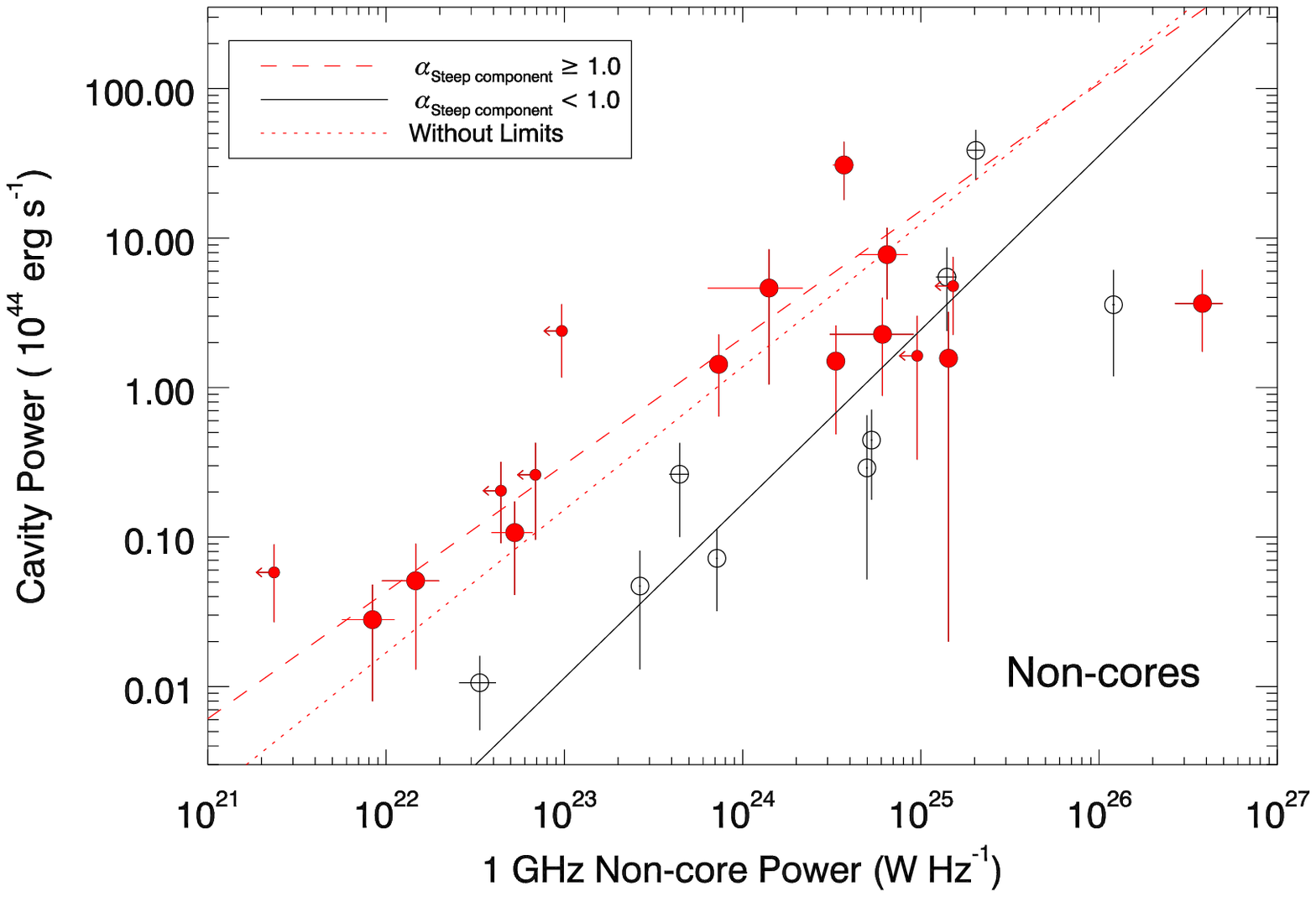}  
  \end{minipage}
  \caption{Correlations between both the non-core radio luminosity and cavity power (top) and core radio luminosity and cavity power (middle). Solid lines are the regression fits to the full data-sets, as found using the buckleyjames algorithm within IRAF and dotted lines are linear fits when the upper-limits on the radio-data are neglected. A correlation with the non-core emission is expected whereas the one with the core is less expected, suggesting that central AGN activity is persistent on bubble-inflation/rise timescales.  Also shown is the non-core to cavity power relationship divided on the radio spectral index of the steep component (bottom).  Filled circles have $\alpha$$_{Steep~component}$$\geq$1.0 whereas open circles have $\alpha$$_{Steep~component}$$<$1.0. A clear index dependent split is seen.  See text for implications. }  
  \label{Cavity_Correlations}
\end{figure}

For statistically relevant samples, BCG radio power correlates well with
the cavity power, albeit with large scatter \cite[e.g.][]{Birzan04, Birzan08, Cavagnolo10}.  These studies
usually depend upon the total radio power at L-band.  Here we attempt to expand upon this by considering
how the cavity power correlates with both the core and non-core radio powers.  A literature search for X-ray cavity systems amongst our Main Sample was performed, and supplemented by recently discovered cavity systems (Hlavacek-Larrondo, {\it private communication}).  Twenty-six of our MS clusters have known X-ray cavities, of which twenty-three are LEs, with the remaining three (A189, RXJ0419.6+0225 and RXJ1522.0+0741) being weak cavity systems in low redshift clusters.  

Cavity powers are estimated using the standard techniques \cite[e.g.][]{Birzan04}. The total enthalpy is given by 4pV, where p is the thermal pressure at the radius of the bubble measured from X-ray observations and V is the volume of the cavity. We assume that the cavities are of prolate shape: $V$ = 4$\pi$R$_{w}$R$_{l}$/3, where R$_{l}$ is the projected semi-major axis along the direction of the jet and R$_{w}$ is the projected semi-major axis perpendicular to the direction of the jet. Errors on the radii are assumed to be $\pm$20\%. Cavity powers are then calculated by dividing the enthalpy with the buoyancy rise time of the cavity \cite[][]{Churazov01}. The latter measures the time it takes a cavity to reach its terminal buoyancy velocity.

Figure \ref{Cavity_Correlations} shows the cavity power both as a function of the non-core and core radio components 
(top and middle panels respectively). Using the IRAF {\it bhkmethod} tool for linear regression, we encouragingly find that the cavity powers strongly correlate with the non-core radio powers (Kendall's $\tau$ = 0.9538, with probability of no correlation 0.02$\%$). This correlation is expected, since non-core emission is usually composed of dying radio emission in lobes that are co-spatial with cavities and hence commonly traces the same outburst as that which evacuated them.  

There is also a reasonably secure correlation (Kendall's $\tau$ = 0.5723, with probability of no correlation $<$1$\%$) between
the cavity-power and the core radio component. This is more surprising, as the
bubble-rise timescale should be long enough to decouple the cavity power from
what is happening in the core unless the core activity is persistent over
timescales of several Myrs. This is therefore indicative of steady AGN fuelling in BCGs over bubble-rise timescales.

Regression fits to the data are performed using the IRAF {\it buckleyjames} algorithm.  As in Section \ref{OIII_section} neither of our variables is truly independent.  We assign the cavity power as the independent variable for fitting purposes since this sub-sample was selected on the basis of having observed cavities.  However, note that in Figure \ref{Cavity_Correlations} we plot cavity power as the ordinate to ease comparison with earlier studies.   We find best fit power-laws: 
\begin{equation}
log_{10}(P_{NC}) = (1.022\pm0.18) \hspace{0.5mm} log_{10}(P_{Cav}) + (24.03\pm0.16)
\end{equation}
for the non-core component and:
\begin{equation}
log_{10}(P_{Core}) = (0.86\pm0.36) \hspace{0.5mm} log_{10}(P_{Cav}) + (22.80\pm0.27)
\end{equation}
for the core component respectively.  Note that a fit to the integrated L-band power gives a consistent result with the fit to the non-core power, as should be expected since the non-core tends to dominate at this frequency.

We further consider the non-core to cavity power correlation, splitting the non-core radio component into those with 
the steepest spectral indices $\alpha$$\geq$1.0 and those with flatter indices $\alpha$$<$1.0 (see Figure \ref{Cavity_Correlations} bottom panel).
This value is chosen so as to include the limits as steeper spectrum objects, which is reasonable for these clusters
as they tend to be well sampled in the GHz range and flatter components would be expected to be detected.  A separation is seen in the lower panel of Figure \ref{Cavity_Correlations}.  For a given cavity power, steeper spectrum non-core components tend to be
lower power than flatter spectrum non-cores. Note that the apparent steep-spectrum outlier lies in a system (Abell 115) with a prominent extended radio relic \cite[][]{Govoni01}, which may be contaminating and inflating the non-core flux measurement.  If limits are removed from this bottom panel then such 
a split is still observed. We propose that this apparent split is suggestive of evolution of the radio emission within individual cavities.  

The regression fits return best fit power laws to our split data of:
\begin{equation}
log_{10}(P_{NC}) = (0.86\pm0.18) \hspace{0.5mm} log_{10}(P_{Cav}) + (24.67\pm0.03)
\end{equation}
for when the non-core component has index $\alpha$$<$1.0 and:
\begin{equation}
log_{10}(P_{NC}) = (1.22\pm0.21) \hspace{0.5mm} log_{10}(P_{Cav}) + (23.51\pm0.03)
\end{equation}
for when the non-core component has index $\alpha$$\geq$1.0 respectively. 

When cavity inflation ceases and the inflated cavity detaches from the BCG, its power is virtually fixed.  Increasing volume of the cavity as it propagates is counteracted by the lowering of the surrounding pressure.  However the synchrotron emitting electron population within the cavity will begin to age, hence the radio spectrum will steepen over time. This will contribute significant scatter within the radio power to jet power
ratio.  Isolating the non-core radio emission and including the low-frequency radio spectral index of this as an additional parameter in such studies could potentially eliminate much of the scatter.  

\section{Discussion} \label{DISCUSSION}

That BCGs exhibit increased likelihood of hosting radio-loud AGN is well established \cite[e.g.][]{Burns90, Best05}.  The trends presented above explore beyond this, showing that there is a great range of behaviour beyond generic `radio-loudness' that appears to be driven largely by environment.

\subsection{LE to NLE Differences} \label{LEtoNLE_differences}
Striking differences are seen in both the frequency of radio-loudness and in the resulting radio luminosity of the AGN between those in LEs and those in NLEs.  Considering Figure \ref{fig:L_LFs} we see that $\sim$50$\%$ of LE-BCGs host AGN with radio powers above 10$^{24}WHz^{-1}$ at L-band as opposed to only $\sim$15$\%$ equivalent for NLEs.  Our overall detection fractions for radio-loudness at L-band are equivalent to those of other X-ray selected cluster samples.  However, they are above those for otherwise detected samples (see Section \ref{LE_NLE_Match_Rates}).  When we remove LE-BCGs from the sample we recover a detection fraction similar to that of otherwise selected massive galaxies \cite[][]{Best05}.  This suggests that the AGN activity of NCC-hosted BCGs is largely governed by the same processes as isolated galaxies of similar mass. That is they affected mainly by their own circumgalactic envelopes.  That the detection fraction for LE BCGs jumps significantly above this shows unequivocably that the cooling flow must be causing the increased duty cycle of radio loudness, therefore it seems natural that this same cooling is, directly or indirectly, fuelling the AGN.

\begin{figure}
  \begin{minipage}[b]{0.5\linewidth}
  \centering  
  \includegraphics[width=9cm]{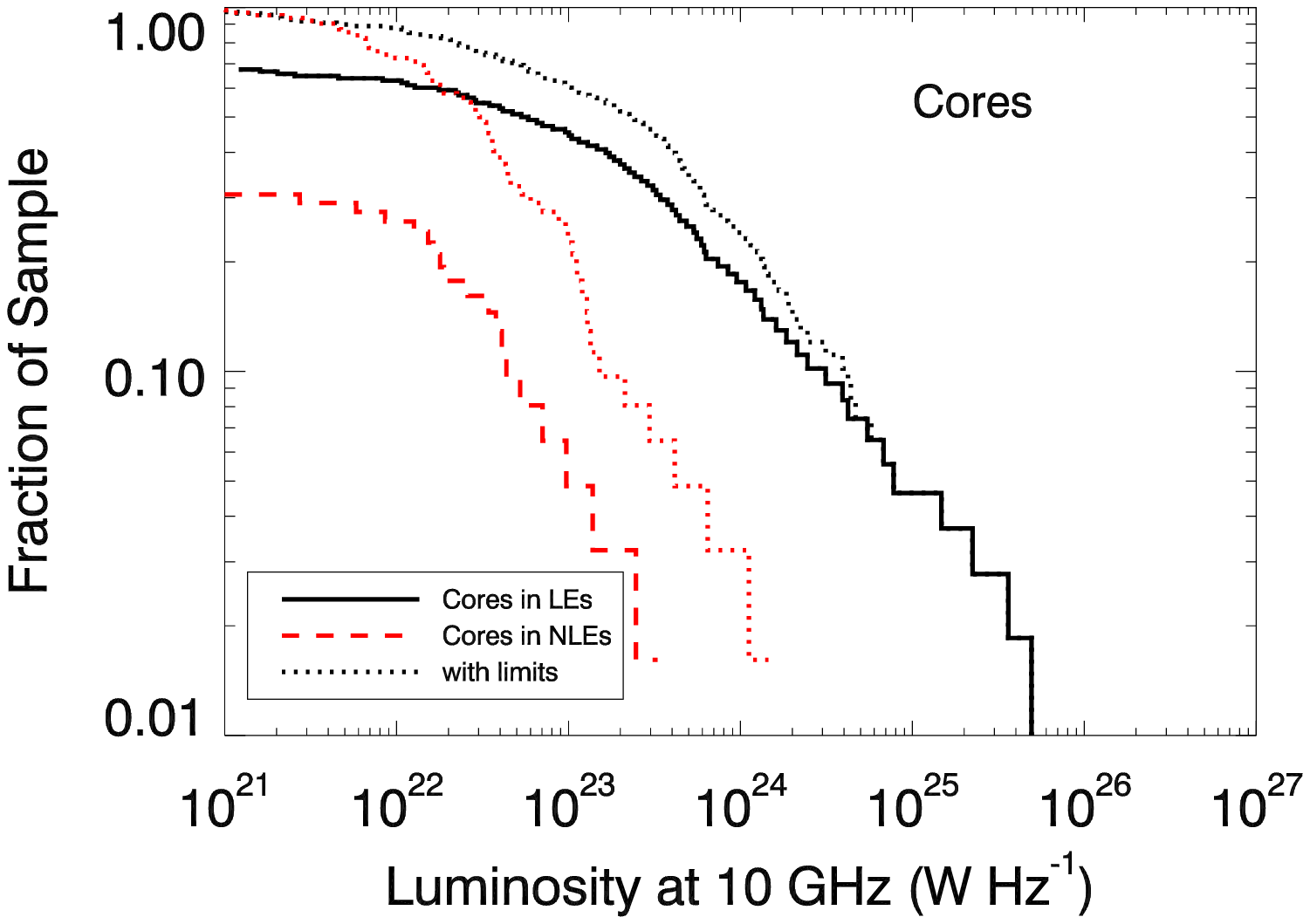}  
  \end{minipage}
  \begin{minipage}[b]{0.5\linewidth}
    \includegraphics[width=9cm]{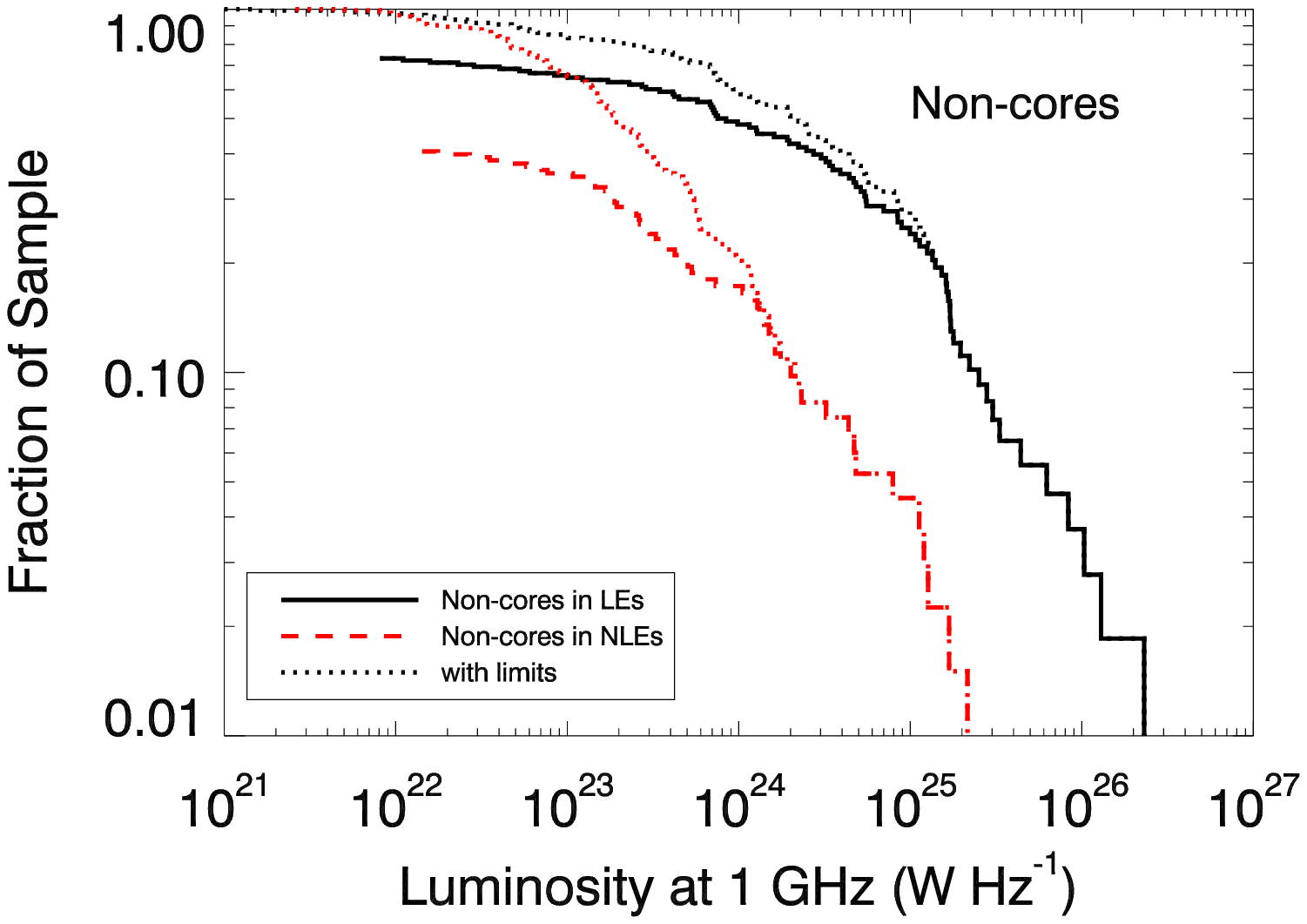}  
  \end{minipage}
  \begin{minipage}[b]{0.5\linewidth}
    \includegraphics[width=9cm]{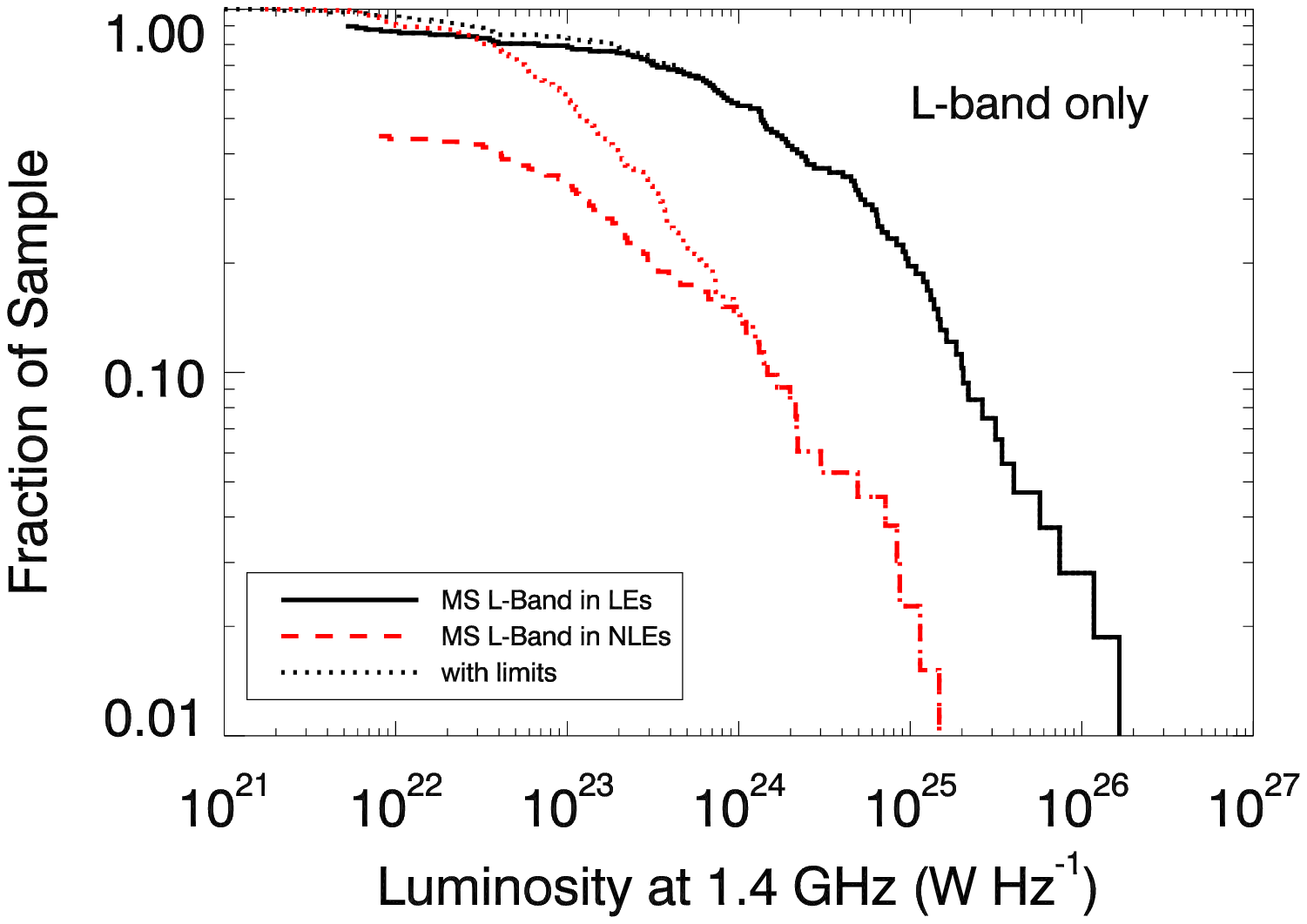}  
  \end{minipage}
  \caption{As for Figure \ref{fig:L_LFs} but now using only the core components at 10~GHz (top), the non-core components at 1~GHz (middle) and the 1.4~GHz monochromatic flux for only those sources in the Main Sample.  Note that the overall spectrum at 1.4~GHz is most often dominated by non-core emission, as is reflected in the similarity between the split at L-band and that of the non-core component.  However, the core activity is apparently even more environmentally dependent, with core powers in NLEs typically being up to two orders of magnitude less than those in LEs.}  
  \label{core_ped_LFs_MS}
\end{figure}

\begin{figure}
\begin{center}
\includegraphics[width=9cm]{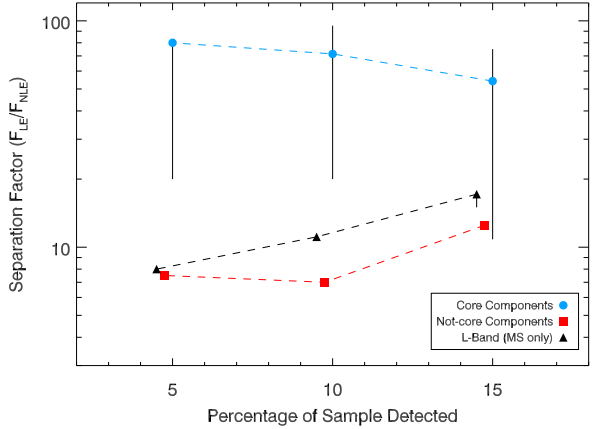}
\caption{Separation of the LF-distributions for LEs and NLEs in terms of the ratio of their detection fraction, for the core components, non-core components and the equivalent 1.4~GHz monochromatic LF for only those sources in the Main Sample, as shown in Figure \ref{core_ped_LFs_MS}.  Error bars are the maximum and minimum separation possible accounting for limits and hence are not truly `errors'.  The large lower uncertainties on the cores shown here are expected to be hugely over-estimated by this method as it would require all sources with a core limit to have an active core so far overlooked from observation, which we deem highly improbable.  Considering this, a clear difference in core activity as a function of cluster environment is claimed.  Note that X-positions have been shifted slightly for clarity.}
\label{fig:Compare}
\end{center}
\end{figure}

\begin{figure}
  \begin{minipage}[b]{0.5\linewidth}
  \centering  
  \includegraphics[width=9cm]{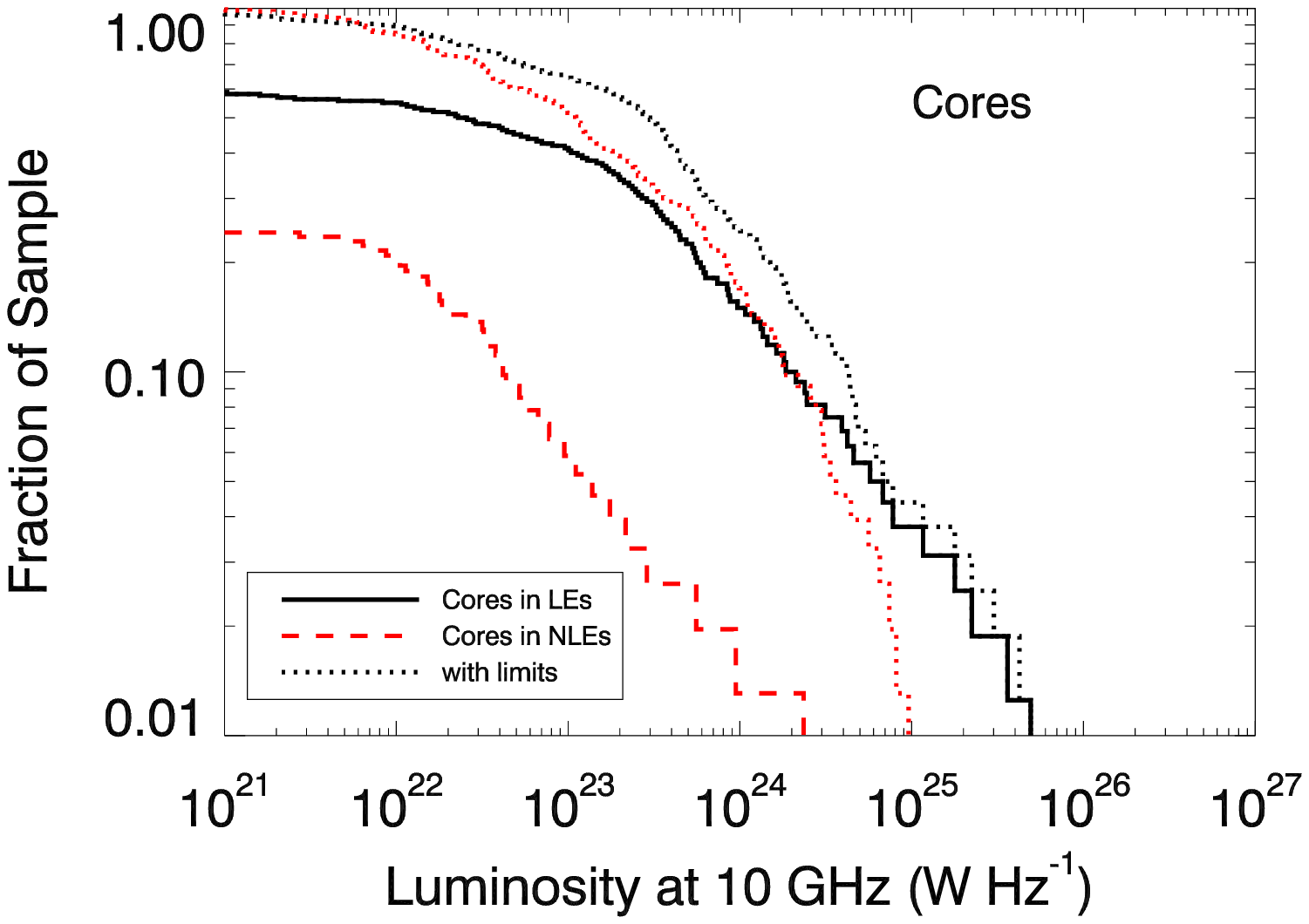}  
  \end{minipage}
  \begin{minipage}[b]{0.5\linewidth}
    \includegraphics[width=9cm]{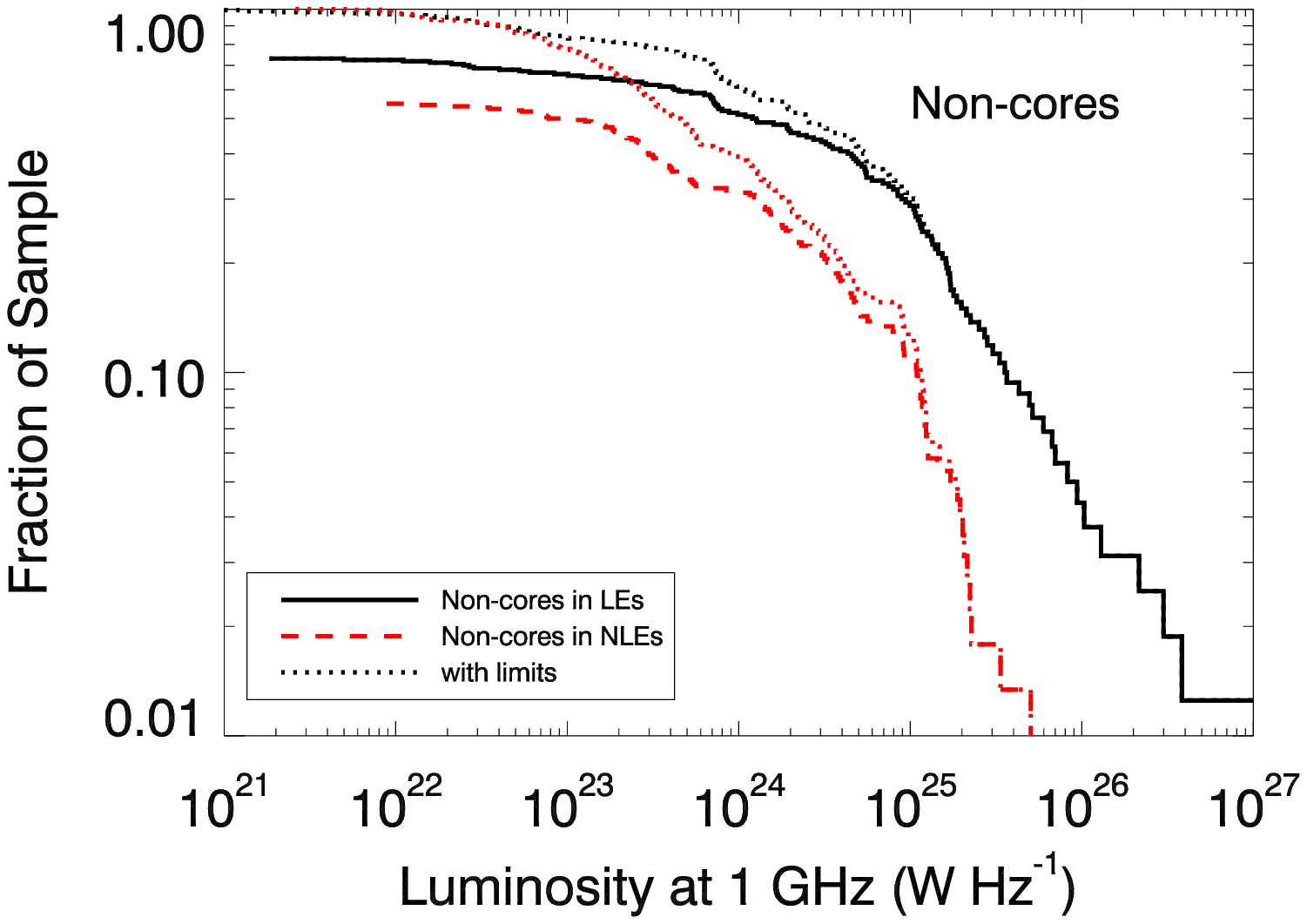}  
  \end{minipage}
  \caption{As for Figure \ref{core_ped_LFs_MS} but using the MS+.  Note that unlike the MS, which was largely unbiased on radio priors (see Section \ref{MS_SAMPLE}), the MS+ extension is flux limited at L-Band radio and hence contains a smaller fraction of radio-quiet objects than is truly representative of the full parent sample.  The LF extrema given when limits are included are hence essentially useless for inference purposes and stand only to highlight our ignorance of the full picture when limited data is available.  Notwithstanding, we see here that the larger split is seen amongst the core values (it stands to reason that all cores should be detected within the radio loud rather than radio quiet sample!).  The smaller split in non-cores (as compared to Figure  \ref{core_ped_LFs_MS}) can be attributed to this bias, since with the plot being cumulative the relatively lower detection amongst NLEs in the parent sample would serve to separate the lines further.  That the core split remains large amongst the entirety of the radio loud BCG sample (to given limit) further strengthens the case that this is a genuine example of the environment affects AGN behaviour in a systematic way throughout the Universe.}  
  \label{core_ped_LFs_MSp}
\end{figure}

\subsubsection{Decomposed Luminosity Functions}
We plot in the top and middle panel of Figure \ref{core_ped_LFs_MS}, separate luminosity functions for the core and non-core components, considering only sources in the Main Sample.  In the bottom panel of Figure \ref{core_ped_LFs_MS} we show the equivalent plot to Figure \ref{fig:L_LFs} but incorporating only the sources in the Main Sample.  We see that the total split at L-band is driven by the non-cores, as expected since this is the component that tends to be most dominant at $\sim$1~GHz and is usually brightest overall.  However, we see that the LE to NLE dichotomy is enhanced when considering only the cores.  Core components in LEs are typically two orders of magnitude more powerful than those in NLEs.  This split is equivalently shown in Figure \ref{fig:Compare}, where limits are conservative. Error bars show the range of the maximum and minimum separations of the distributions in Figure \ref{core_ped_LFs_MS} if all cores lay infinitely or infinitesimally below the limits.  

In Figure \ref{core_ped_LFs_MSp} we show the equivalent plot to Figure \ref{core_ped_LFs_MS} but now include the full MS+. Again, a larger split is seen for the core components.

\subsection{Dual-Mode Feedback - A Duty Cycle within a Duty Cycle}

Whilst LE-BCGs have a higher duty cycle of radio-loudness than NLEs, what is really striking is that the fraction of radio-loud sources that contain an {\it active} core component is also much higher in LEs.  In effect this is an {\it inner duty cycle} - the portion of radio-loud time for which an AGN is actively injecting radio-emitting populations - and it appears to be this factor that really characterises the difference between LEs and NLEs.

Several studies have shown that direct Bondi accretion of the hot gas on scales $<$1kpc from the nucleus can provide adequate fuel {\it on average} \cite[see][]{McNamara07}. This rate is however too low for the most extreme outburtsts \cite[e.g.][]{McNamara11}.  \cite{Allen06} claim a strong correlation between the Bondi rate and their measured cavity powers in a small sample of well-observed, local systems.  However, \cite{Russell13} find a much weaker relationship.  

In their recent review \cite{Heckman14} argue that a plausible explanation for these disparate results may be that Bondi accreting gas goes through an intermediate stage where it cools prior to finally being accreted onto the black hole.  Alternatively, the AGN may be fuelled by cold phase accretion whereby blobs of cold gas condense non-linearly at scales $\sim$5-30kpc from the black hole and infall on less than the cooling timescale \cite[e.g.][]{Pizzolato05}. In a recent paper considering the origin of multi-phase gas in cool-cores, \cite[][]{Voit14} find that thermal conduction and cold-mode accretion are complementary processes in regulating cluster cores.  At higher core entropies conduction prevents the core from cooling radiatively whereas at lower entropy, thermal instabilities cause the formation of cold clouds that subsequently precipitate onto the central black hole and power AGN feedback.
These scenarios all require that in some systems there should be large amounts of cold gas near to the nucleus.  Indeed there is evidence for rotating disks of cool gas in some BCGs \cite[e.g.][]{Simkin79, Ekers83, Lim00, Wilman05, Hamer14, Russell14}.  The resultant cold clouds could then power the sporadic periods of high activity whilst the AGN activity could still relate to the wider Bondi rate {\it on average}.

Such a scenario appears likely within cluster cores, where the cooling flow deposits large amounts of molecular gas within the BCG.  The steady build-up of such cold clouds ensures that the replenishment time is low and hence naturally explains the high duty cycle of radio loudness in both NLEs and LEs.  The high {\it inner duty cycle} of {\it active} core components in LEs is slightly trickier to comprehend purely by sporadic processes.   We postulate that within the LE clusters both ready cooled material and direct hot phase gas fuel the AGN.  The clumpy cold clouds would be responsible for the brighter periods of activity \cite[][]{Werner14}.  In between these periods, a steady trickle of hot gas would still be able to reach the nucleus ensuring the persistency of core emission.

A similar picture to this is found in the simulations of \cite{Li14} who find that under idealised conditions, cool clumps condense from the ICM due partly to the actions of jetted outflows.  It could be that the presence of a weak Bondi-powered jet catalyses the formation of cold clouds that can then fuel more powerful activity.  Such a process may be partly responsible for the so called `molecular fountains' seen in some BCGs \cite[e.g. A1835, ][]{McNamara14} where cool material appears to condense behind rising cavities before falling back towards the central engine. 

\subsubsection{Accretion Mode Considerations}
We have not in this paper explicitly classified our BCGs as low-excitation or high-excitation radio galaxies (LERGs or HERGs respectively), a distinction that appears strongly related to the accretion state \cite[inefficient/efficient respectively for LERGs/HERGs:][]{Best12}.  However, the vast majority of BCGs are LERGs, typified by accretion efficiences L$<$0.01L$_{Edd}$ (where L$_{Edd}$ is the Eddington luminosity).  At these low rates, accretion is believed to be via an advection dominated accretion flow (ADAF).  Whatever the coupling efficiency of the mechanical energy output by a BCG, this is not likely to substantially change over the range of luminosities we see in our sample as they always remain in the LERG, ADAF-stage. Changes in radio luminosity are therefore most likely to correspond to changes in mechanical feedback power (as corroborated by Figure \ref{Cavity_Correlations}), rather than large changes in the accretion structure such as the collapse of an ADAF into a classical Shakura-Sunyaev thin-disk \cite[e.g. see review of accretion structures by][]{Done07}.

Additionally, in analogy with the scenario seen in black hole binaries, the transition from an ADAF to a thin-disk accretion mode may be expected to coincide with a switch from magnetically arrested to magneto-rotationally supported accretion and hence the launching of a short-term ballistic jet \cite[][]{Mirabel94, Remillard06, Dexter14}.  Such an event would cause the collapse of the large scale-height magnetic fields present in the inner accretion zone that are required for the launching of a typical, steady jet \cite[e.g.][]{Mirabel92, Fender01}.  The apparently highly persistent activity we infer from our sample would oppose such a magnetic restructuring, certainly for our LEs.  That the periods of higher activity correspond to a difference in accretion structure (namely, a transition to a thin disk) is therefore considered highly unlikely and instead more likely correspond to something more akin to a greatly scaled-up analogue of the flaring seen in Sgr A*. 

fusion timeframes can be constructed to test the outline feasibility of the above scenario.  If we take the Alfv\'{e}n speed of electrons in a typical cluster to be $\sim$100kms$^{-1}$ \cite[][]{Ferrari08} then this translates to a dispersal velocity of roughly 0.1 kpc Myr$^{-1}$.  For a typical electron emitting lifetime of 100Myr then a structure $\sim$10kpc can be expected from simple diffusion alone. If we consider that even in strongly confused cluster cores, the jets will provide some bulk transport away from the central black hole then it is clear that amorphous haloes are fully consistent with this toy model, assuming the dispersion is reasonably homogenous.

\subsection{Considering Relativistic Beaming}

For the size of our parent sample, geometry causes us to expect $\sim$20 sources to be aligned within $\sim$5$^{\circ}$ of the line-of-sight.  Such a population may present itself as a population of flat spectrum, apparently powerful cores but would have their intrinsic core powers overestimated by potentially large factors.  However, note that we purposefully removed BLLac objects from our parent sample so as not to be overtly affected by a handful of strongly beamed sources.  There is also no logical reason why LE-BCGs would more commonly have aligned cores than NLEs.

Relativistic beaming of the inner-jet in an existing radio source could account for some seemingly strong, flat spectrum components of otherwise steep spectrum sources if the jet-axis has precessed into favourable alignment.  Our sample does include a couple of sources that have previously been considered to be mildly beamed.  For example, the BCG of A2055 appears to be a head-tail source on few-arcsecond scales, but the bright core may be beamed \cite[][]{Pimbblet06}.  Similarly, A2627(a) only appears to have marginally resolved jet structure on arcsecond scales but its AGN is X-ray dominated, hence the flux may be boosted by the presence of a beamed BLLac-like nucleus \cite[][]{Rector99}.  However, for this to account for all our split spectrum sources would require a very high precession speed in a significant number of sources.    Additionally, the long timescales between some of the archival data we use in compiling our SEDs shows that the flat and GPS-like behaviour is persistent over more than decade timescales.  This is therefore unlikely to be purely the result of beaming, which would give significant, short-term variation.  Furthermore, VLBA imaging of the central regions shows symmetrical two-sided structures on few parsec scales in many of our active BCGs hence showing that the source is not favourably aligned for strong beaming. This VLBA data will be presented in an upcoming paper. 

Overall, we concede that some blurring of results due to relativistic effects are essentially unavoidable for any large sample of radio-AGN, however due to the arguments stated here we do not believe beaming to overtly affect our results.

\subsection{Survey Considerations and Mimicked Evolution} \label{SurveySection}
The presence of two very different radio components in the SEDs of BCGs, means that the source population sampled in a flux-limited radio survey of AGN is highly dependent upon the observed frequency; unlike the case for a population of single spectral index sources.  Higher frequency samples (roughly comparable to a horizontal split on Figure \ref{THE_DIAGRAM}) will contain a higher proportion of core-dominated systems whereas low frequency selected samples \cite[more equivalent to a vertical split on Figure \ref{THE_DIAGRAM}:][]{vanHaarlem13, Tingay13} will favour those dominated by steeper spectrum emission \cite[see also][]{Sadler14}.  Whilst it is already well established that high frequency surveys will preferentially detect young sources, for BCGs the effect is slightly more subtle in that high frequency samples will preferentially detect those sources with the highest {\it inner duty cycle} of core activity (i.e. BCGs that are more persistent).

As the youngest and most recent AGN activity will peak spectrally at higher frequencies \cite[e.g.][]{O'Dea98, Sadler14}, samples taken at frequencies below the characteristic turnover will preferentially detect older outbursts.  Samples at progressively higher frequencies will therefore contain a greater proportion of younger sources.  Conversely, for a single observing frequency, matching samples at increasing redshift will show the same effect such that the population of radio sources to higher redshift, particularly for redshifts greater than 2 (where a restframe $\sim$15~GHz characteristic GPS peak would be redshifted below 5~GHz), will appear younger and more active - effectively mimicking evolution in the source population.  

Much increased spectral coverage is required before improvements to the radio k-correction above a simple power-law extrapolation will be confirmed possible/necessary or not. However, the effect should be considered when studying the redshift evolution of the radio population.

The powerful core components in LE-BCGs, especially in instances where these cause spectral peaks to higher radio frequencies, may be an important contaminant in Sunyaev-Zel`dovich (SZ) cluster surveys.  The ramifications of these strong core components are explored in a sister paper, Hogan \etal 2015b (submitted).

\section{Conclusions}

We have studied the radio properties in BCGs drawn from a large sample of X-ray selected galaxy clusters. The sample is split by the presence of optical emission lines (H$\alpha$+[NII]), using the proxy that when these lines are present the cluster most likely corresponds to being a strong cool core and when absent the dynamical state of the cluster is markedly less relaxed.  We present and collate a large radio data volume for this sample and consider the overall radio properties of the sample as a function of environment.  Furthermore, we decompose the BCG radio emission into two components - a core, attributable to ongoing injection AGN activity and a non-core comprising everything else.  We then consider these components in order to better understand the activity.  Chiefly we find that:

\begin{itemize}
\item The duty cycle of radio-loudness is significantly higher for BCGs in strong cool-core clusters ($>$85\%) than for those in non strong cool-core clusters ($>$52.5\%).
\item LE BCGs are typically much more radio-powerful AGN at than those in NLEs - $\sim$50\% of LE BCGs hosting a radio source $>$10$^{24}$WHz$^{-1}$ at 1.4~GHz whereas only $\sim$15\% are similarly radio-powerful in NLEs.
\item The prevalence, and typical power, of an {\it active} radio core is highly dependent on cluster state.  60.2\% of LE BCGs in our sample contain a distinguishable core of which $\sim$83.1\% are $>$10$^{23}$WHz$^{-1}$.  Conversely, only 11.6\% of our NLE BCGs contain a distinguishable core, with only $\sim$5.0\% of our NLE BCGs containing an active core of $>$10$^{23}$WHz$^{-1}$.
\item This core component shows better correlation with the canonical AGN tracer, [OIII]5007\AA{}, than the ageing non-core component.  It is thus likely that significant scatter often found in studies between radio emission (particularly from low resolution surveys) and other AGN tracers is due to long timelapses between emission periods.
\item  The brightest cores are only hosted by the most X-ray luminous clusters, although bright non-cores can be hosted by both high and low X-ray luminosity clusters.
\item  Both non-core and core radio emission correlates with cavity power in LEs.  It appears that BCGs are persistently radio-active over bubble-rise timescales in these environments.  
\item The radio-power versus cavity-power correlation appears to evolve with the spectral index of the radio source.
\item  An increasing fraction of flat and inverted spectrum sources in samples selected at increasing radio frequency may mimic luminosity evolution in the total radio population if not fully accounted for.  
\end{itemize}

Overall, we find that BCGs are a varied population with a wide variety of radio behaviour.  Future studies in the so-called `new golden age' of radio astronomy that we are now entering with existing and upcoming facilities should uncover many more facets to these intriguing behemoths.

\section*{Acknowledgments}

We thank the anonymous referee for useful comments and suggestions that have greatly improved this work. MTH acknowledges support from the Science and Technologies Funding Council (STFC), studentship number ST/I505656/1.  ACE acknowledges support from STFC grant ST/I001573/1.  ACF and HRR acknowledge support from ERC Advanced Grant Feedback.  HRR also acknowledges support from a COFUND Junior Research Fellowship at the Durham University Institute of Advanced Study. This research has made use of the NASA/IPAC Extragalactic Database (NED) which is operated by the Jet Propulsion Laboratory, California Institute of Technology, under contract with the National Aeronautics and Space Administration.  This research has made use of data, software and/or web tools obtained from NASA's High Energy Astrophysics Science Archive Research Center (HEASARC), a service of Goddard Space Flight Center and the Smithsonian Astrophysical Observatory. This research work has used the TIFR GMRT Sky Survey (http://tgss.ncra.tifr.res.in) data products.


\pagebreak

\appendix

\section[]{Justification of Fits - Main Sample} \label{APPENDIX_NOTES}

{\bf \subsubsection{VLA - AE125}}
{\bf A1991}  Unresolved at both C and X bands.  Fit to get a measurement of the non-core flux then extrapolate for a core upper limit. Likely to be a Compact Steep Spectrum (CSS) source. \\
{\bf A2146}  Flat spectrum, two observations at C-band show that it is variable which suggests we are seeing a dominant core.  Fit this using the most contemporal C-band observation to get a core measurement.  Get an upper-limit for the non-core by taking the lowest frequency data-point (WENSS) and extrapolating to 1~GHz.  Do this as expect the lowest frequency point to be non-core dominated but not directly observed near 1~GHz so a limit.\\
{\bf RXJ1657.8+2751}  Best fit using a dropline model.  Use this to get a measurement of the non-core.  Clear that there is no dominant core.  Large difference between the peak and integrated fluxes at C-band.  Use the peak of this with a core index of $\alpha$=-0.2 to get an upper-limit for the core flux. \\
{\bf RXJ1720.1+2638}  Core and non-core can be clearly spectrally separated in the radio SED.  There is a clear core component at C-band and L-band (using VLA-A) in addition to being consistent with the peak of the FIRST flux and with a BIMA flux for the core.  These points show a flat core component which can be fitted to get a clear core at 10~GHz.  Shows that this source is non-core dominated at low frequencies.  VLSS, NVSS and integrated flux at C-band (AE125) can be used to get a measure of this component at 1~GHz. Central core appears surrounded by amorphous diffuse emission.  Extrapolation would suggest a contributing non-core flux component at 10~GHz, resolved out at high resolution. \\
{\bf Z7160}  Shows a flat core component, consistent with the unresolved X-band observation, unresolved C-band and BIMA observation.  Fit this using the unresolved VLA-A flux measurement as the 1.4~GHz flux.  At L-band,  it is clear that lower resolution observations are picking up more diffuse emission from the non-core component.  Hence, we know that NVSS is picking up the non-core so connect this to lower frequency GMRT obtained flux to get a measurement of the non-core component. \\
{\bf Z8193}  Seeing a clear flat core component at higher frequencies that transitions to being dominated by a clear steep, non-core component at lower frequencies.  Extrapolating this steeper component back, it is consistent with being the difference between the interferometric flux measured at C-band and the single-dish measures which would pick up this extended emission.  Fit the flat component to get a core measurement then fit only the lower frequency points and extrapolate to 1~GHz to get a limit for the non-core flux. \\
{\bf Z8276}  See a clear split power law, that is well sampled enough to be fit directly.  There is a flat component for the core, which appears slightly variable at C-band (see also Hogan \etal 2015b.  Appears potentially slightly resolved at C-band but well-fit with unresolved gaussian.  See a clear, steep non-core component. \\
{\bf A407} Large, resolved source.  Fit the extended non-core with a power law and then use the VLBA point \cite[][]{Liuzzo10} to extrapolate for a measure of the core (negligible at 1~GHz).  Potentially could include a third component, non-core exhibits spectral turn-up at lower frequencies.  For the purposes of this current statistical study, we can ignore the lowest frequency data as this component, if present, would only be important at frequencies much below 1~GHz. This does however suggest that there have been multiple, powerful periods of activity in this source.  Some of the steepest emission may potentially be attributable to inner-relics.  Note that the position given is that of the high peak and hence most likely to be a hot-spot; it does not appear as if the core is resolved at the resolution of our imaging. \\
{\bf A621} Appears to be a classic steep spectrum source, not showing any spectral curvature.  Elongated but not clearly resolved into components at C-band.  There is a clear split between peak and integrated fluxes at both C and X-bands but morphologically this appears to be due to partially resolved double-lobe structure and hence the peak is likely just the brighter hotspot.  Fit for non-core value but then extrapolate from the peak at X-band to derive a limit on the core. \\
{\bf A757} Only sampled at L and C bands although the sources is clearly flat and unresolved, which is suggestive of core dominance.  Fit for a core value and then extrapolate for a non-core limit. \\
{\bf RXJ1053.7+5450} Weak source.  Sampled at L and C bands only.  Appears slightly resolved but not enough to clearly see a core.  Fit the steep spectrum integrated component then extrapolate from the peak at C-band for the core limit. \\
{\bf A1423} There is a head-tail source nearby but this is centred on a non-BCG galaxy.  The BCG itself is undetected in our C-band imaging.  \\
{\bf A1366} Consistent steep spectrum between 38~MHz and 43.3~GHz.  Appears to be a head-tail in C-band map but clearly not resolving the core, just a hot spot.  Fit with a single power and then extrapolate back from the 43.3~GHz observation to get a limit on the core. \\
{\bf A1190} Appears to be slightly diffuse although not truly `resolved'.  There is a clear peak-to-integrated deficit seen from the fit but morphologically the source appears unresolved. The source has published limits at 23.5 and 43.3~GHz \cite[][]{Lin09} showing that there is no strong, flat component.  Fit with a steep spectrum for the non-core and extrapolate back from 43.3~GHz limit to get a limit on the core.  There is a large, bright (250 mJy) Wide-Angled Tail (WAT) source only $\sim$80'' to the north-west. \\
{\bf A1773} CSS sampled only at L and C bands.  Source is unresolved. Fit an index to retrieve a measure of the non-core and extrapolate for a core limit. \\ 
{\bf A1775(SE)}  Large, messy source. Hint of spectral flattening towards higher frequencies, however it is only sampled up to C-band and so a true split-power model is not able to be robustly performed. Clearly losing some flux at higher resolution but not adequately resolving the source for a direct decomposition. Additionally, it could be argued that instead, there is actually spectral steepening towards lower frequencies.  This appears more believeable from the morphology and so best to neglect lowest frequency points, fit the extended emission with a power law and then extrapolate from the peak (most likely to be a hot spot rather than the active core) to get a limit on the core. \\
{\bf A1775(NW)} Diffuse source, there is resolution but no clearly defined core.  Appears to be reasonably steep ($\alpha$$\sim$0.7).  Fit the diffuse component.  There are limits at 22.5 and 43.3~GHz \cite[][]{Lin09} that show that there cannot be a flat core component attributable to the peak at C-band.  Extrapolate back from the limit at 43.3~GHz to get a limit on any core. \\
{\bf RXJ1442.2+2218} Sampled at just L and C bands.  NVSS picks up excess flux over FIRST peak.  Unresolved at C band.  Extrapolate the flat component to get a measure of the core, then extrapolate from NVSS to get a limit on the non-core. \\
{\bf A2009}  Resolved core at L band, unresolved core/resolved-out extended emission at C and X bands.  Fit the flat component.  Integrated flux at L-band consistent with unresolved NVSS flux, which along with VLSS suggests the presence of an ultra-steep spectrum (USS: $\alpha$$_{Steep}$ $>$ 1.5) component.  Interpolate for the USS component, accounting for the expected core contribution at 1~GHz when inferring non-core component. \\
{\bf A2033} Resolved double-lobed system.  Low resolution literature values sample only the integrated flux which is steep spectrum.  Fit this with a power law.  Able to get a core measurement at C-band, extrapolate from this with generous uncertainty to retrieve a core value. \\
{\bf A2055} Resolved head-tail.  Core is very bright, flat spectrally ($\alpha$$\sim$0.2) from L to C bands whereas extended/integrated emission is relatively steep spectrum ($\alpha$$\sim$0.63; see also Hogan \etal 2015b).  Split-power won't adequately decompose due to small coverage of core, however can fit a power law to each component and derive non-core component contribution.  \\
{\bf RXJ1522.0+0741} Commonly known as MKW3s.  Literature values at low frequencies show that this is a past-outburst source with ageing emission \cite[e.g.][]{Kellermann69, Giacintucci07}.  Fit a dropline to this.  Unresolved core seen at C-band, extrapolate with appropriate uncertainties from this to get a measure for the core flux at 10~GHz.  \\
{\bf A2063} Unresolved at X band and a limit at 22.5~GHz \cite[][]{Lin09}.  Peaks at L and C band consistent with there being an active, flat core component.  Extended emission at C-band and L-band shows there is a steeper, non-core component.  Fit the components separately, accounting for core when deriving non-core contribution. \\
{\bf A2064} Extended source, possible head-tail or perhaps uneven lobes.  No clearly resolved core, integrated spectrum consistent with classic steep spectrum lobe emission, peak most likely a lobe rather than the core so can only realistically get a limit on the core.  Note that this position is from fit of small double, hence position may better reflect a hot-spot rather than the true nucleus.  There is a reasonably bright ($\sim$25 mJy), partially resolved source only $\sim$80'' to the west.  \\
{\bf A2175} CSS source.  Fit for non-core, extrapolate for core limit. \\
{\bf Z1953}  Busy field.  Cluster is at high redshift (z$\approx$0.374) and appears to have blended double BCGs.  Several literature observations match radio-sources to the BCGs but higher resolution imaging shows that the BCG(s) is(/are) radio undetected. \\
{\bf A1132} Another busy field, some literature fluxes flagged as likely being confused.  Appears as a steep, double lobed source.  No clearly resolved core, extrapolate from the peak (likely lobe) to get core limit. Note that the given position is from a fit of the double and hence may better reflect a lobe position rather than the true nucleus.  Fit for non-core. \\
{\bf A761} Resolved source.  Measured core flux consistent with VLA-A \cite[][]{Jetha06} observation of the core.  Fit the integrated emission for the non-core component and then extrapolate from the core measurement with appropriate uncertainty for the core. \\
{\bf A1023} Only sampled at L and C bands, unresolved CSS.  Fit for non-core and obtain a limit on the core. \\
{\bf RXJ1206.5+2810} CSS.  Only sampled at L and C bands.  Fit for non-core and then extrapolate for a core limit. \\
{\bf Z4673} CSS.  Only sampled at L and C bands.  Fit for non-core, extrapolate for the core limit. \\
{\bf Z4803} Weak detection at C-band, no additional detections in the literature, including NVSS limit.  Extrapolate for limits on both components. \\
{\bf A1668} Remove single dish GB6, which appears to be confused with other nearby sources, and then consistent with a steep power law for the non-core. Source is unresolved at C-band but becomes resolved into two lobes ($\sim$0.94'', PA 54.6) at VLA-A array X-band.  VLBA observation (Hogan \etal 2015c, in prep) shows that there is no active core component.  \\
{\bf A1763} Large, head-tail source.  Not resolving out the core here but there is a high resolution observation \cite[][]{Hardcastle04} of the core.  Extrapolate from the resolved core component of this to get a value for the core and then fit the extended emission for the non-core component. Note that position quoted herein is the peak of emission, which here constitutes a hot-spot as opposed to the SMBH core.  \\
{\bf A1677} See a deficit from peak-to-integrated measured flux at C band (and FIRST) but does not appear to be a separated core, the source appears unresolved.  Integrated spectrum consistent with being non-core dominated steep spectrum.  Fit this and extrapolate for a limit on the core. \\
{\bf A2241} Appears to be unresolved CSS although a slight peak-to-integrated flux deficit. Fit for non-core, extrapolate for core limit. Companion source only $\sim$25'' to south-west and also a brighter $\sim$30 mJy head-tail only $\sim$2' to the north.  \\
{\bf A2390} Very powerful core, well covered by \cite{Augusto06}. Spectrum steepens at low frequencies indicative of there being a non-core component.  Fit with a split power model. \\
{\bf RXJ2214.7+1350} Powerful (3C442) radio source.  Much extended emission, steep spectrum to be fit for the non-core.  High resolution C-band data resolves out the extended emission revealing the core.  Extrapolate from this for the core component, negligible compared to integrated emission but consistent when applied to split power model. \\
{\bf A2572A} Slightly resolved source although there is no clearly distinguishable core component.  Integrated emission suggests a dominant steep spectrum non-core and so the flux deficit is more suggestive of the peak being due to one of two lobes rather than a strong core.  Fit for the non-core, extrapolate from the peak to get a core limit. \\
{\bf RXJ0000.1+0816} Only sampled at L and C bands but well-sampled at these.  Variability shows that it is core dominated.  Choose most co-temporal observations, fit for the core (allowing for variability in the uncertainty) and extrapolate to get a limit on the non-core component. \\
{\bf RXJ0021.6+2803} Weak CSS, only sampled at L and C bands. Undetected in AE110 but weak detection in AE125. Fit the non-core and extrapolate for the core. \\
{\bf Z235} Flat-spectrum, core dominated source. Potentially curves above 100~GHz but large uncertainty on the highest radio-frequency data means this cannot be confirmed hence best to fit only with a power-law (Hogan \etal 2015b).  Fit for the core and extrapolate for the non-core component. \\
{\bf RXJ0058.9+2657} Beautiful looking double lobed source.  Core peak in reasonable agreement with VLBA measurement of the core \cite[][]{Liuzzo09}.  Fit the extended emission for the non-core component and then extrapolate from the core flux measurement with appropriate uncertainties to infer a value for the core component. \\
{\bf A160} Large, clearly resolved source.  Spectrally see a difference between the steep (potentially sometimes contaminated in low resolution observations) extended emission and the flat, apparently active core.  Fit these separately for the respective components. \\
{\bf A2634} Another large resolved source, commonly known as 3C~465.  Literature observations splits between higher resolution core observations and low resolution `total' observations.  Can fit these separately, accounting for the very bright core when determining the non-core contribution.  C-band observation (AE125) does hint at recovering some emission from hot-spots although this would not account for all `missing flux'.  Only core component is reported.  \\
{\bf RXJ0107.4+3227} NGC383.  Associated with the strong FRI source 3C~31.  Majority of literature fluxes appear to be integrated, showing the total emission to be dominated by the steep spectrum non-core component.  Fit this, removing fluxes that resolve out significant sections of the total flux.  Resolved core flux from AE125 agrees well with the core flux reported by the Combined Radio All-Sky Targeted Eight GHz Survey \cite[CRATES,][]{Healey07}.  Extrapolate from these to get a measure for the core value. \\
{\bf RXJ0123.6+3315} NGC507.  Another powerful extended source.  Extended emission is resolved out at C-band observations leaving just the unresolved core.  Fit the lower frequency emission for the non-core component and then extrapolate with reasonable uncertainties for a measure of the core component. \\
{\bf A262} Slightly resolved source.  Extended emission dominating integrated spectrum, clear steep component down to $\sim$10~GHz.  VLBA observation \cite[][]{Liuzzo10} of the core at 5~GHz shows that there is an unresolved core component.  Spectral flattening to higher frequencies suggests this component is flat spectrum and the source is fit with a split-power model. \\
{\bf RXJ0740.9+5526} Double source, resolved lobes but no clearly resolved core hence only a limit can be determined for this at C band.  No VLBI observations so can only extrapolate from the limit for a limit on the core component and fit the steep spectrum lobe emission for the non-core component.  Position given is estimated nuclear position. \\
{\bf A115} Large, fuzzy double lobed source (3C~28).  No clear core so only a limit on this, which can be used in conjunction with limits at 22.5 and 43.3~GHz \cite[][]{Lin09} to get a limit on the core with the integrated emission being fit to retrieve the non-core component. Position given is estimated nuclear position. \\

{\bf \subsubsection{VLA - AE110}}
{\bf A75}  Weak unresolved detection at C-band but no additional literature fluxes.  Extrapolate from this point to get limit on the core, no match in NVSS so extrapolate from this limit for a limit on the non-core. \\
{\bf A76}  Only two fluxes.  Unresolved at C-band.  Steep spectrum, fit to get non-core flux and extrapolate from C-band flux with $\alpha$=-0.2 to get upper limit on the core. \\
{\bf A77}  Source unresolved at C-band.  Appears to become diffuse at X-band with VLA-A array. Moderately flat spectrum although diffuse nature and rollover suggests best fit with a dropline model to get the non-core flux and then extrapolate from the X-band flux with $\alpha$=-0.2 to get upper limit for the core. Integrated X-band measurement consistent with \cite{Lin09}, suggests rollover is real. A limit at 22.5~GHz from this paper adds further weight that this is relatively compact yet ageing emission rather than a flat spectrum core.  \\
{\bf A147} Unresolved faint detection at C-band.  No literature fluxes.  Taking the NVSS snapshot image of the region there is a faint source at the position that is not in the NVSS catalogue.  The peak of this source is taken as the flux at 1.4~GHz.  These flux measurements suggest a steep power-law.  Fit to get a measure of the non-core flux and extrapolate C-band point with $\alpha$=-0.2 to get upper limit on the core. \\
{\bf A168}  Only got data at L- and C-band.  However, see that NVSS picks up a lot of extra flux in comparison with the peak of the FIRST detection so extrapolate this NVSS flux with $\alpha$=-1.0 to get a non-core flux limit.  Taking the peak of the FIRST detection with the effectively unresolved (peak $>$70$\%$ integrated) C-band flux then we see a flat core component so use this to estimate the core contribution at 10~GHz. \\
{\bf A193}  USS source.  Could potentially be resolving out flux at C-band.  There is slight extent seen at C-band (peak $>$70$\%$ integrated).  Use $\alpha$=-0.2 from the peak of the C-band observation to get upper-limit for the core and interpolate the integrated fluxes for a reasonable estimate of the non-core. \\
{\bf A189}  Appears slightly elongated at C-band and is a known cavity system \cite[][]{Dong10}.  No clear structure although is a $\sim$35\% flux deficit also suggesting that this source is marginally resolved.  Hint of spectral steepening for the extended emission although only got three points so fit these just with a single power and then extrapolate the C-band peak to 10~GHz with $\alpha$=-0.2 to get an estimate of the core flux. \\
{\bf RXJ0419.6+0225}  Faint unresolved at C-band.  Steep spectrum, fit a single power law for a non-core measurement and then extrapolate for a core limit. \\
{\bf Z6718}  Fairly steady steep spectrum from which the non-core can be determined.  Potentially slightly extended at C-band.  Extrapolate from the peak of the Cband observation for a limit on the core. \\
{\bf A1902} Observed only at L and C-bands, this source is nevertheless unresolved with a steep index.  From this, estimate the non-core and then extrapolate from C-band for a core limit. \\
{\bf A1927} Appear to be seeing unresolved, steep component.  Use this to get a measurement of the non-core flux and then extrapolate with $\alpha$=-0.2 from the unresolved C-band flux to get an upper-limit for the core.  There is a brighter (12 mJy) source $\sim$2' to the north-east as well as diffuse emission $\sim$5.3' to the north-west that may be a phoenix-relic.  These independent sources may be confused with the BCG in low resolution observations. \\
{\bf A1930} There is a flat component between the unresolved FIRST and C-band observations that can be fitted for a core estimate.  NVSS shows additional flux to FIRST, perhaps indicative of there being a weak non-core component being better sampled at this lower-resolution. The low frequency observation from WENSS confirms that a steeper component is present. Extrapolate from this point using $\alpha$=-1.0 to get an estimate for the non-core. \\
{\bf RXJ1440.6+0327}   Observed at only L and C -bands.  Unresolved at C-band, steeper than would be an unambiguous core ($\alpha$=0.62) but appears consistent with being attributable to an active core component between the FIRST peak and unresolved C-band observation.  Measure core from this, allowing for large uncertainty.  Seeing the more diffuse emission from NVSS so extrapolate this point with $\alpha$=-1.0 to get a measure of the non-core flux. \\
{\bf A1978}  The unresolved C-band flux is consistent with a steep component if linked to the NVSS observation.  Take a measurement of the non-core from this and then extrapolate the C-band peak to 10~GHz with $\alpha$=-0.2 to get a limit on the core component. \\
{\bf MACS1532.9+3021}   CSS, unresolved at both C- and X-band.  Fit with a single power to take a measurement of the non-core. Extrapolate back from the BIMA measurement at 28~GHz \cite[][]{Coble07} using $\alpha$=-0.2 to get an upper-limit for the core. \\
{\bf A2110}  Composed of an USS component ($\alpha$=-1.9) and then a flatter ($\alpha$=-0.25) core component that can be directly fit with a split power model.  Consistent with it being unresolved at C-band. \\
{\bf A2108 (a) \& (b) } No additional literature fluxes for sources at either of these positions.  Brighter source nearby could have swamped catalogue detections.  No SED fit possible, urge caution using given fluxes, potentially spurious.  \\
{\bf RXJ1604.9+2356}  Resolved source.  Appears disturbed, shows potential FRI-like lobes or maybe head-tail morphology.  Have a steady source that hints at synchrotron ageing although steep component adequately fit with a single power law to get the non-core flux.  Very resolved at C-band and clear that the FIRST point is resolving out much of the lower spatial frequency flux compared to NVSS.  Use the C-band peak to get a measure of the core flux from this flat component.  Note that given position is estimate of nuclear position, falling between dominant lobes.  \\
{\bf A2204}  Source well-fit with a split power model, showing spectral flattening to higher frequency.  Appears unresolved at C-band although a gaussian fit suggests a flux decrement from the peak, maybe hinting that it is beginning to be resolved at VLA-C resolution. \\
{\bf A2228}   Weak unresolved detection at C-band but no additional literature fluxes.  Extrapolate from this point to get a limit on the core.  Non-detection in NVSS, use this for a non-core limit. \\
{\bf A2244} Faint CSS.  Fit for non-core and extrapolate for a limit on the core. There are two brighter sources within 2' seen at C-band - one an unresolved $\sim$4.1 mJy source to the north, the other a $\sim$30 mJy double to the east.  These may be confused with the BCG at lower resolution. \\
{\bf RXJ1715.3+5725}  NGC~6338. Flat, core dominated source, hint of mild variability.  Consistent with WENSS flux still being core dominated.  Fit with a single power to get core flux then extrapolate with $\alpha$=-1.0 from the WENSS flux to get a non-core upper-limit. See also Hogan \etal 2015b \\
{\bf Z8197} CSS.  Unresolved at C- and X-bands.  Consistent with a steep, single power.  Get the non-core from this and then extrapolate from the X-band flux with $\alpha$=-1.0 to get an upper limit on the core.\\
{\bf RXJ1733.0+4345 } There appear to be inner-relics nearby judging by low resolution, low frequency observations.  The BCG itself is unresolved at C-band.  Fit a flat component to this and the FIRST peak to get a measurement of the core. NVSS is picking up non-core flux above that recovered in FIRST so extrapolate (using $\alpha$=-1.0) from this to estimate the non-core flux.\\
{\bf A2443 } A reasonably well-sampled field although there is a large relic in the cluster \cite[][]{Cohen11} that causes confusion in some images.  Spectral decomposition has used only observations with the necessary resolution to isolate BCG.  The BCG looks largely resolved at C-band, although may be jetted to north-east.  Fit steep component from NVSS to integrated C-band flux to get non-core then extrapolate using $\alpha$=-0.2 from the C-band peak to get an upper limit on the core.  This is a very dynamic field.  In addition to the relic there is a diffuse, $\sim$5 mJy source $\sim$2' to the south as well as a WAT ($\sim$23 mJy) and a head-tail source ($\sim$10 mJy) both 2-3' to north-west.  There is therefore strong potential for confused observations of this BCG.  \\
{\bf A2457 } Weak unresolved detection at C-band but no additional literature fluxes.  Extrapolate from this to get limits on both the non-core and the core. \\
{\bf A2495 } Only two observations.  Slightly resolved at C-band.  Fit steep component from NVSS to integrated C-band flux to get non-core then extrapolate using $\alpha$=-0.2 from the C-band peak to get an upper limit for the core.\\
{\bf A2626 } Sources contains and extremely steep ($\alpha$=-2.2) component at low frequencies and is well sampled over wide frequency range.  Unresolved at C- and X-band.  Fit the two components separately. Good example of a young, flat core but with extreme USS component. \\
{\bf A2627 a} Got an active core component as seen by GISMO/CARMA (see Hogan \etal 2015b).  Resolved into fairly amorphous jetted structure at C-band, no extent at X-band.  Can fit the flatter, high frequency component to the unresolved X-band flux for a measure of the core component.  There is a hint of extended emission at lower frequencies.  Still relatively flat but now what is classified as non-core so fit these seperately and subtract the expected core component to infer a non-core contribution.  \\
{\bf A2627 b} This source is not particularly faint, however there appear to be no additional fluxes for it in the literature.  This may be due to confusion with A2627~(a).  This could also contribute to over-estimating non-core flux for A2627(a).  This source could not be robustly independently fit with current data.  Consider the two components when studying this source. \\
{\bf A2665} Spectrum shows a higher frequency turnover that is consistent with an ageing synchrotron population.  Resolved at C-band and unresolved at X-band suggests seeing the edge of the non-core and then the underlying core component.  Fit a dropline model to get the non-core component and then extrapolate from the unresolved X-band flux to get an estimate of the core component. \\
{\bf A2675} VLSS to NVSS interpolation gives a very steep spectrum ($\alpha$=1.55). Unresolved C-band observation suggests there is a core component contributing at higher frequencies although extrapolation to 10~GHz would be uncertain. Cautiously just fit lowest two fluxes to get a reasonable estimate for the non-core then extrapolate from the C-band flux to get an upper limit on the core.  There are no other sources apparent in field at C-band to suggest that the VLSS is confused. \\
{\bf A2622}  SED has a turnover to high frequency and can be well fit with a dropline model.  No active core is apparent in this system.  Only a diffuse envelope is detected at C-band and there is only an upper-limit from the X-band observation.  Fxtrapolate from this upper limit to get a very stringent upper limit on the core flux.  Note that this interpretation is in agreement with \cite{Giacintucci07} who considered this galaxy to be a dying radio source that is not currently being fed by nuclear activity. \\
{\bf A566}  There is no detection in the C-band VLA imaging at the position of BCG.  There are two nearby radio galaxies.  The BCG position is coincident with what appears to be inner core-relic emission, resolved so as to be able to isolate it in NVSS.  Clearly not active but does appear to be directly related to activity of the BCG within emitting lifetime hence include as a non-core component. \\
{\bf RXJ0751.3+5021} Only got detections at L and C bands, unresolved at both and the non-detection in WENSS shows this source to have a flat index.  Fit for the core and extrapolate from 1.4~GHz with index $\alpha$ = 1.0 to get a limit for the non-core. \\
{\bf Z2089}  Only got detections at L and C bands, unresolved at both with a steep index.  Fit for the non-core and extrapolate from 4.86~GHz with index $\alpha$ = 0.2 to get a limit for the core. \\
{\bf A763} BCG is a large, tailed source.  Resolve out the components and there is a published high-resolution VLA core flux \cite[][]{Jetha06} which agrees with the recovered core flux from our map.  Fit the core as a single power law from these, the FIRST resolved peak and higher frequency observations of the core.  Appears that the interferometric observations may be resolving out some of the more extended flux.  Fit a dropline model to the non-core. \\
{\bf Z1665}  Only got observations at L and C bands, unresolved at both with flat index showing it to be core dominated.  Fit for the core and extrapolate from 1.4~GHz with index $\alpha$ = 1.0 to get a limit for the non-core. \\
{\bf Z2844}  Flat core dominated source with a hint of extra emission at low frequency from WENSS.  Fit the core with a power law and then extrapolate from WENSS to get a limit on the non-core. \\
{\bf A1035}  Only has fluxes at three frequencies.  Unresolved at L and C bands with a flat index suggesting core dominance.  Fit these to get a measure for the core and then extrapolate back from the WENSS point to get a limit on the non-core. \\
{\bf A1204}  Only got observations at L and C bands, unresolved at both, flat index, core dominated.  There is a slight hint that there may be extra emission being picked up by NVSS over FIRST but equally this could be attributed to variability within an active core.  Hence can only reasonably fit for the core and extrapolate from 1.4~GHz with index $\alpha$ = 1.0 to get a limit for the non-core. \\
{\bf A1235}  This source is in the FIRST region but is not in the FIRST catalogue.  It is however detected at C-band (weakly) and there is a clear source in the FIRST cutout.  This gives a flat index suggestive of a core so take an estimate of this and then extrapolate for a limit on the non-core. \\
{\bf A1553}  Appears to be a CSS.  Slightly elongated at C-band and got a clear peak deficit. A lot more flux picked up by NVSS than by FIRST but restricted to just two frequencies.  See a diffuse, core-relic like structure $\sim$10arcsec to north-west in FIRST, not detected at C-band and confused at lower resolution.  Fit (with suitably large errors) for the non-core and then extrapolate to get a limit on the core. \\
{\bf RXJ1326.3+0013}  Slightly resolved image suggests a one sided source.  Fit the extended emission to get the non-core, extrapolate from the peak at X-band to get a limit on the core. \\

{\bf \subsubsection {VLA - AE107}}
{\bf RXJ0341.3+1524}  Unresolved at C-band and has a flat index consistent with it being core dominated.  Fit this and then extrapolate from NVSS using the usual $\alpha$=-1.0 to get upper limit for the non-core.\\
{\bf RXJ0503.1+0608}  Unresolved at C-band although no additional fluxes in the literature.  Not particularly faint ($\sim$5 mJy) suggesting either that previous observations were confused by other sources (there are several radio-bright sources in the vicinity) or that this is core-dominated and variable.  Indeed, considering the NVSS map suggests that both of these scenarios may contribute.  There is a bright nearby confusing source that would be confused at any lower resolution.  Additionally, the BCG can be isolated in NVSS (below the catalogue limit) and shows that the spectrum is inverted and therefore core dominated. Fit for an estimate of the core and then extrapolate for a non-core limit. \\
{\bf A611} Faint and unresolved at C-band.  No additional detections in the literature with only an upper-limit in NVSS.  Use these to get limits on both components using the representative indices. \\
{\bf RXJ0819.6+6336} Spectrum is well-fit by a fairly steep ($\alpha$=-0.8) single power law.  C-band image hints at lobes which is confirmed by high resolution image at X-band with VLA-A array.  In X-band a core component can be clearly distinguished.  Other observations do not have the necessary resolution to confidently separate this so fit their integrated fluxes as a single power to get a measure of the non-core (which here is shown to be extended lobes).  Extrapolate the core measurement at X-band to 10~GHz, with representative index and conservative uncertainty to get a relatively robust measure of the core flux.\\
{\bf A667} Steep source that appears marginally resolved at C-band although only two fluxes are available.  Extrapolate from the C-band peak for a core upper limit. There is a $\sim$7 mJy double source only $\sim$30'' to the south-east that is likely confused at lower frequencies.  Source appears confused in NVSS imaging. \\
{\bf A971} Source appears in the 7C catalogue \cite[][]{Waldram96} but this appears to be confused, otherwise would suggest incredibly steep ($\alpha$$>$3.25) index that is hard to explain physically.  Excluding this point then the spectrum can be fit with a reasonably steep power law ($\alpha$$\approx$0.82) to get a realistic non-core flux. Extrapolate from the unresolved C-band flux to get an upper limit on the core. \\
{\bf A980} Steep spectrum source ($\alpha$$\approx$1.42) with excess flux picked up by NVSS over FIRST.  A single power law gives a reasonable measurement of the non-core flux.  Appears to be a small double in FIRST and marginally resolved at C-band.  Extrapolate from the peak of the C-band measurement to get an upper limit for the core. Note that the given position is the C-band peak and may best represent a lobe rather than the true nucleus. \\
{\bf A1045} Unresolved at C-band with a hint of a slight peak-to-integrated flux deficit at L-band although no clear extent.  Spectrum continues to be steep down to WENSS.  Appear to be seeing a non-core component dominating at lower frequencies.  The core cannot be unambiguosly separated with available data hence fit for the non-core only and take a limit on core.\\
{\bf RXJ1715.1+0309} Spectrum is a steep power law and source is resolved at C-band showing that the fit gives a good measure of the non-core. Extrapolate from the peak at C-band to get an upper limit for the core.  Note that source appears to be resolved into lobes with no clear core.  Perhaps a small-scale double. Position given is that of the peakflux and most likely represents a hot-spot rather than nucleus.  \\
{\bf A2261}  Steep spectrum ($\alpha$$\approx$1.4) source that appears unresolved at C-band.  Fit to get a measure of the non-core flux and then extrapolate from C-band to get upper limit for the core.  Potential confusing source for low resolution observations $\sim$2.5' to north-west. \\
{\bf RXJ1750.2+3504}  Flat spectrum, appears to be core dominated.  Fit with a single power-law to obtain a measurement of the core then extrapolate back from the lowest frequency point \cite[7C;][]{Waldram96} to get an upper limit on the non-core.\\
{\bf RXJ2129.6+0005}  Unresolved at C-band, X-band and has a BIMA point \cite[][]{Coble07}.  Fitting these returns a flat spectrum from which a core component can be inferred.  There is a clear flux difference between the peak of the VLA-A observation at L-band and NVSS that suggests that NVSS is picking up excess flux which can be associated with the non-core. Extrapolate from NVSS using $\alpha$=-1.0 to get an estimate of the non-core flux, accounting for the inferred core. \\

{\bf \subsubsection {VLA - AE099}}
{\bf A291} Only got observations at L-band and C-band.  Unresolved at both but interpolation is flat and there is a hint of variation between repeat observations at L-band suggesting that source is core dominated.  Take an estimate of this and then extrapolate from NVSS to get a limit on the non-core. \\
{\bf Z1121} Head-tail source where the tail is bent, wide and diffuse.  There is only a marginal variation in the power law index of the SED.  The peak-to-integrated flux and the resolved morphology, both at C-band enable the core to be isolated, however a direct split power fit to the integrated SED gives an unreasonable core estimate.  Instead, fit the integrated fluxes with a single steep power law to get the non-core flux and then extrapolate from the resolved core component at C-band with $\alpha$=-0.2 and conservative uncertainty to obtain an estimate of the core flux. \\
{\bf RXJ0821.0+0752} Appears to be a CSS although only observed at L- and C-bands.  Unresolved at both although the index is steep ($\alpha$=-0.8).  Extrapolate to get an upper limit on the core.  Note that there is a strong ($\sim$30 mJy at C-band) head-tail only $\sim$30'' to the south-west.  The BCG is confused with this source in NVSS although FIRST does separate them, hence only FIRST and AE099 can be used. \\
{\bf Z2701} Source is an USS ($\alpha$$\approx$2.0). There appears to be a slight kink in the higher frequency tail.  Source is unresolved at C-band (highest frequency observed at) although this source is also observed with the VLBA (Hogan \etal 2015c, in prep.) that shows there is no active core component on milli-arcsec scales.  Therefore fit for the non-core and then extrapolate for an upper-limit on the core. \\
{\bf Z3146} Unresolved at C-band.  L-band fluxes suggest that NVSS is picking up extra flux from the non-core compared to FIRST whilst the peak of the FIRST detection is consistent with being attributable to an active core, giving a flat ($\alpha$$\approx$0.4) index when combined with the C-band and BIMA fluxes \cite[][]{Coble07}.  Fit this flat core component and then extrapolate from NVSS flux to get a measure of the non-core. \\
{\bf Z3916} Taking the integrated fluxes then the SED is well fit by a dropline model with a steep power-law component ($\alpha$=-1.5).  From the L-band and C-band imaging there are hints that the peaks are partially separating the core but this is not unambiguosly well-resolved as to be able to make a confident core measurement hence extrapolate from the C-band peak for an upper limit on the core.  In the radio imaging alone the source appears to be an FRII at C-band (AE099) and FIRST.  However, overlaying these maps onto optical images of the region shows that there are actually three, roughly aligned radio-galaxies in the vicinity.  The BCG is the brightest (30 mJy at L-band) and then two other unresolved sources, (3 mJy and 10 mJy at L-band) lie $\sim$14'' and $\sim$28'' to the north-west respectively.  These sources are confused in NVSS and therefore note that this source likely suffers from confusion and hence the non-core figure may be an overestimate. It should be noted though that using only the fluxes from AE099 and FIRST would still classify this source as steep spectrum.  \\
{\bf A1651} Consistent with being a CSS although only observed at L- and C-band.  Fit to get estimate of the non-core flux and then extrapolate with a flat index to get an upper limit on the core flux. \\
{\bf A1664}  Source is unresolved at both C- and X-band, showing a strong core with a reasonably flat ($\alpha$=0.5) index from which a measurement of the core can be derived (note that this is confirmed by a VLBA observation, Hogan \etal 2015c).  The spectrum steepens below 1~GHz ($\alpha$=1.4).  A direct split-power fit fails and combing the two power-law fits to the individual components over-estimates the flux in the $\sim$1~GHz range. This suggests that the core component likely turns over (self-absorbs) in ~GHz range.  However, fitting the lower frequency points separately but accounting for the inferred core component gives a reasonable value for the non-core. \\
{\bf A1084} Unresolved at L- and C-bands with the interpolation between these suggesting a flat core component ($\alpha$$\approx$0.1).  The source falls within the TGSS region, with the recovered flux suggesting the presence of a steeper spectrum component at lower frequencies.  Extrapolate with $\alpha$=1.0 from this 150~MHz observations to get an upper limit for the non-core at 1~GHz. Appears to suggest a very core dominated source at $\sim$~GHz range frequencies, perhaps with a very steep spectrum amorphous halo of material to low frequency. \\
{\bf A1885} Multi-epoch coverage shows this source to be variable at C-band suggesting core dominance.  WISH \cite[][]{DeBreuck02} observation confirms this core dominance. There is no significant non-core component, the source appears to be either a flat spectrum sources or a GPS.  Fit to get the core flux and extrapolate from the WISH detection to get an upper limit on the non-core. \\
{\bf A1682}  Cluster has twin BCGs.  A1682(b) is a non-detection in our C-band imaging.  A1682(a) is initially matched as a powerful double-lobed source however further investigation with the optical overlays shows this to be a projected background source and that the BCG itself is radio undetected. \\

{\bf \subsubsection {VLA - AE117 L-Band}}
{\bf Z808} Source contains inner-relics that are clearly visible in the C-band imaging.  These appear to be associated with the BCG and hence dominate the integrated spectrum, which can be well-fit with a dropline model to give a measure of the non-core.  A distinguishable, flat spectrum core is apparent in the L-, C- and X-band imaging.  Fitting this seperately enables a measurement of the central, active core to be obtained. \\
{\bf A478} This source was not covered by our VLA-C observations, however it was observed at both L- and X-bands as art of project AE117, showing one-sided morphology at L-band with the structure resolved out at X-band leaving only an isolated core.  Interpolating between the high resolution L-band peak and the unresolved X-band flux returns a flat spectrum.  There is also a BIMA observation \cite[][]{Coble07} that suggests this source may contain a variable component, thus reinforcing that it is core dominated at higher frequencies.  The measured core index is marginally steeper than usual when skewed by the BIMA measurement but the resolved nature suggests this is still attributable as a separate core component.  NVSS appears to recover the non-core component above the flux expected from the active core so extrapolate from this using $\alpha$=-1.0 to get an estimate of the non-core component. \\
{\bf Z1261} The BCG is associated with the radio source 4C+53.16, which is a clearly resolved FRI/II.  There are distinguishable inner jets at C-band but also clear hot-spots with spherical cap-like lobes.  The integrated SED can be well-fit with a dropline model, highlighting that there is likely synchrotron ageing in the lobes.  Use this fit to retrieve a measure of the non-core.  Observations at L-, C- and X-band show a clearly resolved core, the isolated SED for this component only displaying an inverted index from which a measure of the core can be obtained.  There is also a Very Small Array \cite[VSA;][]{Cleary05} observation at 33~GHz.  This array only has 0.5degree resolution so unlike other high frequency observations does not provide a good measurement of the core flux.  However, the flux measured is consistent with the dropline component so restrains the magnitude of the synchrotron ageing. \\
{\bf RXJ1320.1+3308} Clearly resolved diffuse emission at both L- and C-band (using VLA-A and VLA-C array respectively) with only an unresolved core recovered at X-band (using VLA-A array).  The integrated high-resolution L-band flux is consistent with the flux given by NVSS suggesting we are not losing emission at intermediate scales.  Fit the SED for the resolved core seperately to get a measurement of the active component.  The SED for the diffuse emission alone has a relatively flat spectrum ($\alpha$$\approx$0.56) suggesting this component is resolved out rather than below the detection limit at X-band.  Fit the lower frequency components to get a measure of this diffuse emission, accounting for the core as this is non-negligible at 1~GHz. \\
{\bf A1835} Source is a fairly classic CSS.  Unresolved at all observed frequencies and has a BIMA point at 28.5~GHz \cite[][]{Coble07} that shows the steep spectrum is consistent to higher frequencies.  Fit this to get a measurement of the non-core and then extrapolate from the BIMA flux with index  $\alpha$=0.2 to get an upper limit on the core.\\

{\bf \subsubsection {VLA - AE117 X-Band}}
{\bf MACS0159.8-0850}  Source is undetected at low frequency \cite[e.g. in VLSS,][]{Cohen07} and displays an inverted spectrum in the GHz-range, showing it to be core dominated. Spectrum is steady to 20~GHz \cite[AT20g,][]{Murphy10} so a GPS-fit is not-limiting although a turnover is expected above 20~GHz. Fit a single power-law to get a measurement of the core contribution at 10~GHz and then extrapolate from the VLSS non-detection to get an upper limit on non-core flux. \\
{\bf RXJ0352.9+1941}  Unresolved at C- and X-band and has a mildly inverted component that can be fitted to get a measure of the core.  Appears to be flat spectrum to low-frequency, extrapolate from NVSS with $\alpha$=-1.0 to get a limit on the non-core.  See also Hogan \etal 2015b. \\
{\bf RXJ0439.0+0520}  This source contains a very strong, self-absorbed core component that dominates above $\sim$1~GHz and a steep spectrum tail below this.  Fit the steep spectrum component to lower frequencies for a measure of the non-core and fit the peaked component with the GPS-like model to get a measure of the core. \\
{\bf A646} Source is core dominated with an inverted/peaked component that can be fit to measure the core flux.  Extrapolate from the WENSS flux to get an upper limit for the non-core.  See also Hogan \etal 2015b. \\
{\bf A795} Unresolved at C-band but shows a steep single power law spectrum. Fit with a single power to get non-core flux. There is slight resolution at X-band.  Extrapolate from the X-band peak to get an upper limit on the core. \\
{\bf Z3179} SED can be well fit using a dropline model to get the non-core.  Extrapolate from the X-band flux to get an upper limit on the core. \\
{\bf A1068} Appears marginally resolved at C-band and part of a double, however the second component appears un-associated and is not present at X-band.  Fitting to L-band suggests continuation of steep index, indicative of a CSS.  Fit to get a measure of the non-core and then extrapolate from X-band to get an upper limit on the core. \\
{\bf A1361} BCG associated with the extended radio-source 4C+46.23 that has two-sided morphology at C-band.  The integrated spectrum can be fit with a dropline model to get the non-core component.  The extended emission is resolved out with the VLA-A at X-band to leave an isolated core.  Extrapolate from this to get a reasonable estimate of the core component at 10~GHz. \\
{\bf RXJ1504.1-0248}  Flat spectrum source to high frequency (see Hogan \etal 2015b), and unresolved on milli-arcsecond scales at C-band as seen by \cite{Bourda10} and Hogan \etal 2015c with the VLBA.  Fit this for a measure of the core.  Source contains a known mini-halo \cite[][]{Giacintucci11b}.  This emission can be isolated from the BCG and the radio-power of this is not believed to be provided directly by the AGN hence fit the flat-component to get a core flux and then extrapolate from the NVSS point to get an upper limit for the non-core. \\
{\bf 4C -05.84}  Large, extended source with a spectrum that can be fit with a dropline model.  VLA-A array X-band observation appears to resolve out all the large scale emission leaving just the unresolved core which is two orders of magnitude below what a straightforward extrapolation of the dropline shape would predict.  Fit the integrated SED with a dropline model to get the non-core then extrapolate from the unresolved X-band core flux to get the measurement at 10~GHz for the core.\\
{\bf Z9077}  The centre of this cluster contains two radio-sources that are close enough in projection ($<$45'') to be confused in NVSS and at lower frequencies.  Extrapolate from the peak of the isolated BCG to get an upper limit on the core flux.  Give two values for the non-core: one for if only higher resolution data are used where the source is clearly separated and a significantly higher non-core estimate taken by just fitting the low resolution literature fluxes.  The BCG is indistinct at C-band, appearing as a slightly elongated but faint source whereas the other source appears to have a short northern jet. \\

{\bf \subsubsection {ATCA - C1958 - 2011}}
{\bf A2734} Unresolved at C- and X-band with the ATCA, in conjunction with the NVSS measurement this source has a reasonably steep ($\alpha$$\approx$0.8) power law spectrum.  It has a marginally matched VLSS detection however inspection of this overlaid onto optical imaging of the BCG suggest this is not associated.  Fit the steep component to measure the non-core and then extrapolate with $\alpha$=0.2 to get upper limit on the core.\\
{\bf A2746} Unresolved at C- and X-bands. These fluxes give a flat ($\alpha$=0.4) index consistent with core emission.  SUMSS \cite[][]{Mauch03} detection suggests there may be a steeper component at lower frequencies although this is only constrained enough to be able to get an upper-limit for the non-core by extrapolating with $\alpha$=1.0. \\
{\bf A3112} This source has multi-epoch observations suggesting evidence of variability at C-band and higher frequencies.  Additionally, the source appears potentially slightly resolved with a small jet to the south and there is a larger peak-to-integrated flux discrepancy at X- than C-band.  However, whilst there is evidence of core activity in this source the integrated spectrum is steep and appears to be dominated by the non-core.  There is no clearly resolved core at the current resolutions, hence interpolate between the peak of X-band observation and the 20~GHz observation \cite[][]{Murphy10} to get an upper limit on the core.  Fit the steep component to get a flux at 1~GHz. Note that there is a 95~GHz observation of this source \cite[][]{Sadler08} that is consistent with the integrated spectrum continuing to high frequency and highlights that the core limit derived herein is not expected to be stringent.  The source appears to be similar to an FRI/II in behaviour and may perhaps an inadequately resolved FR-type source. \\
{\bf RXJ0528.9-3927} Faint, unresolved detection at C-band and only an upper limit at X-band.  Extrapolate from the X-band limit to get an upper limit on the core and then fit the C-band observation with lower frequency detections that return an USS ($\alpha$$\approx$1.7) to get a measure of the non-core. \\
{\bf MACSJ0547.0-3904} One of the 34 most luminous MACS clusters \cite[][]{Ebeling10}.  Unresolved at C- and X-band (potential hint of extent at C-band to the north-west but remapping suggests this is most likely an artefact), the inferred flat ($\alpha$$\approx$0.1) index is consistent with the SUMSS flux.  However, the source appears as an NVSS object with almost twice the flux expected flux from the spectral fit.  Overlaying the NVSS and SUMSS maps shows a consistent unresolved source.  Possible that this is a highly variable source or that perhaps the NVSS flux is inflated.  The position (Dec$\approx$-39) is just above the NVSS horizon.  We exclude the NVSS flux from our fit, recovering a core measurement and then extrapolate from the SUMSS detection for a limit on the non-core. \\
{\bf AS701}  Unresolved at both C- and X-bands and see a spectral inversion between these frequencies showing the presence of a core component.  At lower frequencies the spectrum steepens as shown by a TGSS detection, fit therefore for a measure of the non-core. \\
{\bf A3605} Unresolved at C- and X-bands with a flat ($\alpha$$\approx$0.1) spectral component from which a core component can be inferred.  Extrapolating from the 150~MHz TGSS detection (with $\alpha$=1.0) gives a flux at L-band that is consistent with the flux difference at 1.4~GHz between that expected from the core fit and that recovered in NVSS hence it appears reasonable to claim a non-core flux from TGSS only.  This approach constrains the position of the self-absorption turnover of the core component to $<$1.4~GHz. \\
{\bf RXJ1539.5-8335} Source is flat spetrum and appears to be core dominated. Unresolved at C- and X-bands, the SUMSS detection is more suggestive of there being a self-absorption turnover between L and C bands rather than any additional non-core flux component at lower frequencies. Fit with a single power law to get a measurement for the core and extraolate from the SUMSS detection to get a limit on the non-core. \\
{\bf AS805} Source has a consistently steep power-law index between 843~MHz and 9.0~GHz and is unresolved at all observed frequencies, classifying the object as a likely CSS.  Fit the spectrum to obtain a non-core flux and extrapolate from the X-band peak to get an upper-limit on the core component. \\ 
{\bf RXJ1931.6-3354} BCG associated with the radio source PKS 1928-34. Unresolved at C- and X-band with ATCA-6km but considering the SED it appears as if there is likely a steep spectrum component that is being resolved out.  Fit this steep component seperately, then fit the unresolved core.  \\
{\bf RXJ1947.2-7623} Steep spectrum source, fit with single power-law to get a measure of the non-core and then extrapolate from the X-band peak to get an upper-limit on the core. \\
{\bf A3992} Weak detection at C-band, undetected at X-band and no literature detections.   Detection limits do not limit the spectrum to be sufficiently flat or steep and hence from the available information it cannot be claimed to be dominated by either core or non-core dominated.  Extrapolate therefore from the C-band detection with flat index to get core limit and then extrapolate from the SUMSS detection limit for an upper-limit on the non-core component. \\

{\bf \subsubsection {ATCA - C1958 - 2008}}
{\bf A11}  Multi-epoch C-band observations suggest this source may be slightly variable.  There is a distinct steep component in the SED from which to get a measure of the non-core.  The spectrum flattens to high frequency, interpolating between the X-band peak and 20~GHz flux \cite[][]{Murphy10} gives a reasonable estimate of the core component. \\ 
{\bf A3017}  Unresolved at C- and X-bands.  20~GHz detection along with the X-band flux shows that the spectrum flattens to higher frequency and hence is expected to be core dominated at 10~GHz.  The spectrum steepens below this and there appears to be an ultra steep component ($\alpha$$>$1.5) to SUMSS.  These components can be fit separately for reasonable estimates of the core and non-core respectively. \\
{\bf RXJ0331.1-2100} The integrated spectrum can be well fit with a dropline model, indicative of it being non core dominated.  Therefore fit to get a measure of the non-core and then extrapolate from the 20~GHz detection \cite[][]{Murphy10} to get an upper limit on the core. Source is perhaps best classified as a CSS with a spectral rollover to lower frequencies \cite[see e.g.][]{O'Dea98}, perhaps indicative of extended diffuse emission at $<$100~MHz. \\
{\bf A496} VLSS \cite[][]{Cohen07} and CULGOORA \cite[][]{Slee95} detections show there to be an USS component to this source, whereas higher frequency observations show a strong, variable core component that flattens the spectrum and is dominant above a few~GHz.  Fit to fluxes at frequencies above 1.4~GHz to get a measure of the core and then separately fit to the lowest frequency observations to get the non-core.  NVSS (1.4~GHz) is in the `crossover zone' where the core and non-core are believed to contribute roughly equally and hence is used to check that the relative proportions of the components are consistent. \\
{\bf RXJ0543.4-4430} Unresolved at both C- and X-bands. SED shows both a flat and a steep component that canbe fit with a split power-law model to obtain measures of the core and non-core components respectively. \\
{\bf AS555} Source is extended in the C- observations although the core cannot be reliably isolated.  Fit SED with a single power law to get the non-core flux.  There is a peak-to-integrated flux decrement at X-band although again, the core cannot be isolated and hence extrapolate from the peak at X-band to get an upper limit on the core component at 10~GHz. \\
{\bf A3378} The SED has a well sampled steep component from which the non-core flux can be measured.  The peak of the X-band detection in conjunction with the two 20~GHz \cite[][and Mahony, {\it private communication}]{Murphy10} fluxes suggest that there is an active core in this system that becomes more important to higher frequencies although this is not clearly being isolated.  An unambiguous measure of the core contribution cannot therefore be claimed, hence extrapolate from the lower 20~GHz flux to get an upper limit for the core. \\
{\bf RXJ0747.5-1917} This source is perhaps best classified as a CSS with a low frequency turnover.  Unresolved at C- and X-bands although the imaging suffers from an elliptical beam shape.  Fit a dropline model to the SED to get a measure of the non-core component and then extrapolate from the X-band peak to get an upper limit on the core. \\
{\bf A1348}  Source contains a variable core component that can be well-fit with a GPS-like model. This core dominance is confirmed by VLBA imaging (Hogan \etal 2015c).  There is a steeper component in the SED at lower frequencies as seen in VLSS \cite[][]{Cohen07} and WISH \cite[][]{DeBreauck02}, with the shape consistent with the steep component just catching the turnover of the GPS-like core component.  Fit a GPS model to the core and then fit a power law to the steep component to obtain a measure of the non-core, ensuring that the core contribution is accounted for. \\
{\bf RXJ1304.3-3031}  Source is unresolved at C- and X-bands, consistent with the GPS-like shape of the SED showing that this source is core dominated.  Source does not currently match to the TGSS catalogue but does appear in the thumbnail image, below the catalogue limit.  Nonetheless, the positions match and hence extrapolate from this TGSS detection to get an upper limit on the non-core. \\
{\bf RXJ1315.4-1623} This source is unresolved and has an apparently self-absorbed spectrum, consistent with it being core dominated.  There is a WISH detection at 352~MHz \cite[][]{DeBreuck02} that shows it to be peaked and a tentative TGSS map detection at 150~MHz.  This map detection may be indicative of a low-level non-core component although it should be stressed that this flux is recovered just above the noise.  Hence, extrapolate from this to get only an upper limit on the non-core. \\
{\bf A3581 } Source is flat spectrum ($\alpha$$\approx$0.5) down to low frequency \cite[e.g. 74~MHz with the VLSS][]{Cohen07} and shows signs of variability in repeat observations at both C- and X-bands. The source is resolved with the VLBA (Hogan \etal 2015c), with the integrated flux of this roughly consistent with the unresolved C-band flux thus showing this source to be highly core dominated.  Fit the resolved core component then remove this from the recovered flux of the lower resolution observation for an estimate of the non-core component.  Also see Hogan \etal 2015b.  \\
{\bf AS780} Source is variable (Hogan \etal 2015b) and has a flat spectrum showing it to be core dominated.  Fit the flat spectrum to get a core flux and then extrapolate back from WISH \cite[][]{DeBreuck02} to get an upper limit on the non-core.  Non-detection in VLSS \cite[][]{Cohen07} suggests that this source must turnover to low frequency.  \\
{\bf  RXJ1524.2-3154} The detected fluxes imply a flat spectrum although this source is undetected by TGSS at 150~MHz and the upper limit implies that this source must be a GPS .  Fit the core component and then extrapolate from the TGSS limit to get an upper-limit on the non-core. \\
{\bf RXJ1558.4-1410}  This source contains a very dominant, self-absorbed core component along with a steep spectrum power-law component to low ($<$1~GHz) frequencies.  Fit the core with a GPS-like model and then fit the steep component seperately with a power-law to obtain a value for the non-core.  See Hogan \etal 2015b for a more in-depth look at this source. \\
{\bf A3638} This source is unresolved and has a steep spectrum ($\alpha$$\approx$0.93) between C-band, X-band and 20~GHz \cite[][]{Murphy10} but no other literature fluxes are available.  However, there is a nearby bright source that is confused with the BCG at low resolutions and hence could explain why this source is apparently missing in some low frequency catalogues.  From the high resolution data the source can be classified as a CSS.  Fit the steep spectrum to obtain a measurement of the non-core and then extrapolate from the 20~GHz detection to get an upper-limit on the active core component.  \\
{\bf A3639} Steep spectrum source although there is a detection at 20~GHz in AT20g \cite[][]{Murphy10} that suggests the spectrum may flatten to higher frequencies, which is indicative of this system containing an active core.  There are insufficient data points to adequately fit using a direct split power model although the steep component can be fit with a single power-law for a non-core component and then interpolating between the 20~GHz and X-band data with conservative uncertainty allows for the core component to be estimated. \\
{\bf AS851} Source has a flat spectrum and multi-epoch observations at C-band suggest that it varies, indicating that this is core dominated.  Fit to get a measure of the core flux and then extrapolate from the SUMSS detection for an upper limit on the non-core. \\
{\bf RXJ2014.9-2430} A steep spectrum source that is potentially slightly extended to the West as seen with the ATCA-6km array at C-band although this is not apparent at X-band.  Fit the SED to get a non-core flux and then extrapolate from the X-band detection to get an upper limit on the core. \\
{\bf RXJ2213.1-2754} The SED for this source suggests significant variability at GHz frequencies.  Additionally, it appears as if some low resolution observations may suffer from confusion.  Fitting a power-law to the SED returns a flat index ($\alpha$$\approx$0.5). Use this to obtain an estimate of the core component and then extrapolate from the NVSS detection (lower frequency observations appear to be confused and hence are excluded) to get an upper limit on the non-core. \\
{\bf A3880} The BCG is associated with the radio-source PKS~2225-308, and shows extended single-sided structure at both C- and X-band.  From lower resolution data at C-band \cite[e.g. with Parkes,][]{Wright90} it appears as if the interferometric observations are still missing extended emission.  Fitting a single power-law to the SED comprising the integrated fluxes returns a steep index ($\alpha$$\approx$0.99). Use this to obtain a measure of the non-core flux and then extrapolate from the resolved core's peak flux at X-band to obtain an estimate of the active core component. \\
{\bf AS1101} Extended FRI-like structure seen at C-band although this is split into a more one-sided FRII-like morphology comprising a core and a hot spot at X-band.  The integrated fluxes are consistent with a steep ($\alpha$$\approx$1.4) index for the integrated emission that can be fit this to get a measure of the non-core.  Extrapolate from the resolved X-band core with $\alpha$=0.2 to get an estimate of the core flux at 10~GHz. \\
{\bf A2580} Source can be fit with a dropline model suggesting that it is non-core dominated.  Extrapolate from the X-band flux to obtain a core limit.  The WENSS flux \cite[][]{Rengelink97} suggests that this source may turnover and flatten to lower frequency which could be suggestive of it being wrongly interpreted and actually being a core-dominated source.  Multi-epoch of milli-arcsecond scale observations are required to unambiguously characterise this source. \\
{\bf A2667} The SED is too undersampled to be reasonably directly fit using a split power.  However, both C and X-band images show it to be unresolved, which combined with the 20~GHz flux \cite[AT20g,][]{Murphy10} suggest that this source is core dominated and perhaps variable.  The NVSS (1.4~GHz) flux is above the flatter index fitted at higher frequencies, which suggests there could be additional diffuse emission in the system although there are no lower frequency detections to reliably constrain this.   Therefore, fit the flat spectrum component with a power-law to get a measurement of the core component and then extrapolate from the NVSS flux to get an upper limit on the non-core. \\
{\bf A4059} Source is resolved into two-sided structure at C-band, with a clearly distinguishable core component.  The SED of the integrated fluxes is well fit with a dropline model and confirms that this system is dominated by ageing emission in the lobes.  The core component can be estimated by extrapolating from the peak of the X-band emission. \\
{\bf \subsubsection {ALL THE NON-DETECTIONS}}
All the non-detections were cross-matched with NVSS and/or SUMSS, depending on their position in the sky.  The relative detection thresholds of these surveys and our C-band observations are such that where there was a match (confirmed by inspection of visual overlays to optical data) then this was taken as evidence of the system being dominated by a steep component.  The catalogue flux was in these instances extrapolated to obtain an estimate of the non-core component and a core limit derived by extrapolating from the higher frequency limit.  In most instances there was no detection in either catalogue.  In these situations, an extrapolation was made from the NVSS/SUMSS limit for the non-core and again, the higher frequency limit was extrapolated from to obtain a a core limit, using the usual indices as described in the main text.

\section[]{Data Tables - Main Sample} \label{APPENDIX_TABLES}

\onecolumn
\begin{longtable}[l]{@{}llllrrrrcr@{}}
\caption{Recovered C-band fluxes. Note that the value given in the core column is the peak of the emission if unresolved or the peak of the core component for resolved sources.  Redshifts are from Crawford \etal (1999) unless otherwise stated.} \label{C_BAND} \\
  \hline\hline
        &            & \multicolumn{2}{c}{Position(J2000)} & \multicolumn{3}{c}{Fluxes (mJy)} & & &  \\
   Cluster & BCG & RA & DEC &  Core  & Total & RMS & Redshift & Lines? \\
  \hline\hline
\endfirsthead
\multicolumn{2}{l}{{\tablename} \thetable{} -- Continued} \\
  \hline\hline
        &            & \multicolumn{2}{c}{Position(J2000)} & \multicolumn{3}{c}{Fluxes (mJy)} & & &  \\
   Cluster & BCG & RA & DEC &  Core  & Total & RMS & Redshift & Lines? \\
  \hline\hline
  \\
\endhead
  \multicolumn{3}{l}{{Continued on Next Page\ldots}} \\
\endfoot
  \\ \hline \hline
\endlastfoot
 \\
 \bf{VLA-C Array}  \\
 \hline
\\
\multicolumn{2}{l}{\bf{\emph{AE125 : 1999 : Bandwidth 50~MHz}}}   \\
RXJ0000.1+0816  & UGC12890                  & 00:00:07.09 & 08:16:45.2  &  41.90 &  46.62 & 0.06 &   0.040 & $\surd$ \\
RXJ0021.6+2803  & IV Zw 015 NOTES01         & 00:21:36.82 & 28:03:02.6  &   0.39 &   0.37 & 0.05 &   0.094 & x \\ 
Z235            & 2MASX J00435213+2424213   & 00:43:52.17 & 24:24:21.1  &  27.32 &  30.28 & 0.27 &   0.083 & $\surd$ \\ 
RXJ0058.9+2657  & NGC326/4C+26.03           & 00:58:22.63 & 26:51:58.7  &  11.21 & 558.63 & 0.26 &   0.048 & x \\
\end{longtable}
\twocolumn


\onecolumn
\begin{longtable}[l]{@{}llllrrrrcr@{}}
\caption{ L Band fluxes.} \label{L_BAND} \\
  \hline\hline
        &            & \multicolumn{2}{c}{Position(J2000)} & \multicolumn{3}{c}{Fluxes (mJy)} & & &  \\
   Cluster & BCG & RA & DEC &  Core & Total & RMS & Redshift & Lines? \\
  \hline\hline
\endfirsthead
\multicolumn{1}{l}{{\tablename} \thetable{} -- Continued} \\
  \hline\hline
        &            & \multicolumn{2}{c}{Position(J2000)} & \multicolumn{3}{c}{Fluxes (mJy)} & & &  \\
   Cluster & BCG & RA & DEC &  Core \footnote{Peak if unresolved, peak of core component if resolved.}  & Total & RMS & Redshift\footnote{Crawford {\it{et al}} (1999) unless otherwise stated.} & Lines? \\
  \hline\hline
  \\
\endhead
  \multicolumn{3}{l}{{Continued on Next Page\ldots}} \\
\endfoot
  \\ \hline \hline
\endlastfoot
 \\
 \bf{VLA-C Array}  \\
 \hline
 \\
\multicolumn{2}{l}{\bf{\emph{AE125 : 1997 : Bandwidth 50~MHz }}}   \\
 RXJ0341.3+1524   & III Zw054                & 03:41:17.50 & 15:23:47.0  & 17.51 & 18.14 & 0.35 & 0.029  & x \\  
 A646             & 2MASX J08220955+4705529  & 08:22:09.50 & 47:05:52.7  & 48.68 & 49.86 & 1.10 & 0.1268 & $\surd$ \\  
 Z2844            & NGC 3099                 & 10:02:36.54 & 32:42:24.3  &  9.84 & 10.03 & 0.16 & 0.0502 & x \\
 \\
\end{longtable}
\twocolumn


\onecolumn
\begin{longtable}[l]{@{}llllrrrrcr@{}}
\caption{ X Band fluxes.} \label{X_BAND} \\
  \hline\hline
        &            & \multicolumn{2}{c}{Position(J2000)} & \multicolumn{3}{c}{Fluxes (mJy)} & & &  \\
   Cluster & BCG & RA & DEC &  Core & Total & RMS & Redshift & Lines? \\
  \hline\hline
\endfirsthead
\multicolumn{1}{l}{{\tablename} \thetable{} -- Continued} \\
  \hline\hline
        &            & \multicolumn{2}{c}{Position(J2000)} & \multicolumn{3}{c}{Fluxes (mJy)} & & &  \\
   Cluster & BCG & RA & DEC &  Core   & Total & RMS & Redshift & Lines? \\
  \hline\hline
  \\
\endhead
  \multicolumn{3}{l}{{Continued on Next Page\ldots}} \\
\endfoot
  \\ \hline \hline
\endlastfoot
 \\
 \bf{VLA-C Array}  \\
 \hline
 \\
\multicolumn{2}{l}{\bf{\emph{AE125 : 1997 : Bandwidth 50~MHz }}}   \\
RXJ0341.3+1524    & III Zw054                       & 03:41:17.54 & 15:23:47.6  & 10.82 & 11.09 & 0.09 & 0.029  & x \\
A646              & 2MASX J08220955+4705529         & 08:22:09.54 & 47:05:52.7  & 72.09 & 74.57 & 0.06 & 0.1268 & $\surd$  \\
A621              & 2MASX J08111245+7002284         & 08:11:12.38 & 70:02:30.4  &  5.26 & 14.73 & 0.09 & 0.223  & x \\
Z2844             & NGC 3099                        & 10:02:36.55 & 32:42:24.2  &  6.95 &  7.14 & 0.09 & 0.0502 & x \\
\end{longtable}
\twocolumn

\section[]{SED Decompositions - Main Sample} \label{APPENDIX_DECOMPOSITIONS}

\onecolumn
\begin{longtable}[l]{@{}lllllcll@{}}
\caption{ Flux contribution of the core component at 10~GHz and the non-core component at 1~GHz as inferred from model fitting when fit best using two power law model.  Note that radio SEDs are almost invariably too under sampled to perform a complete, unequivocal spectral breakdown.  These fits were conservatively performed using a simple yet consistent, physically motivated model.  See Section \ref{SED_Fits} for description of the fitting and model.  Also see Appendix \ref{APPENDIX_NOTES} for comments and justification for individual sources.  Fit labels 1: Resolved core component seen in images fit separately.  2: Core component fit separately, inferred non-core flux for lower resolution literature fluxes from SED shape. 3: Two components fit direct from SED using a split power model. 4: Only a single component reliably extractable SED, extrapolate to get limits. 5: Single observed flux or two fluxes with indeterminate index, extrapolate for both limits. (i) When core component considered negligible at 1~GHz hence total flux at 1~GHz taken as non-core flux.  (ii) Core expected to contribute significant fraction of measured flux at 1~GHz, non-core contribution taken as the difference between total fitted flux at 1~GHz and core contribution at this frequency. (*See notes) } \label{Model_Fitted} \\
  \hline\hline 
  Cluster & Non-core (at 1~GHz)  & Core (at 10~GHz)    & Steep Index & Flat Index & Flat          & Fit Label & Lines? \\
          & (mJy)                &  (mJy)              &             &            & normalisation &           &        \\
  \hline\hline
\endfirsthead
\multicolumn{1}{l}{{\tablename} \thetable{} -- Continued} \\
  \hline\hline
  Cluster & Non-core (at 1~GHz) & Core (at 10~GHz)   & Steep Index & Flat Index & Flat          & Fit Label & Lines? \\
          & (mJy)               & (mJy)              &             &            & normalisation &           &        \\
  \hline\hline
  \\
\endhead
  \multicolumn{3}{l}{{Continued on Next Page\ldots}} \\
\endfoot
  \\ \hline \hline
\endlastfoot
 \\
\multicolumn{3}{l}{ \bf{Core and Non-core Values}}  \\  
 \hline
 \\
 A11            & 77.4   $\pm$ 22.7  & 21.7  $\pm$ 10.7 & 0.85 $\pm$ 0.12 & 0.35  $\pm$ 0.22 & 126.2 $\pm$ 7.4    & 2(ii) & $\surd$  \\  
 RXJ0058.9+2657 & 2022.1 $\pm$ 592.1  & 6.5   $\pm$ 3.1  & 0.79 $\pm$ 0.16 & 0.2  $\pm$ 0.2  & 10.35  $\pm$ 3.39  & 1(i)  & x \\
 RXJ0107.4+3227 & 4895.7 $\pm$ 1006.0 & 103.1 $\pm$ 5.4  & 0.74 $\pm$ 0.09 & -0.29 $\pm$ 0.03 & 52.88 $\pm$ 2.78  & 1(i)  & $\surd$  \\  
 A160           & 1002.4 $\pm$ 96.2   & 8.6   $\pm$ 3.4  & 0.89 $\pm$ 0.09 & 0.33 $\pm$ 0.18 & 18.49  $\pm$ 7.32  & 1(i)  & x \\
 RXJ0123.6+3315 & 103.4  $\pm$ 49.8   & 1.27  $\pm$ 0.61 & 1.30 $\pm$ 0.35 & 0.2  $\pm$ 0.2  & 2.02   $\pm$ 0.65  & 1(i)  & x \\
\end{longtable}
\twocolumn

\onecolumn
\begin{longtable}[l]{@{}lcccrrrrrrr@{}}
\caption{ Flux contribution of the core component at 10~GHz and the non-core component at 1~GHz as inferred from model fitting when fit best using a dropline model.  Fit label 1: Clearly separated core and non-core components in map. 2: Inferred non-core flux contribution for lower resolution fluxes by considering SED. } \label{Model_Fitted_dropline} \\
  \hline\hline 
   Cluster & Non-core (at 1~GHz) & Core (at 10~GHz) & Dropline Parameters & Flat Index & Flat Norm. & Fit Label & Lines? \\
           &  (mJy)              &  (mJy)           &                     &            &            &           &        \\    
  \hline\hline
\endfirsthead
 \\
\multicolumn{3}{l}{ \bf{Core and Non-core Values}}  \\
 \hline
 \\

 Z808       & 765.2  $\pm$ 11.8  & 1.3 $\pm$ 0.8  & 1235.2 1.13 1.10 1.06 &  0.2  $\pm$ 0.2 & 2.10 $\pm$ 1.12      & 1     & $\surd$ \\
 Z1261          & 1959.8 $\pm$ 29.7  & 32.7 $\pm$ 8.1  & 1962.1 0.61 0.63 6.28 & -0.23 $\pm$ 0.15 & 19.1 $\pm$ 4.7 & 1(i)  & x \\
 A763           & 213.1  $\pm$ 9.8   & 1.6  $\pm$ 1.2  & 213.1  0.77 0.99 4.35 &  0.10 $\pm$ 0.34 & 2.0  $\pm$ 1.5 & 1(ii) & x \\
 A1361          & 1342.3 $\pm$ 29.0  & 1.9  $\pm$ 1.1  & 1857.9 0.57 0.85 0.75 &  0.2  $\pm$ 0.2  & 3.1  $\pm$ 1.4 & 1(i)  & $\surd$ \\
 RXJ1522.0+0741 & 266.7  $\pm$ 145.8 & 2.1  $\pm$ 1.7  & 2500.9 1.24 1.06 0.17 &  0.2  $\pm$ 0.2  & 3.3  $\pm$ 1.1 & 1(ii) & x \\
\end{longtable}
\twocolumn

\onecolumn
\begin{longtable}[c]{@{}lcccrrrrr@{}}
\caption{ Flux contribution of the core component at 10~GHz and the non-core component at 1~GHz when using the GPS model. Fit label 1: Clearly separated core and non-core components in map. 2: Inferred non-core flux contribution for lower resolution fluxes by considering SED. } \label{Model_Fitted_NEW_GPS} \\
  \hline\hline 
   Cluster & Non-core (at 1~GHz) & Core (at 10~GHz) & GPS Parameters & Steep Index & Fit Label & Lines? \\
           &    (mJy)            &  (mJy)           &                &             &           &        \\
  \hline\hline
\endfirsthead
\multicolumn{1}{l}{{\tablename} \thetable{} -- Continued} \\
  \hline\hline
   Cluster & Non-core (at 1~GHz) & Core (at 10~GHz) & GPS Parameters & Steep Index & Fit Label & Lines? \\
           &    (mJy)            &  (mJy)           &                &             &           &        \\
  \hline\hline
  \\
\endhead
  \multicolumn{3}{l}{{Continued on Next Page\ldots}} \\
\endfoot
  \\ \hline \hline
\endlastfoot
 \\
\multicolumn{3}{l}{ \bf{Core and Non-core Values}}  \\ 
 \hline
 \\
 RXJ0439.0+0520      & 62.3  $\pm$ 10.0 & 250.8 $\pm$ 29.5 & 1.32 1.89 -0.81 & 1.07 $\pm$ 0.10 & 2 & $\surd$ \\
 A1348               & 110.2 $\pm$ 12.0 & 113.5 $\pm$ 11.3 & 1.96 0.62 -0.52 & 0.92 $\pm$ 0.08 & 2 & $\surd$ \\
 RXJ1558.4-1410      & 186.7 $\pm$ 34.0 & 351.2 $\pm$ 6.6  & 2.59 0.86 -0.90 & 0.90 $\pm$ 0.09 & 2 & $\surd$ \\
\multicolumn{3}{l}{ \bf{Core Values with Non-core Upper Limits}}  \\ 
 \hline
 \\
 RXJ1304.3-3031  & $<$3.0   & 12.6  $\pm$ 0.9  & 1.58 -0.03 -0.05   & 1.0             & 2  & $\surd$ \\
 RXJ1315.4-1623  & $<$5.6   & 12.8  $\pm$ 3.8  & 1.56 -0.09 -0.35   & 1.0             & 2  & $\surd$ \\
\end{longtable}
\twocolumn

\section[]{Justification of Fits - Main Sample+ Extension} \label{APPENDIX_MSP}

These clusters do not have targeted observations presented within this study and so do not appear in the flux lists. We therefore give in the notes below the name of the BCG for each cluster.  Fluxes were taken from the NED and HEASARC databases as well as being retrieved from FITS image downloaded from the NRAO's Image Archive (https://archive.nrao.edu/archive/archiveimage.html).

\subsubsection{\bf BCS - LEs}
{\bf Hercules A}  A well studied classic FRII, bright object with esily distinguishable core and lobes.  There are a large number of literature data points of integrated emission available in NED from which to get a measure of the non-core.  Extrapolate from a 14.9~GHz VLA observation \cite[][]{Dicken08} of the core region for an estimate of the core component. This cluster is also known as RXJ1651.1+0459, and the radio source associated with the BCG is often called 3C~348. \\
{\bf A2052} UGC 09799, associated with the radio source 3C~317.  Extended source with clearly distinguishable core.  The source is reasonably well studied (see e.g. Hogan \etal 2015b).  Fit to the SED of integrated fluxes for a non-core measurement.  Source has multi-frequency VLBA observations \cite[][]{Venturi04} that isolate the core.  Fit to these separately to retrieve a measure of the active core component. \\
{\bf A2199} NGC6166, with associated radio source 3C~338.  Extended, FR-II like source.  VLBI measurements at several frequencies \cite[][]{Ly04, Gentile07} show a sub-dominant but inverted core component.  Fit this VLBI SED separately for a measurement of the core component and then fit the integrated flux SED for a measurement of the non-core. \\
{\bf A1795} CGCG 162-010.  Extended source, some high resolution observations in NRAO archive from which the core can be isolated.  Create and fit individual SEDs for the core and extended components to enable measurements of both the core and non-core respectively.  See also Hogan \etal 2015b. \\
{\bf RXJ0338.7+0958} 2MASX J03384056+0958119. Appears to be a compact steep spectrum object, hence fit the integrated spectrum to retrieve an estimate of the non-core component.  An NRAO archival observation at X-band shows it as unresolved.  Extrapolate from this to obtain an upper limit on the core component. \\
{\bf A1367} NGC3862.  Powerful extended radio source with a well sampled SED that can be fit directly to obtain an estimate of the non-core component.  There have been a number of high resolution VLBI observations of the core of this system \cite[][]{Lara04, Kharb09} from which a core-only SED can be created.  Fit this for a measure of the core component. \\

\subsubsection{\bf BCS - NLEs}
{\bf A2029} IC 1101.  A wide-angle tailed (WAT) source with a morphologically distinct but energetically sub-dominant core.  Fit the archival SED for a measure of the non-core component and then extrapolate from a high resolution VLA archival image at X-band of the isolated core \cite[][]{Jetha06} to get a core estimate with appropriate uncertainty. \\
{\bf A1656} NGC4874.  Source appears to be a small-scale WAT.  Integrated emission is lobe-dominated, hence fit the archival SED for a measure of the non-core.  There is no clearly discernible core in the majority of observations however there is a VLBA observation at C-band \cite[][]{Liuzzo10} and higher frequency observations that seem to isolate the core \cite[][]{Lin09} that can be fitted separately to get a reasonable measure of the core. \\
{\bf A2312} 2MASX J18540626+6822565.  Steep spectrum source that appears to be an FRI at 38~MHz \cite[][]{Rottgering94}.  There is no archival data above C-band and only the subtlest hint of potential spectral flattening at C-band but not enough to claim unambiguosly the presence of a strong core component.  Fit the lobe dominated integrated SED for a measurement of the non-core and then extrapolate from C-band \cite[][]{Gregory91} for an upper-limit on the core. \\
{\bf A1314} IC 0712.  Source seems to be a CSS with consistent index to high frequency \cite[43.3~GHz,][]{Lin09}.  Fit the SED for non-core and then extrapolate from the 43.3~GHz point for a core limit. \\
{\bf A2147} UGC 10143. Source appears unresolved in available archival images.  The index fitted to the SED is relatively flat ($\alpha$$\approx$0.67), however the source is undetected at 22.5~GHz \cite[][]{Lin09} and hence cannot be dominated by a flat core and is perhaps best classified a CSS.  Take an estimate of the non-core from the fitted SED and then extrapolate from the limit at 22.5~GHz for an upper limit on the core. \\
{\bf A2065} 2MASS J15222917+2742275. Source only appears to have been detected at L-band.  Marginally resolved by FIRST and there is a relatively large flux decrement between the FIRST and NVSS resolutions hence it seems reasonably to believe this source is dominated by its non-core emission.  Extrapolate with appropriate indices and uncertainties for a non-core measurement and an upper limit on the core component. \\

\subsubsection{\bf eBCS - LEs}
{\bf A2396} 4C +12.76.  This is a strong, lobed source that has good archival coverage at lower resolution but no clear measurements of a distinct core.  Fit the total SED for a measurement of the non-core.  The highest frequency detection is at C-band \cite[][]{Gregory91} although the single-dish resolution is too low to isolate the core.  The highest resolution map is from FIRST, hence take the average of the central pixels in this and extrapolate from these to obtain the best limit on the core contribution. \\
{\bf A2036} 2MASX J15110875+1801529.  This source is a WAT although there is no obviously distinct core component.  Single dish C-band observations would suggest flattening, however this is most likely due to confusion as interferometric observations at C-band are consistent with a steady steep index to X-band detection and a non-detection at 22.5~GHz \cite[all][]{Lin09}. This non-detection shows the spectrum does not substantially flatten.  Use only the higher resolution maps to create an SED and fit this for a measure of the steep non-core component and then extrapolate back from limit at 22.5~GHz for core limit. \\
{\bf A923}  B2 1003+26. The available archival observations show the source to be an unresolved CSS up to C-band.  Higher resolution VLA-A image in VLA archive \cite[see also][]{Fanti86} shows the presence of an active core component at $\sim$0.5 mJy.  There is also an X-band image in the NRAO database taken in VLA-D/A array.  Although the automatic pipeline image suffers from phase issues, blindly taking the peak pixel of the recovered source does suggest a $\sim$0.5 mJy source is present.  Fit the integrated lower resolution SED for a measure of the non-core component and then extrapolate from the VLA-A detection at C-band with index $\alpha$=0.0 to recover an estimate for the core. \\
{\bf A580} 2MASX J07255712+4123069.  Higher resolution maps at C and X-band \cite[][]{Parma07} show the radio-source to be associated with the BCG.  Additionally associating the archival radio source B3 0722+414 with the BCG would suggest that the source could be best fit by a dropline model for the ageing non-core (with parameters: 54.69, 1.26, 0.91, 3.85).  Below 2~GHz the SED displays a reasonably steady power-law and so for the MAIN SAMPLE+ estimate the non-core component by fitting only to these lowr frequency fluxes. The source is potentially slightly resolved in the X-band imaging although the core is not distinct.  Extrapolate from the peak of this observation for a limit on the core component. \\
{\bf Z1883} 2MASX J08425596+2927272.  Archival detections are available only at L and C band, which show the source to be steep spectrum.  Fit for a measure of the non-core and extrapolate for an upper limit on the core. \\
{\bf Z1370} 4C +74.13.  The BCG is associated with an unresolved, ultra-steep spectrum radio source.  There are X and U-band observations available in the VLA Image database although neither recovers the source and both limits are consistent with the extrapolated index.  Fit the observations of the USS $>$0.074~GHz to retrieve an estimate of teh non-core and then extrapolate from the limit imposed by the non-detection at U-band to put an upper-limit on the core component.  Note that there are very low frequency (38 and 22~MHz) observations that suggest the USS index may flatten to lower frequencies. \\
{\bf Z3959}  SDSS J111551.90+012955.  Distant cluster (z$\approx$0.35), at FIRST resolution the BCG is consistent with being associated with a steep spectrum radio source.  See that there is a large flux deficit between FIRST and NVSS suggesting that this source is resolved between the respective resolutions of these catalogues. Combined these imply that the source is dominated by its non-core emission.  Therefore fit the spectrum for an estimate of this component and then and extrapolate from the peak of the FIRST detection to get an upper limit on the core. \\
{\bf A853} 2MASX J09421480+1522509.  The radio source is only apparently detected by the FIRST and NVSS surveys, although the FIRST resolution shows that this is associated with the BCG. The source is unresolved in NVSS but is resolved into a small WAT in FIRST showing that it is non-core dominated.  This interpretation is consistent with the limits of other surveys \cite[such as GB6,][]{Condon94} showing that the index must be steep, and that the spectrum does not flatten to show a flat core component.  Extrapolate from the NVSS integrated emission for an estimate of the non-core and then extrapolate from the peak of the FIRST detection with index $\alpha$$_{core}$=0.2 for an upper limit on the core. \\
{\bf RXJ0004.9+1142} UGC 00032.  SED can be fit with a split power law showing the presence of an active core in addition to a steeper component.  Fit the SED therefore to obtain measures of the core and non-core components from the flat and steep spectrum components respectively. \\

\subsubsection{\bf eBCS - NLEs}
{\bf RXJ1844.1+4533} 3C~388.  A bright FRII with a morphologically distinct but energetically sub-dominant core component.  Fit the integrated SED to recover a measure of the non-core.  Retrieve images from the NRAO image archive at L-, C-, X- and U-bands to measure distinct core fluxes.  Use these to create a core only SED from which to determine the active core component. \\
{\bf A1446} 4C +58.23.  A bright WAT source.  Fit the integrated fluxes for a measurement of the non-core component and then fit a separate SED of the isolated core fluxes \cite[e.g.][]{Laurent-Muehleisen97} separately for a core measurement. \\
{\bf RXJ2250.0+1137} NGC 7385.  The integrated SED returns a fairly flat index ($\alpha$$\approx$0.53).  However, morphologically the source is clearly dominated by its extended emission.  Fit the integrated spectrum therefore to obtain a measure of the non-core component.  Additionally there does appear to be an active core component.  From images retrieved from the NRAO Image Archive, a distinguised core is detected at $\sim$118 mJy at both X and U band.  The is a VLBA observation by \cite{Fomalont00}. No map is presented as the source is too structured to recover reasonably.  The authors present the uv data and from the amplitudes it is clear that $\sim$50 mJy is recovered on even the longest baselines.  It appears as if there may be a currently active component or perhaps that this source was very recently shut down.  For a core estimate, fit the unresolved core components at X and U band with suitably large uncertainties. \\
{\bf A278} B2 0154+32.  BCG is associated with a head-tail source.  Fit the integrated flux SED for a measure of the non-core. Archival VLA-B observations at L and C band recover structure but do not fully isolate core component.  For an estimate of the core contribution at 10~GHz, extrapolate from the `head' component on the optical BCG position in the highest resolution map in the VLA Image Archive (at L-band).  The recovered estimate is consistent with the limits imposed by low resolution observations at wavelengths higher than L-band.  Note than an alternative interpretation of this source could be that it is an under-resolved FRII and hence could be a background contaminant rather than associated with the BCG. \\
{\bf A2625} 2MASX J23360272+2032455. Steep spectrum source that is unresolved in NVSS.  There are few archival observations, the highest radio-frequency of which is at C-band where the spectrum is still steep.  Source appears to be a CSS. Fit the spectrum to get an estimate of the non-core component and then extrapolate from the C-band detection \cite[][]{Gregory91} with $\alpha$$_{core}$=0.2 for a core limit. \\
{\bf A2445} 2MASX J22265584+2550094. There are three NVSS \cite[][]{Condon98} sources coincident with the cluster, the brightest of which appears to be associated with the BCG. There are no higher resolution observations in the VLA archive and the source is not within region covered by FIRST.  The source does not appear within the VLSS catalogue but a steep extrapolation from the NVSS detection suggests it should be just below the catalogure limit.  Indeed, an overlay of the VLSS \cite[][]{Cohen07} shows that there is a weak detection below the catalogue limit at the position of the BCG.  This VLSS source is partially blended with the distinct NVSS source to the West.  Take the VLSS peak at the position of the BCG, which along with NVSS and GB6 \cite[][]{Condon94} detections gives a steep index. Classify this source therefore as a CSS and fit for an estimate of the non-core.  There are no higher resolution or higher frequency observations avaliable, so extrapolate from the C-band flux to get a limit on the core. \\
{\bf A2149}  SBS 1600+540A. This cluster contain three potential BCGs.  The most dominant elliptical (16:01:28.14, 53:56:49.7) appears to be associated with a radio source.  This is a steep spectrum source, which can be fitted for measure of non-core. The source is morphologically identified as head-tail object in FIRST imaging although the head still appears confused and thus does not represent a truly naked core.  The source was observed by \cite{Laurent-Muehleisen97} and found to have a core component at 5~GHz.  Extrapolate from this with reasonable uncertainties for an estimate of the non-core.  Other candidate BCGs at $\sim$16:01:28.43, 53:54:15.0 and $\sim$16:01:23.14, 53:51:59.8 are both radio undetected at the FIRST detection limit. \\
{\bf RXJ0027.6+2616} 2MASX J00274579+2616264.  Source is steep spectrum and has no clear core component.  It is undetected in the GB6 catalogue \cite[][]{Condon94} hence see no flat component.  Therefore fit for a non-core estimate and extrapolate from the GB6 limit to put an upper limit on the core component. \\
{\bf RXJ1652.6+4011} 2MASX J16525325+4009130.  BCG is associated with a radio source as matched in both FIRST and NVSS.  It does not appear in the VLSS catalogue \cite[][]{Cohen07} but inspecting an overlay of the region reveals a weak detection at the BCG position.  The derived index suggests that the source could be core-dominated but it is undetected in the GB6 catalogue \cite[][]{Condon94} and similarly undetected with the VLBA \cite[$<$3 mJy,][]{Bourda10}, which shows this source to be non-core dominated.  The index derived using the map detection in the VLSS imaging may therefore be an underestimate.  Fit this spectrum nonetheless for an estimate of the non-core componant and then extrapolate from the VLBA detection limit for a fairly stringent upper limit on the core component. \\
{\bf A1291} MCG +09-19-110. The BCG is revealed to be associated with a small WAT in FIRST \cite[][]{White97}. \cite{Lin09} observed this source at multiple frequencies and showed it to be steep spectrum.  Fit the SED for a non-core measurement then extrapolate from the X-band detection for a core limit.  This is consistent with less stringent limits imposed by extrapolating from non-detections at 22.5 \& 44.3~GHz. \\
{\bf RXJ1852.1+5711} 2MASX J18520859+5711430. Source is resolved but displays amorphous structure. The SED could be fitted with a dropline model and indicates that the source is dominated by ageing emission, perhaps a core relic. Fit the SED for a measurement of the non-core component.  Highest resolution available in the NRAO Image Archive is at C-band.  There is no distinguishable core so take the peak of the emission and extrapolate for an upper-limit on the core. \\
{\bf A2512a} UGC 10187.  Head-tail source. Integrated fluxes to lower frequency give a steep spectrum, which appears to flatten to higher frequency suggestive of increasing core dominance.  Taking the peak fluxes of the nuclear region it appears as if an active core component is present but appears to be confused with the more extended emission in most of the available observations.  Extrapolate back from the seemingly clean core observation at 22.5~GHz \cite{Lin09} for the best estimate of the core component. \\
{\bf A1986} 2MASX J14530832+2153396.  This cluster appears to contain several large ellipticals. The largest of these (and hence the most likely candidate for BCG) is associated with a steep spectrum radio source.  The source is undetected in all of the major C-band catalogues, which confirms this steep spectrum.  Fit the SED for a non-core measurement and then extrapolate from the C-band limits for an upper limit on the non-core component. \\
{\bf A2315} 2MASX J19001653+6956599. BCG appears to be associated with a steep spectrum radio source.  Fit the SED for an estimate of the non-core and then extrapolate from the detection limits at C-band \cite[e.g.][]{Condon94} for a limit on core component. \\

\subsubsection{\bf REFLEX(NVSS) - LEs}
{\bf Hydra A} Little introduction required.  A powerful, extended source that is very well-studied in the literature \cite[e.g.][and references therein]{McNamara00, Hamer14}.  The source is dominated by its extended emission so fit the SED for a measure of the non-core component. Use VLBI observations \cite[e.g.][]{Taylor96, Araya10} of the central core regions only to independently fit for the core component. CLuster is also known as A780 and RXJ0918.1-1205. \\
{\bf A2597} The BCG is associated with the powerful extended radio source PKS 2322-12. The archival SED is dominated by the extended emission hence fit this for a measure of the non-core.  NRAO archival images at X-band with the VLA-A array allow the core to be isolated. The flux of this component agrees reasonably well with the flux recovered on milli-arcsecond scales with the VLBA \cite[][]{Taylor99}.  Use a representative index of $\alpha$$_{core}$=0.2 to extrapolate from the VLBA detection for an estimate of thr core component at 10~GHz. \\
{\bf A3532} PKS1254-30. This is a large extended source, with the lobes looking disturbed.  Clear that the archival SED is dominated by the extended emission hence fit to this to get a measure of the non-core component. There are no available observations at high resolution or frequencies greater than 4.85~GHz.  Extrapolate from the peak of an archival VLA C-band image to obtain an upper limit on the core. \\
{\bf RXJ0132.6-0804} PKS 0130-083. The SED of this source has a flat spectrum and appears to be highly core dominated.  VLBI observations \cite[][]{Bourda10, Bourda11, Petrov11} show it to be slightly extended on $\sim$few parsec scales.  A separate SED for the core can be produced from the VLBI imaging, from which the core component can be estimated.  Extrapolating from the WENSS \cite[][]{Rengelink97} detection (using $\alpha$$_{steep}$=1.00) agrees well with the missing flux at 1~GHz between the VLBI obtained core contribution and the flux recovered in lower resolution maps. Hence, use this extrapolation to infer an estimate for the non-core component. \\
{\bf A2415} 2MASX J22053865-0535330.  This source has a steep spectrum source at low frequencies ($<$1~GHz).  At higher frequencies than this it appears to be dominated by a self-absorbed core component (Hogan \etal 2015b).  VLBA imaging shows extent on $\sim$few pc scales (see Hogan \etal 2015c) the integrated emission of which is consistent with lower resolution fluxes, confirming the source to be core dominated at higher frequencies.  For the purposes of the MS+, fit two separate power laws to the steep and flat components in the SED to obtain estimates of the non-core and core component respectively.  Note that the given indices are measured locally to the normalisation frequencies for each component. \\
{\bf RXJ1206.2-0848} Also known as MACS J1206.2-0847, this is a high redshift cluster (z$\approx$0.44), whose BCG is identified in NED only as as `MACS J1206.2-0847 BCG'.  This is associated with a steep spectrum radio source.  It is undetected in the PMN \cite[][]{Griffith93} and AT20g \cite[][]{Murphy10} surveys showing that there is no significant inverted/flat component.  Fit the SED for a measure of the core component and then extrapolate from the higher frequency non-detections to get a limit on the core contribution. \\
{\bf RXJ0501.6+0110} 2MASX J05012816+0110357. The relatively sparsely sampled SED appears to have a relatively flat index ($\alpha$$\approx$0.7) but is undetected at GB6 \cite[][]{Condon94} and above which suggests the spectrum steepends to higher frequency.  The SED can therefore be attributed to dominant non-core emission.  Extrapolate from the non-detections for an upper-limit on the core component. \\
{\bf A1644} 2MASX J12571157-1724344. Flat spectrum source that appears to be variable considering multi-epoch C-band observations. Not a clear GPS-like spectrum with the available archival coverage but can fit the spectrum with a flat index power law to obtain a measurement of the core component.  There is a faint detection at the BCG position in WENSS that is just below the catalogue limit \cite[][]{Rengelink97}, suggesting that the spectrum turns over in the GHz range.  Extrapolate from this to get a limit on the non-core component. \\
{\bf RXJ0445.1-1551} NGC1650. Appears to be a core dominated flat spectrum radio source.  Fit the SED for a measurement of the core component and then extrapolate from the WENSS detection \cite[][]{Rengelink97} for an upper limit on the non-core. \\
{\bf A85} MCG -02-02-086. The BCG is associated with a CSS source.  VLA images from the NRAO Image Archive at X, U and K bands show that the source remains unresolved and steep out to above 20~GHz.  Fit the SED for a measure of the non-core and then extrapolate from the 22.5~GHz flux to place an upper limit on the core component. \\
{\bf RXJ1347.5-1144} GALEX J134730.7-114509.  High redshift (z$\approx$0.45) cluster with a powerful radio source associated with the BCG.  \cite{Gitti07a} report the presence of a radio mini-halo within this cluster.  Archival, higher resolution (VLA-A) observations at C- and X-band show that there is a morphologically distinguishable core component associated directly with the BCG.  Interpolating between these core-only fluxes returns a flat spectrum, that can be extrapolated for an estimate of the core component at 10~GHz.  The SED steepens to lower frequencies although this is likely contaminated by the mini-halo emission.  Fit the fluxes at frequencies below 1~GHz for an estimate of the non-core component, accounting for the expected core contribution at 1~GHz. Note that this is potentially affected by mini-halo emission.  \\
{\bf A383} 2MASX J02480342-0331447.  The BCG is associated with a radio source that maintains a steady steep spectral index at least as high as 22.5~GHz \cite[BIMA,][]{Coble07}.  Fit for a measure of the non-core and then extrapolate from a non-detection at 31~GHz \cite[OVRO,][]{Mason09} for a core limit. \\
{\bf A1663} 2MASX J13025254-0230590. The BCG appears associated with an NVSS source that is also matched to a source detected at 1.28~GHz with the GMRT \cite[][]{Giacintucci07}.  We are unable to determine structure from these observations.  The source is undetected at both PMN \cite[][]{Griffith93} and AT20g \cite[][]{Murphy10}, however the limits do not rule out the presence of a flat component. Equally the limit from the MRC \cite[][]{Large81} does not rule out a steep spectrum.  The emission cannot therefore strongly be claimed to be attributable to either core or non-core dominated emission and so extrapolate with appropriate $\alpha$$_{core}$=0.2 and $\alpha$$_{non-core}$=1.0 indices for limits on both components. \\
{\bf A3698} NGC6936.  BCG optical position coincident with NVSS, however no other radio detections are available.  Cannot attribute to steep or flat spectrum source by limits at other frequencies hence extrapolate for limits on both components. \\
{\bf RXJ1050.4-1250} USGCS152/NGC3402.  The BCG of this small group appears to be associated with a reasonably active radio source.  The source is well sampled at L-band in the archives, although it is not clearly resolved.  There is a large flux decrement between NVSS and the higher resolution observations indicative potential extended emission on intermediate scales. \cite{O'Sullivan07} observed this source at L-, C- and X-bands with the VLA and showed that it contains a flat spectrum core component.  Fit for this then extrapolate from the NVSS flux for an estimate non-core emission, accounting for the expected core contribution at 1~GHz. \\
{\bf A2402} 2MASX J21582888-0947489. The BCG is optically matched to an NVSS radio source, however there are no additional radio detections.  Limits at other frequencies do not limit the spectrum to be unambiguously associated with a flat core or a steep non-core and hence the best approach is to extrapolate from the NVSS detection with relevant indices for limits on both the core and non-core components. \\
{\bf A2566} 2MASX J23160519-2027528. The BCG is matched to a radio source in NVSS.  There are no other radio detections and survey limits do not clearly distinguish this source to be either core or non-core dominated hence use relevant indices to extrapolate from the NVSS detection to place a limit on both the core and non-core components. \\
{\bf RXJ1506.4+0136}  NGC5846.  This is the brightest member of a relatively local group (z$\approx$0.006) and is a well studied LINER.  The radio SED has a flat index, appears to be variable and is detected on milli-arcsecond scales with the VLBA \cite[][]{Nagar05}, confirming it to be core dominated.  Use the most contemporaneous observations and fit for an estimate of the core component.  Extrapolate from the lower frequency detection at 750~MHz \cite[][]{Heeschen64}, accounting for the expected 1~GHz core contribution, for a limit on any potential non-core component in the system. \\
{\bf RXJ2147.0-1019} 2MASX J21470043-1019118. BCG is associated with a radio source in NVSS (at 1.4GHz), however there are no additional radio detections and limits at other radio frequencies are not limiting on the nature (core or non-core) of the emission.  Therefore make extrapolations from the NVSS detection, using the approriate indices to obtain upper limits on both the core and non-core components. \\
{\bf S0301} IC1860.  The BCG is associated with a radio source that is detected in both SUMSS \cite[][]{Mauch03} and NVSS \cite[][]{Condon98}. However, the index between these detections is not sufficiently steep or flat to unambiguously label the source as core dominated (or otherwise) and hence the only reasonable approach is to extrapolate for limits on both components using appropriate indices. \\
{\bf A281} 2MASX J01570689-0553102.  The BCG is associated with an NVSS source, although it has no other radio detections available.  Survey limits at other radio frequencies do not elucidate whether this emission can be attributed definitely to either core or non-core emission.  Therefore extrapolations with appropriate indices are made from the NVSS flux to place upper limits on both components. \\
{\bf RXJ1501.1+0141} NGC5813. This source was studied by \cite{Nagar05} as part of a survey of nearby LINERs and is also well sampled in the NRAO Image Archive.  There is a flat spectrum component that persists to above 10~GHz that can be fitted to estimate a core component at 10~GHz.  The source appears slightly resolved at L-band, showing that there should be a sizeable non-core contribution at lower frequencies. Extrapolate from L-band, accounting for the expected core contribution to obtain an estimate of the non-core component at 1~GHz. \\

\subsubsection{\bf REFLEX(NVSS) - NLEs}
{\bf A3695} 2MASX J20344470-3549019. BCG associated with a WAT radio source.  The archival SED is well sampled and appears to be dominated by the extended emission, hence perform a fit to this to obtain a measure of the non-core component.  Higher resolution VLA observations at X, K and Q-band are available in the NRAO Image Archive that allow the core to be isolated.  Fitting to the core-only SED derived from these maps returns a relatively steep index ($\alpha$$\approx$0.7). However this core is distinctly resolved and hence this can be used to confidently estimate the core component. \\
{\bf A3560} PKS1329-328. This system appears in the optical to perhaps be a merging BCG, with two clear peaks visible.  There is a large WAT radio source centred on brighter of the optical peaks.  The archival SED is well sampled and dominated by the extended emission and so fit to this to derive the non-core component.  In the NRAO Image Archive there are higher resolution images available at X-band (hybrid VLA C/B-array) and C-band (hybrid VLA B/A-array) that allow measures of the core-only flux.  These suggest that there is a sub-dominant, inverted spectrum core component.  Extrapolate from the fit to these for an estimate of this core component at 10~GHz. \\
{\bf A3528S} ESO 443- G 007. The BCG is associated with a large extended WAT radio source.  The archival SED appears to be dominated by the extended emission and can therefore be fitted directly for an estimate of the non-core component. n-core from.  Several observations of this source appear in the NRAO Image Archive, although most do not appear to clearly isolate the core component.  Obtain a best estimate of the core component by extracting a core-only region flux measurement at X-band (from a hybrid VLA B/C-array observation) and extrapolate from this for the expected core contribution at 10~GHz. \\
{\bf A3165} PKS0344-291.  The BCG appears to be associated with an extended, steep spectrum radio source, the SED of which can be fitted to obtain a measure of the non-core component. The best estimate of the core component is obtained using an X-band VLA map from the NRAO Image Archive that allows the core to be isolated.  Measure the core-only flux from this and then extrapolate with $\alpha$$_{flat}$=0.2 for an estimate of the core component at 10~GHz. \\
{\bf A2717} ESO 349- G 022. The BCG is associated with a powerful WAT radio source.  Supplementing the archival SED with fluxes measured from maps retrieved from the NRAO IMage Archive, the non-core component can be measured by fitting the integrated SED.  The highest frequency observation available is at C-band and the core is not isolated.  Therefore extrapolate from the peak flux at C-band for an upper limit on the core component. \\
{\bf A2399} 2MASX J21573344-0747393. The BCG is associated with a powerful WAT.  Retrieving images from the NRAO Image Archive allows the core to be isolated at several frequencies above C-band and so a core-only SED can be produced.  Fit this for a measure of the core component and then fit the integrated flux SED at lower frequencies to determine the non-core component at 1~GHz.  Note that the non-core has a relatively flat index $\alpha$$\approx$0.62) but is clearly morphologically differentiated and so the estimate should be robust. \\
{\bf A3490} PKS1142-341. This source has a reasonably steep SED and appears to be marginally resolved at C-band in the NRAO Image Archive.  It therefore seems that the SED is dominated by the non-core, and so fit directly for an estimate of this component. The source is undetected by AT20g \cite[][]{Murphy10} so showing it not to contain a strong core component to higher frequencies. Extrapolate from this non-detection for an upper limit on the core component. \\
{\bf S753} NGC5419. This is a peculiar source, potentially host to an amorphous or mini-halo radio source.  Higher resolution imaging from the NRAO Image Archive with the VLA-A array (at L-band) and the VLA-B array (at C-band) show the system to have a morphologically distinguished point source associated with the BCG.  This is also distinguishable in the imaging from the SUMSS survey \cite[][]{Mauch03}.  The archival images show that there is a low surface brightness, amorphous source of $\sim$7kpc extent centred on this point source.  There is no larger scale extent seen in archival VLA-D images which suggests that this cluster does not host a true, powerful mini-halo on $\sim$few hundred kpc scales however the lower resolution data does have higher flux than is resolved in the higher resolution images, which may suggest the presence of very large scale extent.  Alternatively it could be that the source is confused with other nearby galaxies.  Fit the integrated SED for a measure of the non-core and then create a core-only SED from which to estimate the core component.  Note that rather than a core to non-core system, an alternative interpretation of this system is that it may be a CSS dominated BCG sat in the centre of a cluster-centric amorphous halo. \\
{\bf A3528N} ESO 443- G 004. This is the fainter of the two BCGs (see A3528S above), it is however also associated with a reasonably bright radio source.  Excising fluxes that appear to have the two radio source confused, the SED for this source returns a steep index from which the non-core component can be measured.  The source is resolved in an X-band image (with the hybrid VLA-C/B array) from the NRAO Image Archive although the core is not clearly isolated and so a robust measurement of this is unavailable.  Instead, extrapolate from the peak flux at X-band to get an upper limit on the core component. \\
{\bf A1791} 2MASX J13485423-2527242. The BCG is associated with a WAT radio source.  The archival SED is dominated by the extended emission and hence can be fitted to obtain a measure of the non-core component.  The core cannot be clearly isolated in any of the available imaging from the NRAO Image Archive. A VLA-C array observation at C-band provides the highest available resolution although the core is not isolated.  Extrapolate from the peak flux at C-band to put an upper limit on the core component at 10~GHz. \\
{\bf RXJ2043.2-2144} 2MASX J20431459-2144343. The BCG is associated with a compact steep spectrum radio source.  Fit for a non-core measure and then extrapolate for a limit on the core component. \\
{\bf A3376} 2MASX J06020973-3956597.  The BCG is associated with a WAT radio source.  Fit the extended emission dominated SED for a measure of the non-core component and then extrapolate from the C-band peak (from NRAO Image Archive) for an upper limit on the core. \\
{\bf A2984} ESO 298- G 017.  The BCG is associated with a CSS.  Fit the SED for a measure of the non-core component and then extrapolate from the AT20g limit \cite[][]{Murphy10} for an upper limit on the core component. \\
{\bf RXJ1332.9-2519} PMN J1332-2519.  This is an extended source with clear lobes seen at both NVSS \cite[][]{Condon98} and TGSS resolutions.  Fit the SED for a measure of the non-core component and then extrapolate from the AT20g limit \cite[][]{Murphy10} to put an upper limit on the core component. \\
{\bf RXJ0359.1-0320} PMNJ0359-0320.  The BCG is associated with a lobe dominated FRI.  Fit the SED for a measure of the non-core and then extrapolate from the AT20g limit \cite[][]{Murphy10} for an upper limit on the core. \\
{\bf A3856} 2MASX J22183938-3854018.  The BCG is matched to a radio source, however this is only detected in NVSS and SUMSS.  The source is unresolved at these, although the index is steep enough $\alpha$$\approx$1.36) to classify this source as a non-core dominated CSS.  Fit the for a measure of the non-core and then extrapolate from the AT20g limit \cite[][]{Murphy10} for an upper limit on the core component. \\
{\bf RXJ0137.2-0912} MCG -02-05-020.  The BCG is associated with an extended radio source that appears to be dominated by a core relic.  This inner relic appears to have been associated with the BCG in most historical observations and hence the flux of this is included as part of our definition for the non-core.  Fit the integrated SED for a measure of the non-core.  Highest frequency flux available is at C-band \cite[][]{Griffith95}, however there is no image available.  In the snapshot image from the FIRST catalogue \cite[][]{White97} the core is clearly isolated.  For an estimate of the core component extrapolate from the L-band core-only flux with generous uncertainty. \\
{\bf A2355} PMN J2135+0126. The BCG is associated with a steep spectrum radio source.  Fit the archival SED for a non-core measurement.  \cite{Laurent-Muehleisen97} obtain a core only flux measurement at C-band using the VLA-A array.  Extrapolate from this measurement to get an estimate fo the core component at 10~GHz. \\
{\bf A499}  PMN J0437-2027.  This is potentially a head-tail or lobed source, although note that it could be two coincidentally aligned and confused sources. Fit the archival steep SED to get a non-core estimate. There is a slight spectral flattening at C-band although this is a single dish \cite[][]{Griffith94} and non detection at AT20g suggests that there is no strong core component, hence the single dish observation may be partially confused.  Extrapolate for an upper limit on the core component. \\
{\bf RBS0540} MCXC J0425.8-0833.  The BCG is associated with a small double radio source and was studied by \cite{Belsole05}.  Can reasonably claim the SED to be non-core dominated hence fit for a measure of this component.  Extrapolate from the peak flux at C-band to get an upper limit on the core. This cluster is also known as RXJ0425.8-0833. \\
{\bf RXJ1301.6-0650}  MCXC J1301.6-0650.  The BCG is associated with a small, barely resolved double source detected in FIRST.  Fit the archival SED for a measure of the non-core.  Single dish observations at C-band \cite[][]{Griffith95} do not resolve the core and hence the best limit on this component is obtained by extrapolating from the peak of the FIRST emission. \\
{\bf A3027} NVSS J023049-330619.  This source is detected only in the NVSS \cite[][]{Condon98}, SUMSS \cite[][]{Mauch03} and TEXAS \cite[][]{Douglas96} surveys and would be expected to be below the detection limit of other surveys.  Fitting to these three detections only returns an intermediate index of $\alpha$=0.63.  The source is unresolved and undetected in both Parkes \cite[][]{Griffith93, Griffith94, Griffith95} and AT20g \cite[][]{Murphy10} however the detection limits do not restrict the source to being unambiguously core or non-core dominated.  Interpolate for a limit on the non-core component and then extrapolate from the AT20g limit for an upper limit on the core component. \\
{\bf A2331} 2MASX J20581214-0745346. The BCG is associated with a CSS radio source, with no detections above L-band.  Fit the archival SED for a measure of the non-core. The SED appears to flatten out to lower frequencies \cite[e.g. VLSS,][]{Cohen07}, which can be attribuated to being a a low frequency effect rather than evidence of a low peaking core as this source is undetected in both the AT20g \cite[][]{Murphy10} and also PMN \cite[equatorial region,][]{Griffith94} surveys.  Extrapolate from these to obtain an upper limit on the core component.  Note that if the SED was better sampled then perhaps this source would be best fit with a dropline model, hence the reported index here may be underestimating true spectral shape. \\
{\bf S617} ESO 565- G 030.  This cluster could be argued to be a double BCG system with the also radio loud ABELL S0617:[GMT93].  However, ESO 565- G 030 appears to be the dominant galaxy and is matched with a steep spectrum radio source.  Fit the SED for a measurement of the non-core component and then extrapolate from the AT20g limit \cite[][]{Murphy10} for an upper limit on the core. \\
{\bf A3764} 6dF J2125455-344352. The BCG is matched to a radio source that appears in both NVSS \cite[][]{Condon98} and SUMSS \cite[][]{Mauch03} although there appear to be no other radio detections.  Nonetheless, inerpolating between these two detections show the source to be steep enough ($\alpha$$\approx$1.09) to claim that the system is not core dominated.  Use this interpolation therefore to estimate the non-core component and then extrapolate from the non-detections with Parkes \cite[e.g.][]{Griffith93} for an upper limit on the core. \\
{\bf RXJ0340.1-1835} NGC1407. The BCG is associated with a radio source that displays a steep spectrum to C-band \cite[e.g.][]{Disney77}.  There are no detections at higher raido frequencies so fit the archival SED for a measure of the non-core and then extrapolate from C-band for an upper limit on the core component. \\
{\bf RXJ2034.9-2143} 2MASX J20345447-2144019.  The BCG is matched to an NVSS source \cite[][]{Condon98}.  The only other cross-matched radio detection is with PMN J2034-2143 \cite[][]{Griffith94}.  Simple interpolation between these two detections would give a flat ($\alpha$$\approx$0.27) index indicative of a core-dominated source, however the source does not appear in the AT20g catalogue \cite[][]{Murphy10} as would be expected were this index to be persistent although the source could have a GPS-like spectrum.  However, there is another reasonably bright NVSS source within the Parkes resolution limit and hence it is highly likely that these two sources are confused in the 5~GHz map. There is too much doubt therefore to unambiguously claim this to be a core dominated system but likewise not enough data to disprove this, hence only quite high upper limits can be claimed on each component with any degree of certainty. \\
{\bf A2389} 2MASX J21541313-0359527. The BCG is matched to a steep spectrum radio source that is unresolved at NVSS resolution \cite[][]{Condon98}.  Fit the SED for a measure of the non-core and then extrapolate from the C-band detection \cite[highest frequency detected at,][]{Griffith95} for an upper limit on the core. \\
{\bf RXJ1254.7-1526} 2MASX J12544491-1526228. The BCG is matched to a steep spectrum, head-tail radio source.  The steep index of the archival SED persists down to low frequency \cite[e.g. 74~MHz with VLSS,][]{Cohen07}. The source is undetected in the PMN surveys \cite[][]{Griffith93, Griffith94, Griffith95} and AT20g \cite[][]{Murphy10} so fit the SED for a measure of the non-core and then extrapolate for an upper limit on the core. \\
{\bf A907} 2MASX J09582201-1103500.  The BCG is optically matched to a steep spectrum radio source.  The archival SED remains steep to low frequency and the source is undertected in AT20g \cite[][]{Murphy10} and C-band surveys \cite[e.g.][]{Griffith93} showing that the spectrum does not flatten to higher frequency.  Fit the SED for a measure of the non-core and then extrapolate for an upper limit on the core. \\
{\bf RBS1712} 2MASX J21020985-2432011. This is a moderate redshift (z$\approx$0.2) cluster whose BCG is matched to a flat spectrum radio source ($\alpha$$\approx$0.34), that appears to be core dominated.  Fit the archival SED for an estimate of the core component.  The best limit on the non-core contribution is obtained by extrapolating from the NVSS \cite[][]{Condon98} detection. \\
{\bf RXJ1655.9-0113} 2MASX J16555506-0112315. The BCG is radio-loud and matched to an unresolved NVSS source \cite[][]{Condon98}.  There appear to be no archival observations of this source, however it does appear within the (flux corrected) primary beam of C-band (VLA-C array and VLA-B array) observations of another nearby source that are in the NRAO Image Archive.  From the higher resolution VLA-B array image it is apparent that the source has a resolved miniature FRI morphology and hence its SED can be expected to be lobe dominated.  Fit the archival SED for a measure of the non-core and then extrapolate from the resolved core component for an estimate of the core component. \\
{\bf RXJ1139.4-3327}  6dF J1139235-332648.  The BCG appears to be associated with an unresolved steep spectrum source.  Note however that the centroid of the radio source is slightly off the optical position and hence could perhaps be a background source.  Fit the SED for a measurement of the non-core component and then extrapolate from the NVSS detection \cite[][]{Condon98} for the best upper limit on the core. \\
{\bf A2384B} ESO 600- G 014.  The BCG is associated with an NVSS radio source \cite[][]{Condon98} although there are no other radio-observations attributed to it in NED.  PKS 2149-20 is nearby ($\sim$16') and bright ($\sim$1.7Jy at 1.4~GHz in NVSS).  This source is well sampled in the VLA-archive, however ESO 600- G 014 lies just outside the C-band VLA-D primary beam and so unfortunately these observations cannot be used to constrain the BCG further.  Upper limits from non-detections in the major surveys do not restricit the source to be unambiguously steep or flat and so the conservative approach is to extrapolate using the relevant indices for upper limits on both the core and non-core components. \\
{\bf A3694} 2MASX J20344138-3403569. The BCG is matched to a radio source that appears in both the NVSS \cite[][]{Condon98} and SUMSS \cite[][]{Mauch03} catalogues.  There are no other radio detections but the interpolation between these points is steep enough to claim that the source is non-core dominated so take a measurement of this and then extrapolate for an upper limit on the core. \\
{\bf A295} UGC 01525.  The BCG is matched to a radio source, although this is detected only at 1.4~GHz in NVSS \cite[][]{Condon98} and FIRST \cite[][]{White97}.  The source is unresolved in NVSS but appears to have miniature FRII-like morphology in FIRST.  Extrapolate from the NVSS detection for an estimate of the non-core component and then extrapolate from a core-only flux obtained from the FIRST image for an estimate of the core component at 10~GHz.. \\
{\bf A1391} LCRS B114712.7-120204.  There appears to be a distinct NVSS source \cite[][]{Condon98} aligned with the BCG.  However, this could potentially be associated with the nearby bright source PMN J1149-1218, particularly if PMN J1149-1218 is a head-tail (or heavily antisymmetric FRII), which would require higher resolution data to determine. If the BCG is radio-loud in its own right then this is not matched to anything other NVSS so the best approach is to extrapolate from the ambiguous NVSS detection to obtains upper limits on both the core and non-core components, using appropriate indices. \\
{\bf A3934} 2MASX J22533252-3343057. The BCG is associated with a steep spectrum radio source.  Fit the SED for a measure of the non-core component and then extrapolate for an upper limit on the core. \\ 
{\bf A2401} 2MASX J21582246-2006145. The BCG is matched to an unresolved NVSS source \cite[][]{Condon98} however no further radio detections are available.  The nature of this radio emission is therefore ambiguos so extrapolate from the NVSS flux for limits on both components. \\
{\bf A3570} ESO-LV 3250191.  This system has two bright ellipticals, this galaxy appears to be dominant.  The source is marginally resolved in NVSS \cite[][]{Condon98}.  The archival SED has a steep spectrum although it is not sampled above L-band. In the NRAO Image Archive there is a map at L-band using the VLA-B array of a nearby source that has ESO-LV 3250191 within the field and shows it to be marginally resolved.  This therefore all points towards the SED being non-core dominated.  Fit the steep spectrum for a measure of the non-core component and then extrapolate from the resolved core peak at L-band for an estimate of the core component. \\
{\bf A1300} 2MASX J11315413-1955391. This is a moderately high redshift cluster (z$\sim$0.3) with multiple radio-loud constituents.  The BCG is matched to NVSS \cite[][]{Condon98} and VLSS \cite[][]{Cohen07} but there are no other radio detections.  INterpolating between these gives a steep index so use this to obtain a measurement of the non-core component (however note that the source is potentially partially confused in VLSS and so this index could be unrepresentatively steep).  There are no high radio frequency or high resolution observations available, hence extrapolate from the NVSS detection for an upper limit on the core component. \\
{\bf A3814} 2MASX J21490737-3042043. The BCG is matched to a radio source that appears only in the NVSS \cite[][]{Condon98} and SUMSS \cite[][]{Mauch03} catalogues. Interpolating between these show the source to be steep index and hence most likely dominated by its non-core emission.  Obtain a measure for this component and then extrapolate for an upper limit on the core. \\
{\bf A4038} IC 5358.  This cluster appears to contain two large ellipticals, of which this one appears to be dominant. Matched to a radio detection in NVSS \cite[][]{Condon98} but not otherwise detected in the radio.  The source is resolved, thus showing it to not be a core-only system.  Extrapolate from the integrated flux for an estimate of the non-core component.  The resolution of NVSS is insufficient to isolate the core, so extrapolate from the peak flux for an upper limit on the core component. \\
{\bf RXJ1459.0-0843} 2MASX J14590518-0842367. Optical overlay of the NVSS region shows that the BCG is associated with a radio source (although note that this is not associated in NED).  There are no other radio detections available and survey limits do not restrict the nature of the emission so can only extrapolate with representative indices to get upper limit on both the core and non-core components. \\
{\bf S721} ESO 382- G 007. The BCG is associated with a steep spectrum radio source.  Fit the SED to get a measure of the non-core component and then extrapolate to place an upper limit on the core. \\
{\bf RXJ1256.9-3119} ESO 443- G 014. The BCG is associated with a steep spectrum radio source.  Fit the SED for a measurement of the non-core and then extrapolate from the NVSS detection \cite[the highest frequency that the source is detected at,][]{Condon98} for an upper limit on teh core. \\
{\bf A2420} 2MASX J22101878-1210141. Optically matched to an NVSS source \cite[][]{Condon98} although no other radio detections are available an so the nature of the emission is inconclusive.  Extrapolate with representative indices to put upper limits on both the core and non-core components. \\
{\bf RXJ1252.5-3116} 2MASX J12523429-3116186. The BCG is associated with a very steep spectrum unresolved (NVSS resolution) radio source.  Fit the archival SED for a measurement of the non-core component.  There are no detections above L-band and the higher frequency survey limits are not restrictively deep.  Extrapolate from the L-band flux \cite[NVSS,][]{Condon98} to get an upper limit on the core component. \\
{\bf A3497} 2MASX J12000606-3123162. The BCG is associated with a steep spectrum radio source  Fit the SED to get a measure of the non-core and then extrapolate to put an upper limit on any active core component. \\
{\bf A3364} 2MASX J05473773-3152237.  The BCG appears to be associated with a steep spectrum radio source. Fit the archival SED for an estimate of the non-core component and then extrapolate for an upper limit on any core component.  Note that the match for this source is tentative and may represent a projection related mis-match. \\

\subsubsection{\bf REFLEX(SUMSS) - LEs}
{\bf A3526} NGC4696. The BCG is associated with the powerful, extended radio source PKS1246-410.  The archival SED is steep spectrum and dominated by the extended emission, hence fit this to obtain a measure of the non-core component.  The source was studied by \cite{Taylor06} who used a combination of the VLA-A array and the VLBA to uncover its core properties, hence can fit a separate SED to obtain a measure of the core component only.  This cluster is also commonly referred to as the Centaurus cluster. \\
{\bf RXJ1840.6-7709} ESO 045- G 011.  The BCG is matched to a powerful, relatively flat spectrum radio source.  The source is unresolved at 843~MHz in SUMSS \cite[][]{Mauch03} and tagged as a point source in the 8.4~GHz ATCA imaging of the CRATES survey \cite[][]{Healey07}. There is however a large flux decrement between this and a lower resolution single dish observation \cite[][]{Wright90} although the source is unresolved at 20~GHz in AT20g \cite[][]{Murphy10}.  The SED is relatively flat $\alpha$$\approx$0.58) although the source does not appear to be only a core.  Interpolate between the unresolved X-band flux and the AT20g detection for an estimate of the core component and then fit the integrated spectrum, accounting for the expected core contribution for a measure of the non-core. \\
{\bf A3363} 2MASX J05451060-4756596/PKS 0543-479.  Note that this galaxy lies at z$\approx$0.12 (VLT redshift, Edge \etal in prep.) whereas NED has it at z$\sim$0.2768.  The BCG appears to be associated with a strong head-tail radio source.  However, note that in an optical overlay of the SUMSS imaging of this region \cite[][]{Mauch03} the centroid is not exactly aligned and this source may therefore be mis-matched.  The index is ambiguous ($\alpha$$\approx$0.69), however using a MOST observation \cite{Jones92} claim the source to be marginally resolved at 408~MHz and hence the flux can be attributed to non-core emission.  Fit the SED for a measure of the non-core. There is no isolated measure of the core available hence extrapolate for an upper limit on the core component. \\  
{\bf A3396} 2MASX J06284979-4143373. The BCG is matched to a steep spectrum radio source.  Fit the SED for a measure of the non-core and then extrapolate to put an upper limit on the core. \\
{\bf RXJ2151.3-5521} PKS 2148-555. The BCG is associated with a flat spectrum radio-source \cite[see e.g.,][]{Healey07}.  The archival SED is somewhat confusing however an overlay of the SUMSS image shows that the reported SUMSS flux for this source is actually the flat-spectrum core of an FRII and that confusion with the lobes accounts for the confused SED.  \cite{Jones92} show it to be a double in MOST imaging at 408~MHz.  Fit the integrated SED (ensuring inclusion of all source components) for a measurement of the non-core component.  Observed in AT20g at multiple frequencies \cite[][]{Murphy10}, hence can fit a separate core-only SED to get a measure of the active core component.   This cluster is also known as A3816. \\   
{\bf RXJ2124.3-7446} 2MASX J21240939-7445538. The BCG is matched to a reasonably bright SUMSS source \cite[$\sim$75 mJy,][]{Mauch03} although there are no other radio detections. Survey limits do not restrict the nature of the emission and hence this is ambiguous so extrapolate with representative indices for limits on both the core and non-core components. \\
{\bf A2871} SARS 016.37936-36.99021. The BCG is matched to a steep spectrum radio source that appears slightly extended in the SUMSS map \cite[][]{Mauch03}.  Fit the for a measure of the non-core component and then extrapolate for an upper limit on the core. \\
{\bf RXJ1317.1-3821} 2MASX J13171224-3821544.  This is a moderate redshift cluster (z$\sim$0.26) with a fairly weak BCG.  Nonetheless, this source is matched to SUMSS \cite[][]{Mauch03}, NVSS \cite[][]{Condon98} and an ATCA observation \cite[][]{Feain09} and fitting to these gives a flat spectrum and hence implies that the source is core dominated.    Fit the flat component for a core estimate and then extrapolate for an upper limit on the non-core. Note that morphologically, in SUMSS this source appears to look a bit like an FRII although considering the optical overlay the most likely explanation appears to be that this is chance alignment with two background sources.   \\ 
{\bf S540} ESO 306- G 017. The BCG is matched to a SUMSS source \cite[][]{Mauch03} although there are no other radio detections and hence the nature of the emission is ambiguous.  Extrapolate using appropriate indices to put limits on both the core and non-core components. \\
{\bf S301} IC 1860. The BCG is associated with a steep spectrum radio source.  Fit the archival SED for a measure of the non-core component and then extrapolate to put an upper limit on the core. \\
{\bf RXJ0303.7-7752} 2XMM J030345.6-775243.  The BCG appears to be associated with a SUMSS source \cite[][]{Mauch03} although no other radio detections are available.  There are a number of other radio sources nearby.   Extrapolate using appropriate indices to put upper limits on both the core and non-core components. \\

\subsubsection{\bf REFLEX(SUMSS) - NLEs}
{\bf S41} ESO 350- G 015.  The BCG is not associated with the strong radio source PKS 0023-33 in NED, but overlay of the radio map onto the optical shows them to be associated.  This is a powerful, extended steep spectrum source where the archival SED potentially slighly flattens to higher radio frequencies.  Fit the SED for an estimate of the non-core component.  The source is not detetced in AT20g \cite[][]{Murphy10} showing that the spectrum does not flatten out, hence extrapolate from the non-detection to put an upper limit on the core. \\
{\bf A3330} FAIRALL 0790. The BCG is matched to a slightly one-sided radio source in SUMSS \cite[][]{Mauch03}.  The spectral index of the archival SED is on the cusp of being considered steep, which coupled with the observed extent suggests that this source is most likely dominated by its non-core emission.  Fit the SED for an estimate of this component and then extrapolate from the AT20g non-detection \cite[][]{Murphy10} to put an upper limit on the core. \\
{\bf A3911} AM 2243-525.  This cluster appears to contain a merging BCG-pair.  There is a SUMSS source \cite[][]{Mauch03} matched to this that appears slightly elongated.  Note that the centroid of the radio source actually lies between the two optical centres. It could be that this is a double AGN with two radio-loud cores.  The archival SED is steep spectrum, which can be fitted to obtain an estimate of the non-core component.  Extrapolate for an upper limit on the core component. \\
{\bf S861} 2MASX J20184779-5242334. The BCG is associated with an amorphous radio source in SUMSS \cite[][]{Mauch03}. The SED appears to be steep \cite[associating it with a PMN detection,][]{Griffith93} which along with the resolved nature of this source shows it to be non-core dominated, hence fit for this then extrapolate from the C-band detection to put an upper limit on the core component. \\
{\bf A3360} 2MASX J05401053-4323182.  The BCG is associated with a CSS radio source.  Fit the SED to obtain an estimate of the non-core component and then extrapolate from the AT20g non-detection to put an upper limit on the core component. \\
{\bf A3728} 2MASX J21050350-8243367.  This cluster appears to be a double BCG system, this one of which is matched to a steep spectrum radio source.  Fit the SED for an estimate of the non-core component and then extrapolate from the C-band detection \cite[][]{Wright94} to place an upper limit on the core component.  The partner BCG ([GSB2009] J210456.52-824351.6, z$\sim$0.0974) is undetected in SUMSS \cite[][]{Mauch03}. \\
{\bf RXJ2143.9-5637}  2MASX J21435923-5637206. The BCG is matched to a steep spectrum radio source.  This source is classified as a point source at 843~MHz in the SUMSS catalogue \cite[][]{Mauch03} but is extended at 408~MHz with the MOST \cite{Jones92}.  Interpolating between these would then suggest that the SED must flatten towards the higher frequency detection at C-band \cite[5~GHz,][]{Wright94}.  However, visual inspection of the SUMSS imaging overlaid onto the optical image shows the source to actually be extended and that the lobes are classified as distinct source in the SUMSS catalogue.  Including these with the central component gives an integrated flux that is consistent with a single steep index source and consistent morphologies at different frequencies.  Fit the SED for a measure of the non-core component and then extrapolate from the C-band flux to put an upper limit on the core. This source is also known as APMCC699. \\
{\bf S547} ESO 253- G 027. The BCG is matched to a SUMSS radio source \cite[][]{Mauch03}.  Note that there is a small offset in the centroid between the unresolved radio source and the optical position.  However, this is within one resolution element and there appear to be no other potential optical counterparts seen and hence we leave this as a match.  The source is undetected elsewhere in the radio, however a non-detection with Parkes \cite[PMN Southern region,][]{Wright94} restricts the spectral index to be steeper than $\alpha$$_{steep}$$>$0.7 and hence we classify the source as most likely non-core dominated.  Extrapolate from the SUMSS detection for an estimate of the non-core component and then extrapolate from the non-detection to put an upper limit on the core. \\
{\bf A4023} 2MASX J23401245-8511328. The BCG is matched to a one-sided extended radio source at 843~MHz in the SUMSS catalogue \cite[][]{Mauch03}.  The extent coupled to the source not being detected in the PMN catalogue \cite[][]{Wright94} show the source to be steep spectrum and most likely non-core dominated.  Extrapolate from the SUMSS detection for an estimate of the non-core component and then extrapolate from the non-detection to place an upper limit on the core. \\
{\bf RXJ2018.4-4102} IC4991.  The BCG is associated with an elongated radio source in SUMSS \cite[][]{Mauch03}.  The source was observed by \cite{Slee94} with the Parkes-Tidbinbilla-Interferometer (PTI).  They found no parsec-scale core and gave an upper limit of $<$4 mJy on the milli-arcsecond scale structure at 2.3~GHz, with a lower resolution integrated flux for the system of 20 mJy at 5~GHz.  Interpolate between this 5~GHz flux and the SUMSS detection for an estimate of the non-core component and then extrapolate from the PTI non-detection to place an upper limit on the core component at 10~GHz. \\
{\bf S927} FAIRALL 0095.  The optical BCG is mathed to a radio-source in the SUMSS catalogue \cite[][]{Mauch03}.  However, there are no other radio detections available and survey limits at other radio frequencies do not limit the index to be unambiguously core or non-core dominated.  Extrapolate therefore with relevant indices to put limits on both the core and non-core components. \\
{\bf S592} 2MASX J06384515-5358225.  The BCG appears associated with a SUMSS radio source \cite[][]{Mauch03}.  No other radio detections are available and survey limits do not clearly restrict the source to be sufficiently steep or flat spectrum to unambiguously attribute the emission with a core or non-core origin.  Therefore, extrapolate from the SUMSS detection with appropriate indices to place a limit on both the core and the non-core components. \\
{\bf RXJ0658.5-5556} 2MASX J06583806-5557256.  The BCG is matched to the SUMSS radio source SUMSS J065837-555718 \cite[][]{MAuch03} although no other radio detections are available to restrict the nature of the emission.  Extrapolate with appropriate indices to put upper limits on both the core and non-core components.  This source is also known as IES0657 or perhaps more popularly, the `bullet cluster'.  \\
{\bf RXJ0738.1-7506} 2MASX J07380644-7506269. The BCG is matched to a potentially marginally resolved SUMSS source \cite[][]{Mauch03}.  Note that the radio centroid is slightly off from the optical centroid but visual inspection of the overlaid images suggests that the sources are associated.  There are however no other radio detections and the source is not unambiguously resolved.  Extrapolate therefore with appropriate indices to put limits on both the core and non-core components. \\
{\bf RXJ2031.8-4037} SUMSS J203153-403728. This is a reasonably high redshift cluster (z$\sim$0.34) whose BCG is matched to a SUMSS source \cite[][]{Mauch03}. There are no other radio observations available that allow the nature of this emission (core or non-core) to be unambiguously defined so extrapolate with appropriate indices to put limits on both the core and non-core components. \\
{\bf A3718} APMBGC 187+014+002.  This system appears to contain a central galaxy pair.  The redshift of this component (z$\sim$0.13) places it within the cluster.  The southern counterpart galaxy has a higher redshift given to it in NED (z$\sim$0.17) and so it is possible that the pair is only associated in projection.  The named galaxy appears to be associated with a radio source in SUMSS \cite[][]{Mauch03}. There are no other radio detections and survey limits do not allow the nature of the emission (core or non-core) to be clearly defined hence extrapolate with appropriate indices to put upper limits on both the core and non-core components. Note that if the galaxies are physically associated then the companion, 2MASX J20555594-5455493, would appear to be dominant and is radio undetected. \\
{\bf S405} FAIRALL 1274.  The BCG is matched to an unresolved SUMSS source \cite[][]{Mauch03} but has no other radio detections available.  Furthermore, survey limits do not allow the nature of this emission to be unambiguosly attributed to either the core or non-core and so extrapolate with appropriate indices to put upper limits on both the core and non-core components. \\
{\bf RXJ2224.4-5515} 2MASX J22241856-5514513.  The BCG is matched to an unresolved SUMSS source \cite[][]{Mauch03}.  There are no other radio detections and survey limits to not restrict the index to be unambigously steep or flat. Extrapolate with appropriate indices to place upper limits on both the core and non-core components. This source is also known as APMCC772.  \\
{\bf RXJ0217.2-5244} APMUKS(BJ) B021530.61-525927.7.  This is a reasonably high redshift cluster (z$\sim$0.34) whose BCG position appears tentatively matched to a SUMSS radio source \cite[][]{Mauch03}.  The match is within the resolution of SUMSS but the centroid position is more than the catalogue's given positional accuracy from the optical position.  The source could be extended just below the resolution limit but no higher resolution radio data are available to confirm or deny this.  We leave this source as an associated match but conservatively extrapolate to place only limits on both the core and non-core components. \\
{\bf RXJ0322.2-5310} IC 1917. There is a faint radio source centred on the BCG as seen by overlaying the SUMSS map onto an optical image of the region \cite[][]{Mauch03}.  There are no other radio detections and so the nature of this emission cannot be unambiguously attributed to either a core or a non-core.  Extrapolate therefore, with appropriate indices, to place upper limits on both the core and non-core components.  The source is also known as APMCC391.  \\
{\bf A3736} 2MASX J21032863-4320360. The BCG is matched to an unresolved SUMSS radio source \cite[][]{Mauch03}.  There are no other radio detections and so the nature or the emission (core or non-core) cannot be clearly identified. Extrapolate using appropriate indices to place upper limits on both the core and non-core components.  \\
{\bf S849} 2MASX J20090639-5422478.  The BCG is matched to a SUMSS radio source \cite[][]{Mauch03}.  No other radio detecions are available to unambiguosly associate this to either core or non-core emission so the best approach is to extrapolate with appropriate indices to put upper limit on both the core and non-core components. \\
{\bf RXJ2254.0-6315} 2MASX J22540401-6314509.  The BCG is matched to a SUMSS source \cite[][]{Mauch03} but there are no other radio detections and available survey limits do not restrict the source to have an unambiguously steep or flat spectral index.  The exact nature of the emission cannot therefore be classifed and so we extrapolate with the representative indices to place upper limits on both the core and non-core components. The source is also known as AM2250. \\
{\bf A3158} ESO 156- G 008 NED01.  This galaxy is a large elliptical that is matched to a SUMSS source \cite[][]{Mauch03}.  No other radio observation are available to restrict the nature of the emission and so extrapolate with representative indices to put upper limits on both the core and non-core components.  Note that there is another large elliptical, ESO 156- G 008 NED02, whose redshift (z$\sim$0.0581) places it within the cluster.  This galaxy appears equally dominant although is radio undetected and so this cluster may contain twin BCGs, only of of which is currently radio-loud. \\
{\bf RXJ2023.4-5535} 2MASX J20232088-5535495.  The BCG is matched with the radio source SUMSS J202319-553545 \cite[][]{Mauch03}. No other radio detections are available and survey limits do not restrict the spectral index to be unambiguosly steep or flat.  Therefore, extrapolate with appropriate indices to put upper limits on both the core and non-core components. \\

\section[]{SED Decompositions - Main Sample+ Extension} \label{APPENDIX_MSP_DECOMPOSITIONS}

\onecolumn
\begin{longtable}[l]{@{}llllllll@{}}
\caption{Decompositions for the sources in the Main Sample+.  Parameters as for Table \ref{Model_Fitted}.} \label{MAIN_SAMPLE_plus} \\
  \hline\hline 
   Cluster & Non-core (at 1~GHz) & Core (at 10~GHz) & Steep Index & Flat Index & Flat normalisation & Redshift & Lines?  \\
           &    (mJy)            &  (mJy)           &                &             &           &        \\
  \hline\hline
\endfirsthead
\multicolumn{3}{l}{{\tablename} \thetable{} -- Continued} \\
  \hline\hline
  Cluster & Non-core (at 1~GHz) & Core (at 10~GHz) & Steep Index & Flat Index & Flat normalisation & Redshift & Lines? \\
          &    (mJy)            &  (mJy)           &                &             &           &        \\
  \hline\hline
  \\
\endhead
  \multicolumn{3}{l}{{Continued on Next Page\ldots}} \\
\endfoot
  \\ \hline \hline
\endlastfoot
 \\
 \bf{BCS}  \\
 \hline \hline
\\
 \\
\multicolumn{3}{l}{ \bf{Core and Non-core Values}}  \\ 
 \hline
 \\
Hercules-A     & 62484 $\pm$ 4802 & 4.9   $\pm$ 1.1  & 1.01 $\pm$ 0.06 &  0.2  $\pm$ 0.2  & 7.8   $\pm$ 3.7   & 0.155  & $\surd$ \\
A2052          &  7124 $\pm$  107 & 302.6 $\pm$ 79.3 & 1.09 $\pm$ 0.24 &  0.14 $\pm$ 0.03 & 420.4 $\pm$ 104.3 & 0.0355 & $\surd$ \\
A2199          &  4966 $\pm$   14 & 20.3  $\pm$ 0.5  & 1.26 $\pm$ 0.12 & -0.07 $\pm$ 0.04 & 17.3  $\pm$ 1.7   & 0.031  & $\surd$ \\
A1795          &  1295 $\pm$   20 & 26.6  $\pm$ 8.5  & 0.94 $\pm$ 0.03 &  0.51 $\pm$ 0.08 & 86.1  $\pm$ 17.4  & 0.0632 & $\surd$ \\
A1367          &  6860 $\pm$   17 & 111.0 $\pm$ 49.0 & 0.72 $\pm$ 0.01 &  0.35 $\pm$ 0.19 & 250.1 $\pm$ 111.4 & 0.0217 & $\surd$ \\
A2029          & 757.1 $\pm$  6.7 &  2.7  $\pm$ 0.1  & 1.23 $\pm$ 0.04 &  0.2  $\pm$ 0.2  & 4.3   $\pm$ 2.5   & 0.0780 & x \\
A1656          & 235.9 $\pm$ 27.3 &  7.7  $\pm$ 4.7  & 0.72 $\pm$ 0.05 &  0.46 $\pm$ 0.23 & 7.7   $\pm$ 4.7   & 0.0240 & x \\
\multicolumn{3}{l}{ \bf{Non-core Values with Core Upper Limits}}  \\
 \hline
 \\
RXJ0338.7+0958   & 62.5  $\pm$ 0.8 & $<$3.3  & 1.48 $\pm$ 0.01 & 0.2 & 5.2  & 0.0338 & $\surd$ \\  
A2312            & 125.7 $\pm$ 3.4 & $<$17.0 & 1.39 $\pm$ 0.08 & 0.2 & 26.9 & 0.0945 & x \\  
A1314            & 43.8  $\pm$ 8.7 & $<$2.4  & 0.83 $\pm$ 0.12 & 0.2 & 3.8  & 0.0336 & x \\  
A2147            & 20.1  $\pm$ 6.2 & $<$2.1  & 0.67 $\pm$ 0.07 & 0.2 & 3.3  & 0.0355 & x \\  
A2065            & 19.0  $\pm$ 1.4 & $<$1.1  & 1.0  $\pm$ 0.2  & 0.2 & 1.7  & 0.0749 & x \\  
 \\
 \bf{eBCS}  \\
 \hline \hline
\\
 \\
\multicolumn{3}{l}{ \bf{Core and Non-core Values}}  \\  
 \hline
 \\
A923            &  112.0 $\pm$ 5.3  & 0.5   $\pm$ 0.4  & 0.86 $\pm$ 0.02 & 0.00  $\pm$ 0.20 & 0.5   $\pm$ 0.3 & 0.1168 & $\surd$ \\
RXJ1844.1+4553  & 6843.7 $\pm$ 14.8 & 51.2  $\pm$ 3.6  & 0.73 $\pm$ 0.03 & 0.34  $\pm$ 0.03 & 111.6 $\pm$ 7.9 & 0.0917 & x \\
A1446           &  667.1 $\pm$ 18.2 & 8.9   $\pm$ 5.4  & 1.14 $\pm$ 0.02 & -0.36 $\pm$ 0.43 & 3.9   $\pm$ 2.4 & 0.1028 & x \\
RXJ2250.0+1137  & 2335.5 $\pm$ 89.0 & 118.0 $\pm$ 56.0 & 0.53 $\pm$ 0.20 & 0.0   $\pm$ 0.2  & 118   $\pm$ 56  & 0.0262 & x \\
A2149           &   99.4 $\pm$ 9.7  & 7.0   $\pm$ 0.9  & 0.83 $\pm$ 0.05 & 0.2   $\pm$ 0.2  & 11.0  $\pm$ 4.5 & 0.0655 & x \\
A2152           &   86.4 $\pm$ 3.5  & 3.3   $\pm$ 0.5  & 1.39 $\pm$ 0.21 & 0.2   $\pm$ 0.2  & 5.2   $\pm$ 1.9 & 0.0441 & x \\
RXJ0004.9+1142  &   46.1 $\pm$ 11.5 & 16.6  $\pm$ 3.4  & 1.08 $\pm$ 0.09 & 0.36  $\pm$ 0.14 & 37.7  $\pm$ 7.6 & 0.0761 & $\surd$ \\
\multicolumn{3}{l}{ \bf{Non-core Values with Core Upper Limits}}  \\
 \hline
 \\
A2396            & 729.6 $\pm$ 9.3  & $<$41.8 & 1.08 $\pm$ 0.06 & 0.2 & 66.2 & 0.1924 & $\surd$ \\  
A2036            & 482.0 $\pm$ 8.7  & $<$11.0 & 1.01 $\pm$ 0.06 & 0.2 & 17.4 & 0.1159 & $\surd$ \\  
A580             &  53.2 $\pm$ 3.3  & $<$1.3  & 1.24 $\pm$ 0.31 & 0.2 & 2.1  & 0.1113 & $\surd$ \\  
Z1883            &  36.8 $\pm$ 7.9  & $<$3.3  & 1.52 $\pm$ 0.45 & 0.2 & 5.2  & 0.198  & $\surd$ \\  
Z1370            &  61.5 $\pm$ 4.5  & $<$0.3  & 2.24 $\pm$ 0.05 & 0.2 & 0.5  & 0.216  & $\surd$ \\  
Z3959            &  25.4 $\pm$ 6.2  & $<$3.8  & 1.33 $\pm$ 0.09 & 0.2 & 6.0  & 0.3521 & $\surd$ \\  
A853             & 129.2 $\pm$ 9.4  & $<$12.1 & 1.0  $\pm$ 0.2  & 0.2 & 25.4 & 0.1664 & $\surd$ \\  
A278             & 498.4 $\pm$ 9.8  & $<$7.6  & 0.85 $\pm$ 0.09 & 0.2 & 12.0 & 0.0891 & x \\  
A2625            & 146.6 $\pm$ 6.8  & $<$40.7 & 0.85 $\pm$ 0.02 & 0.2 & 64.5 & 0.1005 & x \\  
A2445            & 113.6 $\pm$ 66.7 & $<$36.3 & 0.71 $\pm$ 0.02 & 0.2 & 57.6 & 0.1700 & x \\  
RXJ0027.6+2616   &  88.0 $\pm$ 2.5  & $<$15.5 & 0.71 $\pm$ 0.01 & 0.2 & 24.6 & 0.3668 & x \\  
RXJ1652.6+4011   &  73.6 $\pm$ 6.0  & $<$2.9  & 0.63 $\pm$ 0.04 & 0.2 & 4.6  & 0.1475 & x \\  
A1291            &  56.5 $\pm$ 5.7  & $<$2.5  & 1.12 $\pm$ 0.04 & 0.2 & 4.0  & 0.0500 & x \\  
RXJ1852.1+5711   &  91.3 $\pm$ 3.1  & $<$0.7  & 1.01 $\pm$ 0.02 & 0.2 & 1.1  & 0.1068 & x \\  
A1986            &  58.1 $\pm$ 1.5  & $<$15.5 & 0.97 $\pm$ 0.01 & 0.2 & 24.6 & 0.1108 & x \\  
A2315            &  22.8 $\pm$ 5.0  & $<$11.1 & 0.97 $\pm$ 0.10 & 0.2 & 17.6 & 0.0931 & x \\  
 \\
 \bf{REFLEX - NVSS}  \\
 \hline \hline
\\
 \\
\multicolumn{3}{l}{ \bf{Core and Non-core Values}}  \\ 
 \hline
 \\
Hydra A          & 59472.6 $\pm$ 55.6  & 242.8 $\pm$ 8.6  & 0.92 $\pm$ 0.01 & 0.35 $\pm$ 0.01 & 541.0 $\pm$ 19.2 & 0.0549 & $\surd$ \\
A2597            &  2648.0 $\pm$ 127.0 &  30.5 $\pm$ 5.0  & 1.12 $\pm$ 0.11 & 0.2  $\pm$ 0.2  & 48.3  $\pm$ 28.3 & 0.0830 & $\surd$ \\
RXJ0132.6-0804   &   136.0 $\pm$ 14.1  &  77.7 $\pm$ 3.6  & 1.0  $\pm$ 0.2  & 0.45 $\pm$ 0.03 & 217.8 $\pm$ 10.1 & 0.1489 & $\surd$ \\
A2415            &   239.4 $\pm$ 9.8   & 107.2 $\pm$ 18.6 & 1.15 $\pm$ 0.02 & 0.54 $\pm$ 0.09 & 373.8 $\pm$ 64.8 & 0.0573 & $\surd$ \\
RXJ1347.5-1144   &    24.4 $\pm$ 16.1  &  17.8 $\pm$ 3.0  & 1.08 $\pm$ 0.30 & 0.33 $\pm$ 0.08 & 38.3  $\pm$ 6.4  & 0.4500 & $\surd$ \\
RXJ1050.4-1250   &    46.2 $\pm$ 3.1   &   1.6 $\pm$ 0.6  & 1.0  $\pm$ 0.2  & 0.42 $\pm$ 0.19 & 2.5   $\pm$ 0.9  & 0.0154 & $\surd$ \\
RXJ1501.1+0141   &    18.9 $\pm$ 2.6   &   2.2 $\pm$ 1.5  & 1.0  $\pm$ 0.2  & 0.15 $\pm$ 0.37 & 3.2   $\pm$ 2.1  & 0.0066 & $\surd$ \\
A3695            &  2039.9 $\pm$ 14.8  &  39.0 $\pm$ 3.4  & 0.83 $\pm$ 0.04 & 0.71 $\pm$ 0.03 & 200.0 $\pm$ 17.4 & 0.0888 & x \\
A3560            &  1620.7 $\pm$ 16.0  &   5.7 $\pm$ 1.3  & 0.85 $\pm$ 0.07 & -0.48 $\pm$ 0.12 &  1.9 $\pm$ 0.4  & 0.0480 & x \\
A3528S           &  1485.1 $\pm$ 15.2  &   2.3 $\pm$ 0.1  & 1.18 $\pm$ 0.04 & 0.2  $\pm$ 0.2  &   3.6 $\pm$ 1.8  & 0.0574 & x \\
A3165            &  1089.5 $\pm$ 12.1  &  26.5 $\pm$ 0.9  & 0.84 $\pm$ 0.01 & 0.2  $\pm$ 0.2  &  42.0 $\pm$ 24.6 & 0.1423 & $\surd$ \\
A2399            &   412.2 $\pm$ 15.8  &   7.6 $\pm$ 1.5  & 0.62 $\pm$ 0.02 & 0.20 $\pm$ 0.06 &  12.1 $\pm$  2.5 & 0.0567 & x \\
S753             &   444.4 $\pm$ 9.5   &  13.5 $\pm$ 1.2  & 0.60 $\pm$ 0.02 & 0.71 $\pm$ 0.14 &  68.8 $\pm$  6.1 & 0.0138 & x \\
RXJ0137.2-0912   &   300.1 $\pm$ 9.0   &   7.9 $\pm$ 3.2  & 1.78 $\pm$ 0.01 & 0.2  $\pm$ 0.2  &  12.5 $\pm$  7.3 & 0.0413 & x \\
A2355            &   260.7 $\pm$ 9.3   &   2.6 $\pm$ 0.3  & 0.80 $\pm$ 0.15 & 0.2  $\pm$ 0.2  &   4.1 $\pm$  2.4 & 0.1244 & x \\
RXJ1655.9-0113   &    99.9 $\pm$ 4.2   &   3.0 $\pm$ 0.5  & 1.38 $\pm$ 0.20 & 0.2  $\pm$ 0.2  &   4.8 $\pm$  2.3 & 0.0408 & x \\
A295             &    58.4 $\pm$ 3.9   &   2.5 $\pm$ 1.0  & 1.0  $\pm$ 0.2  & 0.2  $\pm$ 0.2  &   4.0 $\pm$  1.9 & 0.0428 & x \\
A3570            &    76.9 $\pm$ 5.6   &   2.2 $\pm$ 0.7  & 2.48 $\pm$ 0.36 & 0.2  $\pm$ 0.2  &   3.5 $\pm$  1.7 & 0.0375 & x \\
\multicolumn{3}{l}{ \bf{Core Values with Non-core Upper Limits}}  \\
 \hline
 \\
A1644            & $<$14.8  & 103.2 $\pm$ 9.1  & 1.0            & 0.06 $\pm$ 0.05 & 119.5 $\pm$ 10.5 & 0.0475 & $\surd$ \\
RXJ0445.1-1551   & $<$10.9  &  65.0 $\pm$ 7.6  & 1.0            & 0.21 $\pm$ 0.08 & 104.6 $\pm$ 12.2 & 0.0360 & $\surd$ \\
RXJ1506.4+0136   & $<$7.9   &   7.1 $\pm$ 1.9  & 1.0            & 0.06 $\pm$ 0.12 & 8.2   $\pm$ 2.2  & 0.0057 & $\surd$ \\
RBS1712          & $<$95.8  &  33.4 $\pm$ 3.6  & 1.0            & 0.34 $\pm$ 0.09 & 73.6  $\pm$ 7.9  & 0.1899 & x \\
\multicolumn{3}{l}{ \bf{Non-core Values with Core Upper Limits}}  \\
 \hline
 \\
A3532            &  1439.2 $\pm$ 15.5 & $<$59.7  & 0.89 $\pm$ 0.04 & 0.2 & 94.6 & 0.0542 & $\surd$ \\  
RXJ1206.2-0848   &   268.5 $\pm$ 8.9  & $<$46    & 1.23 $\pm$ 0.01 & 0.2 & 72.9 & 0.4413 & $\surd$ \\  
RXJ0501.6+0110   &   162.4 $\pm$ 9.6  & $<$15.5  & 0.69 $\pm$ 0.03 & 0.2 & 24.6 & 0.1245 & $\surd$ \\  
A85              &    74.4 $\pm$ 6.1  & $<$2.5   & 1.05 $\pm$ 0.03 & 0.2 & 4.0  & 0.0557 & $\surd$ \\  
A383             &    52.5 $\pm$ 6.7  & $<$5.4   & 0.74 $\pm$ 0.10 & 0.2 & 6.5  & 0.1883 & $\surd$ \\  
A2717            &   854.2 $\pm$ 20.9 & $<$39.2  & 1.74 $\pm$ 0.06 & 0.2 & 62.1 & 0.0498 & x \\  
A3490            &   493.4 $\pm$ 9.8  & $<$46    & 0.97 $\pm$ 0.15 & 0.2 & 72.9 & 0.0682 & x \\  
A3528N           &   490.5 $\pm$ 14.7 & $<$9.2   & 0.80 $\pm$ 0.01 & 0.2 & 14.6 & 0.0541 & x \\  
A1791            &   401.9 $\pm$ 10.2 & $<$8.8   & 1.22 $\pm$ 0.01 & 0.2 & 14.0 & 0.1263 & x \\  
RXJ2043.2-2144   &   448.8 $\pm$ 13.7 & $<$46    & 1.03 $\pm$ 0.03 & 0.2 & 72.9 & 0.2041 & $\surd$ \\  
A3376            &   381.2 $\pm$ 10.0 & $<$24.8  & 1.10 $\pm$ 0.11 & 0.2 & 24.8 & 0.0456 & x \\  
A2984            &   289.0 $\pm$  7.2 & $<$46    & 0.72 $\pm$ 0.03 & 0.2 & 72.9 & 0.1050 & x \\  
RXJ1332.9-2519   &   277.1 $\pm$ 52.5 & $<$46    & 0.72 $\pm$ 0.02 & 0.2 & 72.9 & 0.1206 & x \\  
RXJ0359.1-0320   &   270.9 $\pm$ 10.0 & $<$46    & 0.92 $\pm$ 0.04 & 0.2 & 72.9 & 0.1195 & x \\  
A3856            &   295.9 $\pm$ 8.8  & $<$46    & 1.36 $\pm$ 0.12 & 0.2 & 72.9 & 0.1423 & x \\  
A499             &   291.0 $\pm$ 8.7  & $<$46    & 0.81 $\pm$ 0.14 & 0.2 & 72.9 & 0.1550 & x \\  
RBS0540          &   119.9 $\pm$ 8.0  & $<$8.7   & 0.71 $\pm$ 0.03 & 0.2 & 13.8 & 0.0397 & x \\  
RXJ1307.6-0650   &   176.9 $\pm$ 8.2  & $<$15.9  & 0.82 $\pm$ 0.02 & 0.2 & 25.1 & 0.0903 & x \\  
A2331            &   169.7 $\pm$ 9.5  & $<$40    & 0.70 $\pm$ 0.03 & 0.2 & 63.4 & 0.0790 & x \\  
S617             &   148.8 $\pm$ 9.2  & $<$46    & 0.72 $\pm$ 0.03 & 0.2 & 72.9 & 0.0336 & x \\  
A3764            &   120.4 $\pm$ 5.6  & $<$45.9  & 1.09 $\pm$ 0.19 & 0.2 & 72.7 & 0.0739 & x \\  
RXJ0340.1-1835   &   114.1 $\pm$ 7.3  & $<$29.3  & 0.75 $\pm$ 0.07 & 0.2 & 46.4 & 0.0059 & x \\  
A2389            &   126.8 $\pm$ 7.2  & $<$39.8  & 0.78 $\pm$ 0.06 & 0.2 & 63.1 & 0.1507 & x \\  
RXJ1254.7-1526   &    98.5 $\pm$ 2.1  & $<$46    & 0.91 $\pm$ 0.01 & 0.2 & 72.9 & 0.1506 & x \\  
A907             &   133.9 $\pm$ 8.1  & $<$46    & 0.88 $\pm$ 0.03 & 0.2 & 72.9 & 0.1620 & $\surd$ \\  
RXJ1139.4-3327   &    93.1 $\pm$ 3.7  & $<$38.9  & 1.23 $\pm$ 0.02 & 0.2 & 61.7 & 0.1095 & x \\  
A3694            &    57.5 $\pm$ 4.4  & $<$27.5  & 0.96 $\pm$ 0.35 & 0.2 & 43.6 & 0.0927 & x \\  
A3934            &    49.5 $\pm$ 4.3  & $<$23.3  & 1.19 $\pm$ 0.03 & 0.2 & 36.9 & 0.2240 & x \\  
A1300            &    56.6 $\pm$ 7.1  & $<$21.3  & 1.53 $\pm$ 0.05 & 0.2 & 33.8 & 0.3077 & x \\  
A3814            &    51.1 $\pm$ 4.3  & $<$21.3  & 1.41 $\pm$ 0.39 & 0.2 & 33.8 & 0.1205 & x \\  
A4038            &    38.5 $\pm$ 3.0  & $<$18.6  & 1.0  $\pm$ 0.2  & 0.2 & 29.5 & 0.0288 & x \\  
S721             &    35.0 $\pm$ 3.0  & $<$18.4  & 0.74 $\pm$ 0.34 & 0.2 & 29.2 & 0.0511 & x \\  
RXJ1256.9-3119   &    35.0 $\pm$ 3.0  & $<$15.6  & 1.08 $\pm$ 0.35 & 0.2 & 24.7 & 0.0565 & x \\  
RXJ1252.5-3116   &    29.1 $\pm$ 2.3  & $<$12.7  & 1.51 $\pm$ 0.03 & 0.2 & 20.1 & 0.0542 & x \\  
A3497            &    25.8 $\pm$ 3.7  & $<$10.7  & 1.65 $\pm$ 0.06 & 0.2 & 17.0 & 0.0683 & x \\  
A3364            &    26.2 $\pm$ 2.6  & $<$10.3  & 1.47 $\pm$ 0.42 & 0.2 & 16.3 & 0.1483 & x \\  
\multicolumn{3}{l}{ \bf{Upper Limits on both Core and Non-core}}  \\
 \hline
 \\
A1663           & $<$52.1    & $<$15.5 & 1.0 & 0.2 & 24.6 & 0.0847 & $\surd$ \\
A3698           & $<$45.2    & $<$15.5 & 1.0 & 0.2 & 24.6 & 0.0195 & $\surd$ \\
A2402           & $<$45.5    & $<$21.4 & 1.0 & 0.2 & 33.8 & 0.0806 & $\surd$ \\
A2566           & $<$38.9    & $<$17.5 & 1.0 & 0.2 & 27.7 & 0.0825 & $\surd$ \\
RXJ2147.0-1019  & $<$29.1    & $<$13.1 & 1.0 & 0.2 & 20.7 & 0.0793 & $\surd$ \\
S0301           & $<$27.4    & $<$12.3 & 1.0 & 0.2 & 19.5 & 0.0229 & $\surd$ \\
A281            & $<$24.4    & $<$11.0 & 1.0 & 0.2 & 17.4 & 0.1276 & $\surd$ \\
A3027           & $<$129.0   & $<$46   & 1.0 & 0.2 & 72.9 & 0.0760 & x \\
RXJ2034.9-2143  & $<$115.4   & $<$46   & 1.0 & 0.2 & 72.9 & 0.1531 & x \\
A2384B          & $<$78.5    & $<$36.7 & 1.0 & 0.2 & 58.2 & 0.0956 & x \\
A1391           & $<$48.5    & $<$23.2 & 1.0 & 0.2 & 36.8 & 0.1555 & x \\
A2401           & $<$47.6    & $<$22.9 & 1.0 & 0.2 & 36.4 & 0.0578 & x \\
RXJ1459.0-0843  & $<$35.7    & $<$18.4 & 1.0 & 0.2 & 29.2 & 0.1057 & x \\
A2420           & $<$27.5    & $<$13.3 & 1.0 & 0.2 & 21.1 & 0.0830 & x \\
 \\
 \bf{REFLEX - SUMSS}  \\
 \hline \hline
\\
 \\
\multicolumn{3}{l}{ \bf{Core and Non-core Values}}  \\ 
 \hline
 \\
A3526           & 5171.1 $\pm$ 15.3  & 66.9  $\pm$ 16.1 & 0.83 $\pm$ 0.01 & 0.50  $\pm$ 0.10 & 212.7 $\pm$ 51.0 & 0.0099 & $\surd$ \\
RXJ1931.6-3354  & 1357.3 $\pm$ 11.9  & 18.2  $\pm$ 2.6  & 0.99 $\pm$ 0.04 & 0.2   $\pm$ 0.2  & 28.8  $\pm$ 13.8 & 0.0979 & $\surd$ \\
RXJ1840.6-7709  & 1093.5 $\pm$ 11.2  & 213.9 $\pm$ 48.7 & 0.58 $\pm$ 0.01 & -0.06 $\pm$ 0.09 & 186.3 $\pm$ 34.9 & 0.0182 & $\surd$ \\
RXJ2151.3-5521  & 1511.9 $\pm$ 20.5  & 101.7 $\pm$ 7.9  & 0.73 $\pm$ 0.11 & 0.06  $\pm$ 0.04 & 117.1 $\pm$ 9.1  & 0.0388 & $\surd$ \\
\multicolumn{3}{l}{ \bf{Core Values with Non-core Upper Limits}}  \\
 \hline
 \\
RXJ1317.1-3821  & $<$18.5  & 11.0 $\pm$ 1.7  & 1.0            & 0.28 $\pm$ 0.53 & 20.9 $\pm$ 3.2 & 0.2567 & $\surd$ \\
\multicolumn{3}{l}{ \bf{Non-core Values with Core Upper Limits}}  \\
 \hline
 \\
A3363          & 770.6 $\pm$ 11.3 & $<$46  & 0.69 $\pm$ 0.01 & 0.2 & 72.9 & 0.1265 & $\surd$ \\  
A3396          & 467.3 $\pm$ 6.5  & $<$46  & 1.60 $\pm$ 0.02 & 0.2 & 72.9 & 0.1784 & $\surd$ \\  
A2871          &  17.4 $\pm$ 2.5  & $<$6.4 & 1.80 $\pm$ 0.70 & 0.2 & 10.2 & 0.1218 & $\surd$ \\  
S301           &  20.2 $\pm$ 3.0  & $<$9.4 & 0.89 $\pm$ 0.56 & 0.2 & 14.9 & 0.0229 & $\surd$ \\  
S41            & 1811.4 $\pm$ 16.3 & $<$46 & 0.90 $\pm$ 0.04 & 0.2 & 72.9 & 0.0498 & x \\  
A3330          & 880.0 $\pm$ 12.1 & $<$46  & 0.69 $\pm$ 0.01 & 0.2 & 72.9 & 0.0918 & x \\  
A3911          & 337.5 $\pm$ 9.4  & $<$46  & 1.08 $\pm$ 0.04 & 0.2 & 72.9 & 0.0965 & x \\  
S861           & 273.3 $\pm$ 14.2 & $<$46  & 0.69 $\pm$ 0.05 & 0.2 & 72.9 & 0.0508 & x \\  
A3360          & 195.4 $\pm$ 10.6 & $<$46  & 0.76 $\pm$ 0.06 & 0.2 & 72.9 & 0.0853 & x \\  
A3728          & 127.8 $\pm$ 6.8  & $<$22.5 & 0.99 $\pm$ 0.08 & 0.2 & 35.7 & 0.0977 & x \\  
RXJ2143.9-5637 & 246.8 $\pm$ 11.4 & $<$33.7 & 1.18 $\pm$ 0.06 & 0.2 & 53.4 & 0.0819 & x \\  
S547           & 120.9 $\pm$ 4.3  & $<$46  & 1.0  $\pm$ 0.2  & 0.2 & 72   & 0.0515 & x \\  
A4023          & 120.6 $\pm$ 4.1  & $<$46  & 1.0  $\pm$ 0.2  & 0.2 & 72.9 & 0.1941 & x \\  
RXJ2018.4-4102 &  93.3 $\pm$ 5.8  & $<$3.0 & 0.96 $\pm$ 0.10 & 0.2 &  4.8 & 0.0188 & x \\  
\multicolumn{3}{l}{ \bf{Upper Limits on both Core and Non-core}}  \\
 \hline
 \\
RXJ2124.3-7446  & $<$63.4    & $<$45.9 & 1.0 & 0.2 & 72.7 & 0.0583 & $\surd$ \\
S540            & $<$16.9    & $<$12.3 & 1.0 & 0.2 & 19.5 & 0.0358 & $\surd$ \\
RXJ0303.7-7752  & $<$16.8    & $<$12.1 & 1.0 & 0.2 & 19.2 & 0.2769 & $\surd$ \\
S927            & $<$56.1    & $<$40.5 & 1.0 & 0.2 & 64.2 & 0.0584 & x \\
S592            & $<$45.7    & $<$33.0 & 1.0 & 0.2 & 52.4 & 0.2236 & x \\
RXJ0658.5-5536  & $<$37.4    & $<$28.0 & 1.0 & 0.2 & 44.4 & 0.2969 & x \\
RXJ0738.1-7506  & $<$37.1    & $<$27.7 & 1.0 & 0.2 & 43.9 & 0.1110 & x \\
RXJ2031.8-4037  & $<$34.3    & $<$25.7 & 1.0 & 0.2 & 40.7 & 0.3396 & x \\
A3718           & $<$26.8    & $<$20.1 & 1.0 & 0.2 & 31.9 & 0.1332 & x \\
S405            & $<$23.2    & $<$17.4 & 1.0 & 0.2 & 27.6 & 0.0616 & x \\
RXJ2224.4-5515  & $<$20.8    & $<$15.5 & 1.0 & 0.2 & 24.6 & 0.0790 & x \\
RXJ0217.2-5244  & $<$19.1    & $<$14.3 & 1.0 & 0.2 & 22.7 & 0.3410 & x \\
RXJ0322.2-5310  & $<$18.9    & $<$14.1 & 1.0 & 0.2 & 22.3 & 0.0775 & x \\
A3736           & $<$17.2    & $<$12.9 & 1.0 & 0.2 & 20.4 & 0.0487 & x \\
S849            & $<$16.5    & $<$12.3 & 1.0 & 0.2 & 19.5 & 0.0531 & x \\
RXJ2254.0-6315  & $<$15.6    & $<$11.7 & 1.0 & 0.2 & 18.5 & 0.2114 & x \\
A3158           & $<$14.1    & $<$10.5 & 1.0 & 0.2 & 16.6 & 0.0577 & x \\
RXJ2023.4-5535  & $<$13.3    & $<$9.9  & 1.0 & 0.2 & 15.7 & 0.2381 & x \\
\end{longtable}
\twocolumn

\section[]{OIII Additional Breakdowns}


\onecolumn
\begin{longtable}[c]{@{}lcccrrrrr@{}} 
\caption{Extra decompositions for the 28 sources which fit the initial selection criteria but for various reasons were not in the Main Sample but do have observations for [OIII]. NGC1275 included twice here to illustrate the huge variation of its core component.  The sources from the Main Sample+ for which [OIII] flux measurements were available are: A2495 ({\it [OIII] $<$0.1 $\times10^{-15}$ erg s$^{-1}$ cm$^{-2}$}), A2052 ({\it 5.9$\pm$0.3}), A2199 ({\it 2.4$\pm$0.3}), Hercules-A ({\it 5.0$\pm$0.4}), A1795 ({\it 2.0$\pm$0.5}), RXJ0132.6-0804 ({\it 4.2$\pm$0.2}), A2415 ({\it 1.3$\pm$0.3}), RXJ2043.2-2144 ({\it 0.4$\pm$0.2}), RXJ1317.1-3821 ({\it 0.25$\pm$0.1}), A2597 ({\it 3.8$\pm$0.3}), RXJ0338.7+0958 ({\it 4.4$\pm$1.6}), A2566 ({\it 0.4$\pm$0.2}), A3934 ({\it 0.3$\pm$0.3}), A3698 ({\it 0.42$\pm$0.2}), RXJ1206.5+2810 ({\it $<$0.5}), MACS1532.9+3021 ({\it 1.4$\pm$0.1}), Z8197 ({\it 0.5$\pm$0.1}), A1835 ({\it 1.5$\pm$0.1}), A795 ({\it 0.7$\pm$0.2}), A1991 ({\it 0.4$\pm$0.3}), Z2701 ({\it 0.1$\pm$0.1}), A115 ({\it 0.2$\pm$0.2}), Z3916 ({\it 0.2$\pm$0.1}), Z3179 ({\it 0.8$\pm$0.3}), RXJ0331.1-2100 ({\it 0.45$\pm$0.1}), A3112 ({\it 6.7$\pm$0.5}), AS555 ({\it 4.75$\pm$0.1}), A1668 ({\it 1.7$\pm$0.2}), A2072 ({\it 0.2$\pm$0.1}), A2580 ({\it 0.2$\pm$0.1}), A3378 ({\it 0.45$\pm$0.2}), RXJ2014.9-2430 ({\it 4.7$\pm$0.2}), A2734 ({\it 0.34$\pm$0.2}), RXJ1947.2-7623 ({\it 0.35$\pm$0.2}), A3992 ({\it 0.05$\pm$0.05}), AS805 ({\it 0.15$\pm$0.1}), A3638 ({\it 0.5$\pm$0.2}), AS384 ({\it 0.4$\pm$0.2}), Z7160 ({\it $<$0.1}), A1930 ({\it $<$0.3 }), RXJ2129.6+0005 ({\it $<$0.2}), Z1121 ({\it $<$0.5}), A1084 ({\it $<$0.2}), Z808 ({\it $<$0.2}), RXJ1720.1+2638 ({\it 0.2$\pm$0.1}), Z8276 ({\it 1.6$\pm$0.2}), A2204 ({\it 3.8$\pm$0.2}), A478 ({\it 0.4$\pm$0.3}), RXJ1320.1+3308 ({\it 1.2$\pm$0.3}), A1664 ({\it 1.5$\pm$0.2}), A2009 ({\it 0.3$\pm$0.2}), A2390 ({\it 1.0$\pm$0.2}), A2634 ({\it 1.9$\pm$0.9}), RXJ0107.4+3227 ({\it 2.1$\pm$0.5}), A262 ({\it 2.2$\pm$0.4}), Z8193 ({\it 1.4$\pm$0.3}), RXJ1750.2+3505 ({\it 0.7$\pm$0.2}), RXJ0352.9+1941 ({\it 1.4$\pm$0.2}), RXJ1715.3+5725 ({\it 1.4$\pm$0.3}), RXJ1733.0+4345 ({\it 0.6$\pm$0.2}), A291 ({\it 0.3$\pm$0.1}), RXJ0751.3+5012 ({\it 1.6$\pm$0.6}), A646 ({\it 0.3$\pm$0.2}), A1204 ({\it 0.1$\pm$0.1}), RXJ1442.2+2218 ({\it 0.7$\pm$0.4}), A1885 ({\it 0.4$\pm$0.2}), Z235 ({\it 0.4$\pm$0.2}), A2665 ({\it 0.1$\pm$0.1}), A1361 ({\it 0.9$\pm$0.2}), RXJ0439.0+0520 ({\it 1.3$\pm$0.1}), A4059 ({\it 3.4$\pm$0.2}), A3017 ({\it 1.95$\pm$0.1}), A496 ({\it 1.65$\pm$0.1}), RXJ1504.1-0248 ({\it 7.0$\pm$0.4}), RXJ1558.4-1410 ({\it 5.35$\pm$0.4}), A1348 ({\it 2.15$\pm$0.1}), A11 ({\it 2.5$\pm$0.1}), A3581 ({\it 1.1$\pm$0.2}), RXJ0543.4-4430 ({\it 0.3$\pm$0.2}), RXJ1524.2-3154 ({\it 1.5$\pm$0.2}), A3880 ({\it 2.0$\pm$0.3}), RXJ2213.1-2753 ({\it 1.15$\pm$0.1}), A2667 ({\it 0.25$\pm$0.1}), A3639 ({\it 1.6$\pm$0.2}), RXJ1304.3-3031 ({\it 0.425$\pm$0.2}), RXJ1315.4-1623 ({\it 0.25$\pm$0.1}), AS851 ({\it 2.2$\pm$0.2}), A2746 ({\it 0.15$\pm$0.1}), AS701 ({\it 0.05$\pm$0.03}), RXJ1539.5-8335 ({\it 0.6$\pm$0.2}), RXJ1931.6-3354 ({\it 1.0$\pm$0.2}) and A3605 ({\it 0.15$\pm$0.1}).} \label{OIII_EXTRA_FITS} \\
  \hline\hline
   Cluster & Non-core (at 1~GHz) & Core (at 10~GHz) & OIII ($\times10^{-15}$ erg s$^{-1}$ cm$^{-2}$) & \\
           &    (mJy)            &  (mJy)           &                &             &           &        \\
  \hline\hline
\endfirsthead
\multicolumn{1}{l}{{\tablename} \thetable{} -- Continued} \\
  \hline\hline
   Cluster & Non-core (at 1~GHz) & Core (at 10~GHz) & OIII ($\times10^{-15}$ erg s$^{-1}$ cm$^{-2}$ & \\
           &    (mJy)            &  (mJy)           &                &             &           &        \\
  \hline\hline
  \\
\endhead
  \multicolumn{3}{l}{{Continued on Next Page\ldots}} \\
\endfoot
  \\ \hline \hline
\endlastfoot
 \\
M87        & 263000 $\pm$ 50000 & 951.1 $\pm$ 47.5     & 21.0 $\pm$ 3.5   \\
Cygnus-A   & 2500000 $\pm$ 500000 & 856.3 $\pm$ 35.3   & 100  $\pm$ 10    \\
NGC1275    & 4000 $\pm$ 400     & 16000 $\pm$ 1600     & 200  $\pm$ 20    \\
NGC1275    & 4000 $\pm$ 400     & 6000  $\pm$  600     & 200  $\pm$ 20    \\
3C295      & 29957 $\pm$ 548    & 4.4   $\pm$ 0.1      & 1.05 $\pm$ 0.05  \\
AS463      & 6870.2 $\pm$ 864.5 & 57.4 $\pm$ 1.9       & 1.3  $\pm$ 0.2   \\
RXJ0745.0+3312 & 289.9 $\pm$ 8.0  & $<$39.0            & 1.65 $\pm$ 0.1   \\
RXJ1000.5+4409 & 6.4 $\pm$ 1.6 &  $<$2.7               & $<$0.1           \\
A2270      &  $<$31.6         & 140.2 $\pm$ 44.2       & 1.06 $\pm$ 0.03  \\  
A2292      &  $<$94.1           & 34.9 $\pm$ 4.1       & 0.2  $\pm$ 0.2   \\
RXJ1350.3+0940 &  $<$10.0         & 264.6 $\pm$ 44.9   & 4.5  $\pm$ 0.3   \\
RXJ1938.3-4748 &  $<$8.2  & $<$6.1                     & 0.06 $\pm$ 0.04  \\
A3866          &  $<$9.9  & $<$3.6                     & 0.05 $\pm$ 0.03  \\
AS1020         &  $<$5.1  & $<$3.8                     & 0.04 $\pm$ 0.03  \\
A3574W         &  $<$13.2   & $<$5.9                   & 0.08 $\pm$ 0.06  \\
A368           &  $<$7.6         & $<$3.6              & 0.07 $\pm$ 0.03  \\
\end{longtable}
\twocolumn


\end{document}